\documentclass{IEEEtran}

\IEEEoverridecommandlockouts

\usepackage[english]{babel}
\usepackage[utf8]{inputenc}
\usepackage[T1]{fontenc}

\usepackage{microtype}
\usepackage{acronym}
\usepackage{setspace}
\usepackage{blindtext}
\usepackage{siunitx}
\DeclareSIUnit{\bel}{B}
\DeclareSIUnit{\belmilliwatt}{Bm}
\DeclareSIUnit{\dBm}{\deci\belmilliwatt}
\usepackage{comment}
\let\tmpsubsubsection\subsubsection
\let\tmpparagraph\paragraph
\usepackage[compact]{titlesec}
\titlespacing{\section}{1pt}{2ex}{1ex}
\titlespacing{\subsection}{1pt}{1.5ex}{1ex}
\def\subsubsection{\tmpsubsubsection}
\def\paragraph{\tmpparagraph}
\usepackage{titlecaps}
\Addlcwords{or the if a an of for on and to -off off in not by via is we it how}
\usepackage{xargs}
\usepackage{soul}
\usepackage{marginnote}

\usepackage[nocompress]{cite}

\usepackage[justification=raggedright,singlelinecheck=false,size=footnotesize]{caption}
\usepackage[justification=raggedright,singlelinecheck=false,size=footnotesize]{subcaption}
\usepackage[pdftex]{graphicx}
\usepackage{epstopdf}
\usepackage{xcolor}
\usepackage{color}
\usepackage{paralist}
\usepackage{pgfplots}
\usepgfplotslibrary{groupplots,dateplot}
\pgfplotsset{compat=newest}
\pgfplotsset{every axis/.append style={
		label style={font=\normalsize},
		tick label style={font=\normalsize}  
}}
\graphicspath{{Figures/}}
\DeclareGraphicsExtensions{.pdf,.jpeg,.png,.jpg}
\usepackage{tikz}
\usepackage{tikzscale}
\usepackage{scalerel}
\usetikzlibrary{patterns,
	arrows,
	plotmarks,
	shapes.arrows,
	spy,
	svg.path,
	shapes.multipart,
	positioning,
    decorations,
    calc,
    decorations.markings,
    matrix
}
\tikzset{add reference/.style={insert path={%
			coordinate [pos=0,xshift=-0.5\pgflinewidth,yshift=-0.5\pgflinewidth] (#1 south west) 
			coordinate [pos=1,xshift=0.5\pgflinewidth,yshift=0.5\pgflinewidth]   (#1 north east)
			coordinate [pos=.5] (#1 center)                        
			(#1 south west |- #1 north east)     coordinate (#1 north west)
			(#1 center     |- #1 north east)     coordinate (#1 north)
			(#1 center     |- #1 south west)     coordinate (#1 south)
			(#1 south west -| #1 north east)     coordinate (#1 south east)
			(#1 center     -| #1 south west)     coordinate (#1 west)
			(#1 center     -| #1 north east)     coordinate (#1 east)   
}}}

\usepackage{amsthm,amsmath,amssymb,amsfonts,mathrsfs}
\usepackage[long,c2]{optidef}
\usepackage{nicefrac}
\usepackage{bbm,bm}
\usepackage{mathtools}
\usepackage{dsfont}

\usepackage{varwidth}
\usepackage{multirow}
\usepackage{booktabs}
\usepackage[linesnumbered,ruled,vlined]{algorithm2e}

\usepackage{array}
\usepackage{enumitem}
\usepackage{enumerate}
\setlist[description]{leftmargin=.5cm}

\usepackage{url}
\usepackage{orcidlink}
\usepackage{hyperref}
\hypersetup{
	colorlinks,
	linkcolor={blue},
	citecolor={blue},
	urlcolor={blue}
}
\usepackage[capitalize,nameinlink]{cleveref}

\let\tmpsection\section
\def\section#1{\tmpsection{\titlecap{#1}}}
\let\tmpsubsection\subsection
\def\subsection#1{\tmpsubsection{\titlecap{#1}}}
\newcommand{\myParagraph}[1]{{\boldmath \bf #1.}}
\makeatletter
\newcommand{\remove}[1]{\@bsphack\@esphack}
\newcommand{\arxiv}[2]{#2\@bsphack\@esphack}
\makeatother

\newcommand{\algoname}{VREM-FL\xspace}
\newcommand{\codesign}{\mbox{computation-scheduling} \mbox{co-design}\xspace}
\newcommand{\eg}{\emph{e.g.,}\xspace}
\newcommand{\ie}{\emph{i.e.,}\xspace}
\newcommand{\vs}{\emph{vs.}\xspace}

\definecolor{comp}{RGB}{255,204,170}
\definecolor{idle}{RGB}{175,198,233}
\definecolor{comm}{RGB}{175,233,198}

\newcommand{\lr}{\left(}
\newcommand{\rr}{\right)}

\newcommand{\lb}{\left\lbrace}
\newcommand{\rb}{\right\rbrace}

\newcommand{\Real}[1]{ { {\mathbb R}^{#1} } }

\DeclarePairedDelimiter\ceil{\lceil}{\rceil}
\DeclarePairedDelimiter\floor{\lfloor}{\rfloor}

\newcommand{\norm}[1]{\left\lVert#1\right\rVert}

\theoremstyle{plain}

\theoremstyle{definition}

\newtheorem{prob}{Problem}

\theoremstyle{remark}
\newtheorem{rem}{Remark}
\newcommand{\bit}{\begin{itemize}}
\newcommand{\eit}{\end{itemize}}
\newcolumntype{?}{!{\vrule width 1pt}}

\newcommand{\rounds}{\mathcal{T}}
\newcommand{\vset}[1][]{\mathcal{V}_{#1}}
\newcommand{\client}{vehicle\xspace}
\newcommand{\clients}{vehicles\xspace}
\newcommandx{\param}[3][1={},2={},3={}]{\theta%
    \ifx#2=={} ^{#1} \else _{#3,#2}^{#1} \fi}
\newcommand{\paramset}{\Theta}
\newcommand{\loss}{\ell}
\newcommand{\cost}{\mathcal{L}}
\newcommand{\location}[2]{x_{#1}^{#2}}
\newcommand{\bitrate}{h}
\newcommand{\bitraterem}[2]{\bitrate_{#1}^{#2}}
\newcommand{\REM}{\gamma}
\newcommand{\vsetsch}[1]{\vset[#1]^S}
\newcommand{\vsetsccard}[1]{M_{#1}^S}
\newcommand{\latencymax}{K_{\text{max}}}

\newcommand{\slots}[1]{\mathcal{K}_{#1}}
\newcommand{\compslot}[2]{a_{#1}^{#2}}
\newcommand{\txslot}[2]{b_{#1}^{#2}}
\newcommand{\compslots}[2]{T_{\text{cpu},#1}^{#2}}
\newcommand{\compslotsmin}{T_{\text{cpu}}^{\text{min}}}
\newcommand{\txslots}[2]{T_{\text{tx},#1}^{#2}}
\newcommand{\latency}[2]{K_{#1}^{#2}}
\newcommand{\txbits}[2]{B_{#1}^{#2}}
\newcommand{\bits}{B}
\newcommand{\bs}[1]{s_{#1}}
\newcommand{\gproxy}[2]{\Gamma(#1,#2)}
\newcommand{\phaseone}{centralized optimization\xspace}
\newcommand{\phasetwo}{local customization\xspace}
\newcommand{\phasethree}{centralized scheduling\xspace}
\newcommand{\vcost}[2]{C_{#2}^{#1}}
\newcommand{\fair}[2]{F_{#1}^{#2}}
\newcommand{\freq}[2]{\phi_{#1}^{#2}}
\newcommand{\aoi}[2]{A_{#1}^{#2}}
\newcommand{\priority}[2]{p_{#1}^{#2}}
\newcommand{\wtx}{w_{\mathrm{tx}}}
\newcommand{\txrate}{\text{tx}_{\text{rate}}}

\addto\captionsenglish{}
\addto\captionsenglish{}

\makeatletter
\let\oldnl\nl% Store \nl in \oldnl
\newcommand{\nonl}{\renewcommand{\nl}{\let\nl\oldnl}}% Remove line number for one line
\makeatother

\newcommand{\blue}[1]{{\color{blue}#1}}

\newcommand{\red}[1]{{\color{red}#1}}
\makeatletter
\newcommand{\review}[2][\@empty]{\@bsphack\@esphack #2\xspace}
\makeatother
\newcommand{\linkToPdf}[1]{\href{#1}{\blue{(pdf)}}}
\newcommand{\linkToPpt}[1]{\href{#1}{\blue{(ppt)}}}
\newcommand{\linkToCode}[1]{\href{#1}{\blue{(code)}}}
\newcommand{\linkToWeb}[1]{\href{#1}{\blue{(web)}}}
\newcommand{\linkToVideo}[1]{\href{#1}{\blue{(video)}}}
\newcommand{\linkToMedia}[1]{\href{#1}{\blue{(media)}}}
\newcommand{\award}[1]{\xspace} % omit awards
\newcommand{\orcid}[1]{\href{https://orcid.org/#1}{\includegraphics[scale=0.03]{orcid}}} % link to ORCID
\addto\extrasenglish{}
\addto\extrasenglish{}
\addto\extrasenglish{}
\addto\extrasenglish{}

\title{
	\titlecap{
		VREM-FL: mobility-aware computation-scheduling co-design for vehicular federated learning
	}
}

\author{Luca~Ballotta\,\textsuperscript{\orcidlink{0000-0002-6521-7142}}, %
	Nicol\`o~Dal~Fabbro\,\textsuperscript{\orcidlink{0000-0002-5325-2792}}, %
	Giovanni~Perin\,\textsuperscript{\orcidlink{0000-0002-7333-3004}},~\IEEEmembership{Member,~IEEE}, %
	Luca~Schenato\,\textsuperscript{\orcidlink{0000-0003-2544-2553}},~\IEEEmembership{Fellow,~IEEE}, %
	Michele~Rossi\,\textsuperscript{\orcidlink{0000-0003-1121-324X}},~\IEEEmembership{Senior~Member,~IEEE}, %
	and~Giuseppe~Piro\,\textsuperscript{\orcidlink{0000-0003-3783-5565}},~\IEEEmembership{Member,~IEEE} %
    \thanks{Copyright \copyright\, 2024 IEEE. Personal use of this material is permitted. However, permission to use this material for any other purposes must be obtained from the IEEE by sending a request to pubs-permissions@ieee.org.}
    \thanks{This work was supported in part by the Italian Ministry of Education, University and Research (MIUR) through the PRIN Project n. 2017NS9FEY ``Realtime Control of 5G Wireless Networks'', by the European Union under the Italian National Recovery and Resilience Plan of NextGenerationEU, partnership on “Telecommunications of the Future” (PE0000001 -- program “RESTART”) and by the EU project ROBUST-6G (GA no. 101139068). 
    Views and opinions expressed in this work are of the authors and may not reflect those of the funding institutions.}
    \thanks{\textit{Corresponding author:} Giovanni Perin (giovanni.perin.1@unipd.it). Luca Ballotta, Nicol\`o Dal Fabbro, and Giovanni Perin contributed equally to this work.}
    \thanks{Luca Ballotta
		is with the Delft Center for Systems and Control, Delft University of Technology, 2628 CD Delft, The Netherlands
		(e-mail: l.ballotta@tudelft.nl).}
    \thanks{Nicol\`o Dal Fabbro
		is with the department of Electrical and Systems Engineering, University of Pennsylvania, 19104, USA
		(e-mail: ndf96@seas.upenn.edu).}
	\thanks{Giovanni Perin,
		Luca Schenato,
        and Michele Rossi
		are with the Department of Information Engineering, University of Padova, 35131 Padova, Italy
		(e-mail: \{giovanni.perin.1; l.schenato; michele.rossi\}@unipd.it). Michele Rossi 
		is also with the Department of Mathematics ``Tullio Levi-Civita,'' University of Padova, 35121 Padova, Italy.}%
	\thanks{Giuseppe Piro
		is with the Department of Electrical and Information Engineering, Politecnico di Bari, 70125 Bari, Italy, 
		and with the Consorzio Nazionale Interuniversitario per le Telecomunicazioni, 43124 Pisa, Italy
		(e-mail: giuseppe.piro@poliba.it).}
}

\acrodef{BS}{base station} 
\acrodef{CPU}{central processing unit}
\acrodef{GD}{gradient descent}
\acrodef{SGD}{stochastic gradient descent}
\acrodef{FL}{\emph{federated learning}}
\acrodef{FedAvg}{\emph{federated averaging}}
\acrodef{ML}{machine learning}
\acrodef{5G}{fifth generation}
\acrodef{REM}{radio environment map}
\acrodef{SUMO}{simulator of urban environment}
\acrodef{LS}{least squares}
\acrodef{AoI}{age-of-information}
\acrodef{mIoU}{mean intersection over union}
\acrodef{SNR}{signal-to-noise ratio}
\acrodef{SINR}{signal-to-interference-plus-noise ratio}
\acrodef{OFDMA}{orthogonal frequency-division multiple access}

\begin{document}
	
	\maketitle
	%!TEX root = ../main.tex

\begin{abstract}
	\boldmath
	Assisted and autonomous driving are rapidly gaining momentum and will soon become a reality. 
	\remove{Among their key enablers, }
	Artificial intelligence and machine learning are \review{regarded as key enablers}  \remove{expected to play a prominent role,} thanks to the massive amount of data that smart vehicles will collect from onboard sensors.
	\remove{In this domain, }
	Federated learning is one of the most \remove{effective and} promising techniques for training global machine learning models while preserving data privacy of vehicles and optimizing communications resource usage.
	In this article, 
	we propose \review{vehicular radio environment map federated learning (\algoname)}, 
	a \codesign for vehicular federated learning that combines mobility of \clients with 5G radio environment maps. 
	\algoname jointly optimizes \review{learning performance of} the global model \remove{learned at the server} and wisely allocates communication and computation resources. 
	This is achieved by orchestrating local computations at the vehicles in conjunction with \remove{the} transmission of their \review{local models} in an adaptive and predictive fashion, by exploiting radio channel maps. 
	The proposed algorithm can be tuned to trade training time for radio resource usage. 
	\review{Experimental results demonstrate that} \remove{the efficacy of utilizing radio maps}  \algoname outperforms literature benchmarks for both a linear regression model (learning time reduced by $28\%$) and a deep neural network for semantic image segmentation (doubling the number of model updates within the same time window).
\end{abstract}

\begin{IEEEkeywords}
	5G, Federated Learning, optimization, \acs{REM}, resource management, scheduling, vehicular networks.
\end{IEEEkeywords}
    \setlength{\textfloatsep}{3pt}
	%!TEX root = ../main.tex

\section{Introduction}
\label{sec:intro}

\IEEEPARstart{V}{ehicular} \review{communications are expected to grow dramatically in the next few years according to the 5G Automotive Association, with applications requiring data rates of several tens of \si{\giga\bit}/\si{\second}~\cite{5gaa2023vehicular}.} 
Siemens estimated that the data generated by autonomous vehicles will range from $3$ to $40$~\si{\giga\bit}/\si{\second} depending on the autonomy level~\cite{siemens2021avdata},
which amounts to up to $19$ \si{\tera\byte} every hour. 

\begin{figure}
    \centering
    \setlength{\fboxsep}{0pt}%
    \subfloat{\fbox{\includegraphics[width=.9\linewidth]{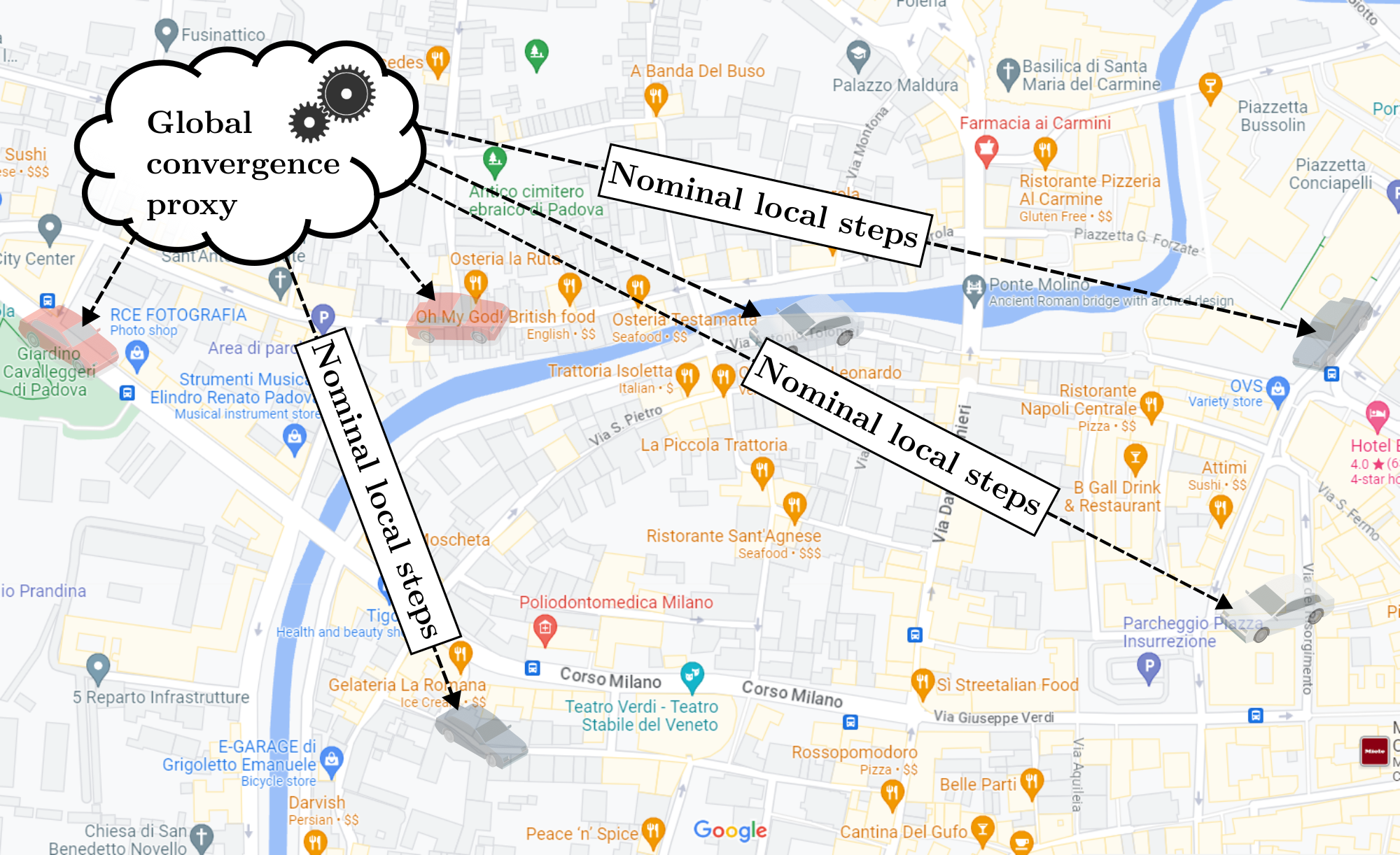}}}\\
    \vspace{1mm}
    \subfloat{\fbox{\includegraphics[width=.9\linewidth]{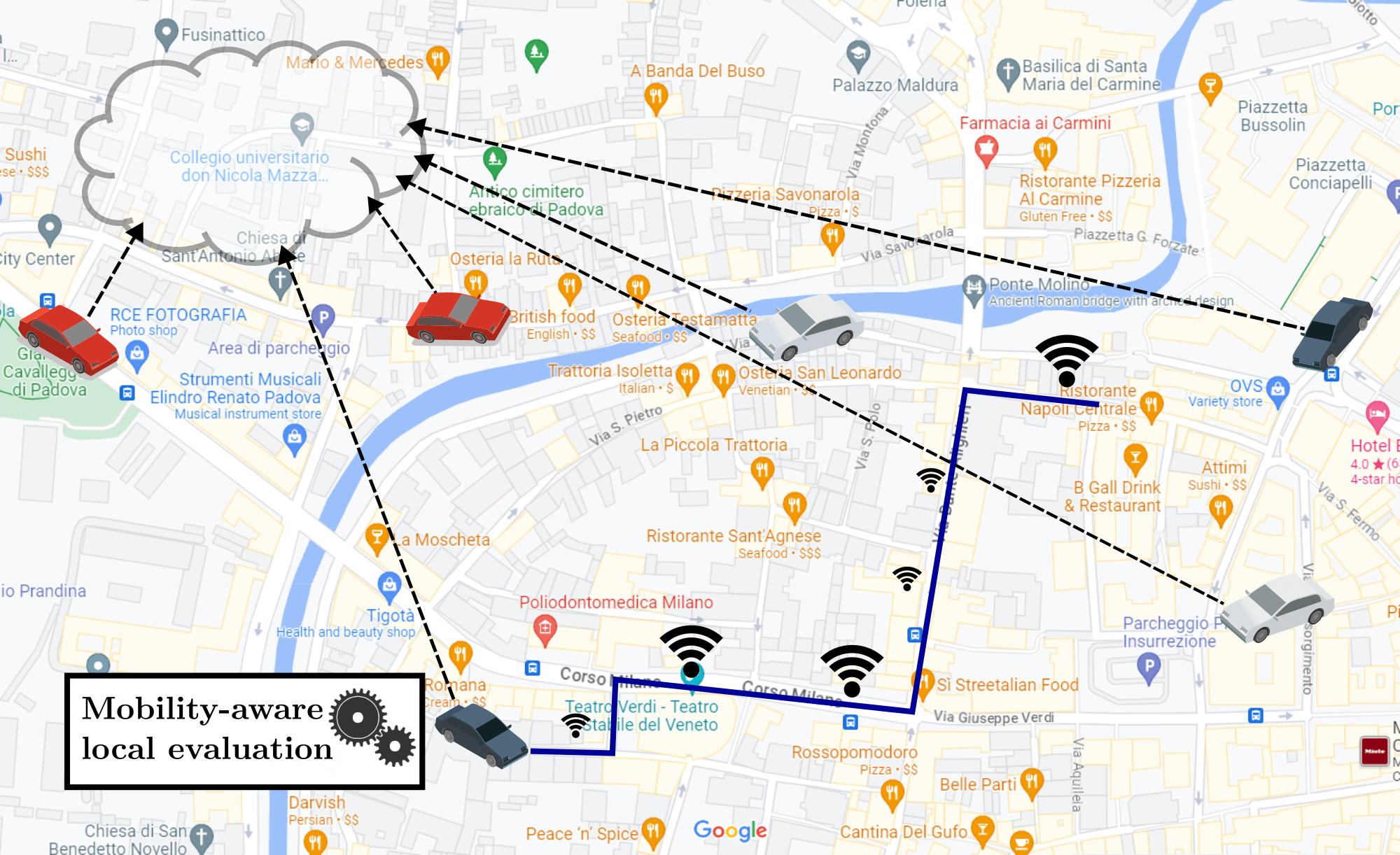}}}\\
    \vspace{1mm}
    \subfloat{\fbox{\includegraphics[width=.9\linewidth]{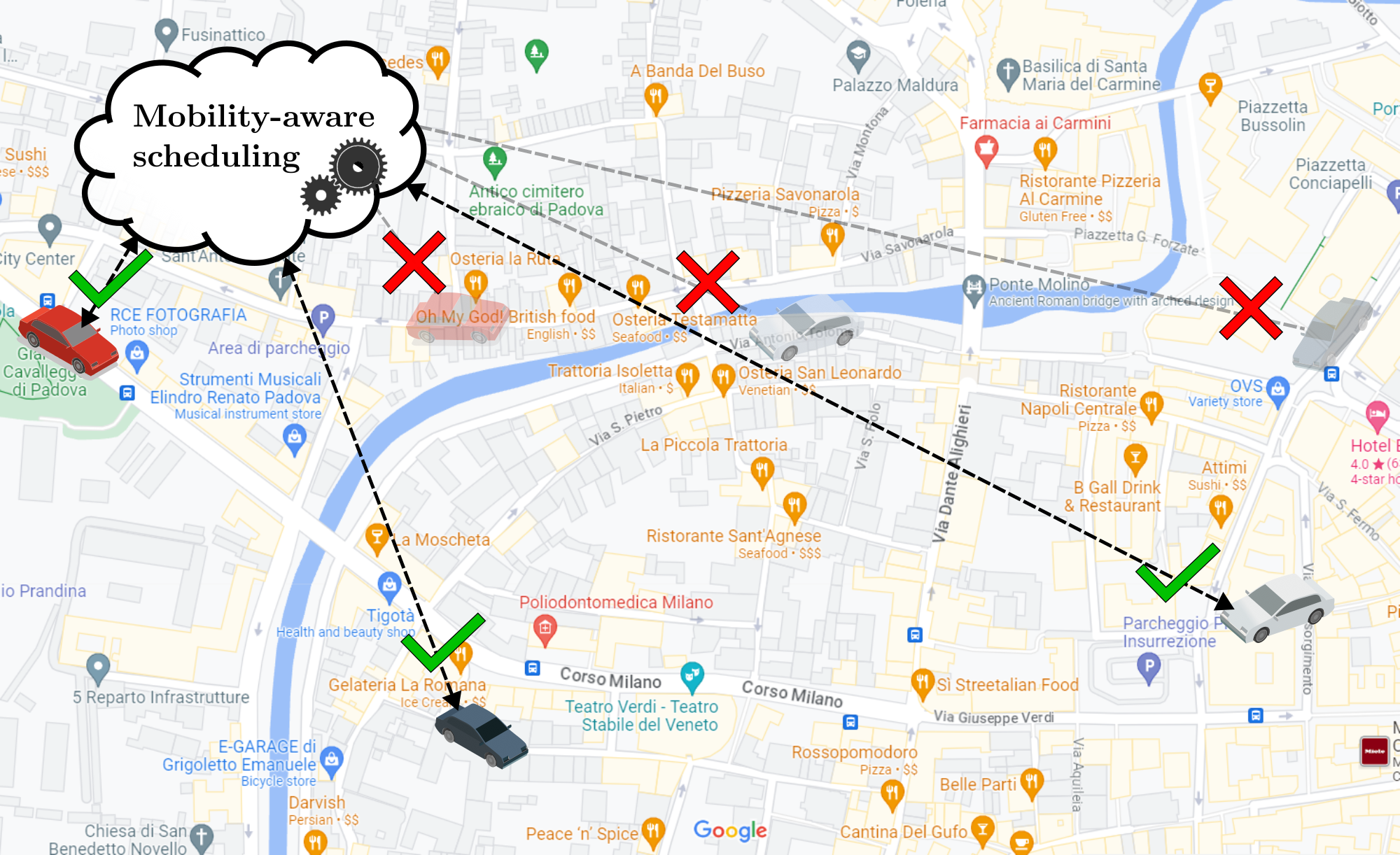}}}
    \vspace{1mm}
    \caption{\review[1]{Illustration of the main phases of \algoname, a mobility-aware co-design for allocation of communication and computation resources in vehicular \acs{FL}.}}
%    	\algoname runs before each learning iteration and comprises three phases. 
    	%\review[1]{In the first phase (top box), the aggregator estimates and broadcasts the optimal number of steps for local updates at the \clients.\remove{ that is computed based on a proxy for global training convergence} In the second phase (middle box), the \clients refine their local steps and, using private mobility information, independently \remove{estimate the experienced channel quality they will experience in the near future and} communicate to the aggregator an estimated \textit{participation cost} that encompasses usage of communication resources and training performance. In the third phase (bottom box), the aggregator combines the received costs with centrally available information about fairness of participation across \clients, scheduling those with the highest priority\remove{ in the current round}.}
    \label{fig:cover}
\end{figure}

The huge availability of data in modern vehicular networks paves the way to training, possibly at runtime, large deep learning models\review{~\cite{MLfor6G}}. 
Nowadays, this is mostly accomplished via \ac{FL}, 
which is the leading solution to train neural networks from decentralized datasets.
\ac{FL} offers several advantages, such as a simple and flexible aggregation phase, and it preserves privacy because end-user data are never transmitted,
but only the model weights are sent from the end-users to the central aggregator. 

Remarkably, despite the extensive amount of work available on communication-constrained~\cite{konevcny2016federated, mitra2021linear} and channel-aware~\cite{chen2020joint, chen2020convergence} \ac{FL} algorithms, scheduling policies for vehicular networks that jointly consider mobility, communication/channel resources, and learning aspects are currently lacking. 

In the present article, these aspects are {\it jointly} tackled for the first time, proposing vehicular \ac{REM} federated learning (\algoname), an \ac{FL} scheduler that orchestrates the computation and the transmission of local models from the end users (the vehicles) to the central model aggregator (at the roadside network). \algoname exploits the availability of \acp{REM} to pick the best instants for the vehicles' local model transmission. This is possible because \acp{REM} are stable over time~\cite {bi2019rem, dalfabbro2022rem} as they depend on static obstacles, such as buildings in urban environments. Hence, their knowledge can be combined with information on the planned user routes to improve \ac{FL} transmission schedules. The optimization criteria correspond to minimizing the channel resources that are wasted due to transmitting when the channel conditions are poor, and to meeting a deadline for the global model update. 

\review[1]{\algoname organizes resource allocation into the three stages shown in \autoref{fig:cover}: 
%\begin{inparaenum}[(1)]
(top box) the central orchestrator, based on channel and computation resources, decides how many \clients \remove{are allowed to} participate in the current round, proposes an initial number of \remove{iterations} steps to update their local models, \remove{(a number of steps for their local updates)} and sets a maximal round latency (deadline);
(middle box) all \clients, based on their private planned routes, learning metrics, and available \acp{REM}, independently estimate the channel quality they will experience in the near future, adjust the number of local steps, \remove{received by the orchestrator} and decide when to transmit their own models. Then, \remove{Here, the \clients independently estimate the channel quality that they will experience in the near future and} they communicate to the aggregator an estimated \textit{participation cost} that encompasses the usage of communication and training resources;
(bottom box) the orchestrator \remove{, upon receiving the \clients' feedback,} combines the received \clients' costs with centrally available information about fairness of participation across \clients, scheduling those with the highest priority. The scheduled \remove{selected} \clients \remove{subsequently their local models and} send their updated local models to the orchestrator, which finally refines the global model.
%\end{inparaenum}
}

\subsubsection*{Contributions}
Our contributions are summarized as follows.
\begin{itemize}
    \item We propose a novel \codesign tailored to vehicular \ac{FL},
    \algoname.
    The knowledge of \clients' routes and \acp{REM} % for scheduling decisions, 
    is leveraged to optimize the communication resources used by vehicles,
    \review{ensuring high learning performance while balancing training time and network resources.
    To the best of our knowledge,
    this is the first co-design of resource allocation in vehicular federated learning that exploits mobility and channel information supplied by \acp{REM} to account for time-varying channel quality, as typically experienced by traveling vehicles.}
%    usage in presence of mobility.
    \item %\marginnote{We may want to change this with reference to new experiment}
    We simulate a realistic environment using the street map of the city of Padova, Italy, and the popular \ac{SUMO}~\cite{sumo}. \review{We also evaluate the performance of VREM-FL on a real-world mobility dataset of taxi cabs in Rome, Italy~\cite{crawdad-roma-taxi}.}
    On these maps, 
    \acp{BS} are deployed according to typical parameters of \ac{5G} cellular systems~\cite{grassi2018massive}, to provide communications services to the \clients. 
    To capture realistic settings where channel quality measures are available at a limited number of locations, 
    \acp{REM} are obtained via Gaussian process regression~\cite{dalfabbro2022rem, muppirisetty2015spatial}.
    \item \algoname is evaluated via an extensive simulation campaign comprising 
    \begin{inparaenum}[(1)]
        \item controlled experiments with a \ac{LS} toy example and
        \item a realistic scenario \remove{involving the execution} 
        \review{on deep learning for a real-world} semantic segmentation task with the popular dataset \texttt{ApolloScape}~\cite{huang2018apolloscape}. 
    \end{inparaenum}
    %\marginnote{Maybe we can remove literature benchmarks from the contribution}
    Also, we compare \algoname with literature benchmarks such as \remove{vanilla} \ac{FedAvg}~\cite{mcmahan2017} and the algorithms in~\cite{GunduzComputationTime}, where clients are selected based on a fairness metric,
    \review{and in~\cite{chen2024efficient}, where client selection is based on channel gain estimates at the beginning of learning rounds.}
\end{itemize}

\subsubsection*{Organization of the Article}
In \autoref{sec:related-work},
the state-of-the-art is presented. 
\autoref{sec:high-level} provides a general overview of \algoname. 
In \autoref{sec:setup}, the system model is presented, including the \ac{FL} setup and the radio environment settings. 
The problem is defined in~\autoref{sec:problem-formulation}, 
while \algoname is detailed in~\autoref{sec:algorithms}. 
Numerical simulation results demonstrating the effectiveness of \algoname are presented in~\autoref{sec:results}. 
Finally, conclusions are drawn in~\autoref{sec:conclusion} along with future research directions.
	%!TEX root = ../main.tex

\section{Related work}\label{sec:related-work}

In this section, we review the state-of-the-art on FL over wireless networks, focusing on existing computation-communication co-design methods and vehicular FL. We also discuss relevant approaches leveraging REMs for wireless resource management, underlining the novelty of our work.
%The emerging applications of machine learning in mobile networks, the bandwidth limitations and the growing concern about data privacy have boosted the interest in adopting the privacy preserving federated learning (FL) framework \cite{mcmahan2017} for the novel mobile edge computing scenarios (refs...)

\subsection{User scheduling for FL in wireless networks}
In the last few years, many research works have investigated scheduling techniques for \ac{FL} over wireless networks \cite{TWC_wirelessFL_SCH_WN4, SAC_WN2Edge, CDC_submodular_WN3, DeviceSchedWN1, IoTBudget}. 
%The objective of the scheduling is reducing the overall FL cost in terms of both training time, network resources and energy consumption.
In~\cite{SchedulingPoor}, scheduling policies were experimented with by simulating users connected to different access points together with different \ac{SINR} thresholds. 
Recent work~\cite{PALORA} proposed PALORA, a scheduling method jointly considering fairness with respect to local models, \ac{SNR} levels, and resource blocks utilization. 
Along the same lines, 
paper~\cite{GunduzComputationTime} considered asynchronous updates and a computation-communication tradeoff, 
and proposed scheduling strategies based on \emph{\ac{AoI}} and a \emph{fairness} metric in wireless networks. In \cite{DataQuality}, scheduling was performed based on data quality metrics to jointly solve a user selection and bandwidth allocation problem. A scheduling approach based on data selection was proposed in \cite{SchedGradientNormDataSel}, where the gradient norm was used to compute a metric that relates learning efficiency with data selected by users for local training. The authors of~\cite{WCLs_chAndLearnSched} proposed a scheduling policy for \acs{FL} over an \ac{OFDMA} scheme where scheduling decisions are based on learning accuracy and channel quality.
\review{Reference~\cite{gradientChannelAwareScheduling} addressed scheduling for over-the-air \acs{FL} based on local gradient,
channel conditions,
and energy consumption to improve learning accuracy and convergence.}

\subsection{FL in vehicular networks}
%The Internet of Vehicles (IoV) is becoming a new paradigm in which vehicles are connected entities, equipped with considerable computing power and connectivity requirements for a variety of applications \cite{IoVMeneghello, IoVMain, IntelligentVehiclesEx}. 
There is a growing interest in machine learning solutions for problems related to vehicular networks, given the increasing need for data-driven algorithms in intelligent transportation systems~\cite{TowardsIntelligentVNs_ML}. Applications examples include wireless resource management \cite{RadioResourceAlloc},  traffic flow prediction \cite{TrafficFlowPrediction}, vehicle-based perception for autonomous driving \cite{FL_IoT_IoVs, AutonomousFL_Conf},
to name a few.

With respect to the federated training setup, some works have focused on \emph{static} vehicular federated learning \cite{Static_1_Survey, Static_2Approach}, with emphasis on scenarios involving parking lots \cite{Static_3PLs} and related issues, like parking space estimation.

Conversely, cooperative training of machine learning models in scenarios involving user mobility, such as vehicular networks, is a key open research topic \cite{SurveyHighMobility, VehFL_bennis}. 
In this context, relevant works have considered vehicular FL for, e.g., image classification~\cite{ImageFLMobility}, proactive caching~\cite{proactiveCaching}, and object detection~\cite{ObjectDetection6G}.

%In \cite{LithiumSOH}, quantization techniques are adopted to perform federated learning in vehicular networks for onboard battery status prediction with experimental data. In \cite{RoadClassification}, a federated learning approach based on vehicle-to-vehicle communications is exploited for  via {Lidar point clouds collected as part of the nuScenes \cite{nuScenes} large-scale autonomous driving dataset}.

Other recent works have investigated the allocation and optimization of network resources in vehicular FL~\cite{FL_veh_opt_1, FL_veh_opt_2, FL_veh_opt_3, FL_veh_opt_4, FL_veh_opt_5}. Specifically, \cite{FL_veh_opt_3} analyzed the impact of vehicular mobility on wireless transmissions, proposing wireless network optimization techniques. The authors of~\cite{FL_veh_opt_5} considered cache queue optimization at the edge server, while other related scheduling techniques were proposed in~\cite{FL_veh_opt_1, FL_veh_opt_2}. Reference~\cite{FL_veh_opt_4} focused on tuning the number of local iterations based on mobility awareness in relation to short-lived connections with the base stations.
The authors of~\cite{SchedResourceAlloc} studied the adoption of FL in an IoV scenario where vehicles communicate with 5G base stations, exploiting context information such as cell association and {mobility prediction} together with the related channel quality for users' scheduling.
\review{The authors of~\cite{vehicleSelectionResourceAllocation} jointly scheduled participating vehicles and optimized computational resources.
While this approach addresses both learning accuracy and latency/energy restrictions,
the decision-making is centralized and may violate the privacy of vehicles. 
A similar centralized approach was proposed in~\cite{chen2024efficient}, 
where the authors select \clients based on a minimum latency criterion by considering the CPU frequency and the latest known/estimated channel gain. 
In this work, model compression was also used to further reduce latency and energy consumption.}

\review{Despite these efforts, the recent survey~\cite{FLvehicularSurvey} remarks that open challenges such as efficient allocation of communication and computational resources, learning-based selection of participating vehicles to enhance training, privacy and security issues, and robustness to noisy training samples are still to be addressed.}
In fact, none of the works mentioned above combines {\it mobility patterns} and {\it radio environmental awareness} to optimize the learning performance of FL, while at the same time optimizing wireless network resources. 
Our work is the first to jointly consider these aspects \review{under private vehicular mobility} which, as we shall see, leads to sizeable advantages over previous solutions \review{and tackles the challenges mentioned in~\cite{FLvehicularSurvey} on resource optimization, vehicle selection, and privacy}.

%A related approach has been proposed in \cite{SchedResourcePlatooning}, where authors consider FL performed in vehicle platooning networks, in which a leader vehicle serves as parameter server aggregating the local gradients of the participating platoon vehicles via V2V communication. A scheduling strategy is proposed to handle the trade-off between the number of participating vehicles and the learning time.

\subsection{REMs for wireless resource management}
A radio environment map (REM) is a geographic database of average communication quality metrics. In recent years, REMs have been proposed as an effective tool to manage wireless resources \cite{bi2019rem}, and 
advocated for predictive resource allocation in~\cite{predictiveREM} and for handover management in 5G networks in~\cite{handoverREM}. 
In 5G systems with massive multiple-input multiple-output (mMIMO) transmission, REMs have been adopted for energy-efficient design~\cite{energyREM}, inter-cell interference coordination~\cite{chikha2022radio}, beam management~\cite{beamManagementREM}, and cell-edge users throughput improvement via dynamic point blanking~\cite{throughputREM}. Although REMs have been used for several resource management applications in wireless communications, the present work is the first to exploit them for network resources optimization in FL and, specifically, in vehicular FL. 
	%!TEX root = ../main.tex

\section{VREM-FL in a nutshell}\label{sec:high-level}

In this article, we are concerned with the optimal resource allocation for \ac{FL} tasks executed by vehicles that travel within an urban environment and by an edge server that acts as the centralized aggregator of their models. In what follows, we interchangeably use the terms ``vehicle'' and ``client''  depending on the role that we would like to emphasize. %to stress the role that we are interested in.
			%!TEX root = ../main.tex

\subsection{The problem: resource allocation for vehicular FL}\label{sec:high-level-problem}
We aim to minimize the objective cost \review{associated with \acs{FL}}
\begin{equation}\label{eq:high-level-objective-cost}
	\mathrm{cost} = \review[1]{\mathrm{cost}_\text{loss} + \mathrm{cost}_\text{latency} + \mathrm{cost}_\text{channel}} \tag{OBJ}
\end{equation}
where $\review[1]{\mathrm{cost}_\text{loss}}$ measures \review{performance of the global model},
$\review[1]{\mathrm{cost}_\text{latency}}$ is the \review{training time}, %according to pre-defined task specifications,
and $\review[1]{\mathrm{cost}_\text{channel}}$ \review{refers to network resources used by the clients to upload their local models.}
%\review{is the time clients use to upload local models through the channel.}

\review{We consider a threefold space of intervention}. 
The first aspect is \textit{scheduling clients} during training.
Due to limited communication \remove{and computation} resources,
\review{all clients cannot transmit their local models at every learning round.}
\remove{it is typically infeasible for all the clients to simultaneously transmit their local model at every learning round. }
Hence, at each round, a subset of clients is selected to update the global model, and a \textit{scheduling strategy} is utilized \review{to choose} which ones.

\review{The second aspect is the \textit{local computation} performed by the scheduled clients\remove{ (vehicles)}.}
To ensure convergence of an FL algorithm to an accurate global model, 
the clients must carefully choose \review{the number of} \remove{\ac{SGD} steps} \review{descent steps to update} their local models. 
\remove{\review{Under heterogeneous clients,}
\remove{and if the clients independently choose their number of local steps at each round,}
\review{optimally tuning local steps is challenged} by the combinatorial nature of the \remove{associated optimization} \review{problem}.}
%\red{In fact, both a complex learning task and the client scheduling make it challenging to retrieve an optimal choice, because the problem typically requires to tune the number of local steps in a time-varying and client-specific fashion according to the subset of scheduled \clients and their local loss functions.}

The third aspect \review{of our design} is the \textit{transmission of local models} from the \clients to the edge server. 
%While the former two aspects mainly affect the $\mathrm{loss}$ and the $\mathrm{latency}$ of the FL task, the $\mathrm{channel \ usage}$ during training mostly depends on how network resources are used by \clients when they upload their models to the server. 
\review{Sending the local models} as soon as the local \review{updates are done (\emph{greedy transmission behavior}) yields} the fastest training. 
However, this strategy \review{ignores} the channel status, which depends on mobility and radio channel.
It descends that a greedy behavior \review{need not make the best} use of the available channel resources. The proposed policy is mobility and channel aware \review{and allows for a more profitable usage of channel resources.}
\remove{through its use, channel resources are more profitably exploited}
\review{It leads to \remove{benefits in terms of} reduced transmission energy and channe usage}, thus releasing resources for other users that also need to exploit wireless transmissions.
			%!TEX root = ../main.tex

\subsection{The solution: VREM-FL}\label{sec:high-level-solution}

To minimize the cost~\eqref{eq:high-level-objective-cost} in a vehicular scenario, we propose Vehicular \ac{REM}-based Federated Learning (\algoname), a co-design that jointly optimizes the threefold decision-making introduced above. An intuitive description of our co-design algorithm is illustrated in~\autoref{fig:cover}. In this section, we provide a high-level overview of how \algoname works and defer the detailed explanation to~\autoref{sec:algorithms}.

\algoname runs \review{before} each learning round and consists of three phases, which correspond to the boxes in~\autoref{fig:cover}. 
%\begin{inparaenum}
%\item\label{phase1} 

During the \textit{\phaseone} phase (top box), the edge server (\ie the orchestrator) computes and broadcasts the number of local steps that all the \clients should perform to achieve the fastest training convergence. % and broadcasts this information. 
This computation uses a proxy for global convergence assuming that all the scheduled \clients run the same number of local steps~\cite{li2019convergence}. %While this is clearly suboptimal in general, it largely simplifies the design and 
For this, the server does not need to know the \clients' local cost functions. This phase is formalized in~\autoref{sec:centralized-problem-1}.

%\item\label{phase2}
In the \textit{\phasetwo} phase (middle box), each \client adjusts the number of local steps based on a local convergence criterion, to trade local training speed for global accuracy. Hence, the \client optimizes the communication of the local model by opportunistically delaying its transmission, seeking the best trade-off between the components $\review[1]{\mathrm{cost}_\text{latency}}$ and $\review[1]{\mathrm{cost}_\text{channel}}$ in the cost~\eqref{eq:high-level-objective-cost}.
To perform this optimization, the \client leverages knowledge of (1) the channel quality via the availability of an estimated \ac{REM} and (2) its planned trajectory, as explained in~\autoref{sec:decentralized-problem}. 
No training is performed at this stage.
Instead, the choice of computation and communication is used in the following phase to discern which \clients are the best candidates to be scheduled at the current round.

%\item\label{phase3}
In the \textit{\phasethree} phase (bottom box), the edge server receives the estimated costs for participating in the round from all the \clients. These costs depend on the local decision-making performed during the previous phase and are combined with global information available at the server that measures fairness of updates among the \clients. 
This, in the form of a combination of \ac{AoI} and scheduling frequency metrics, 
ensures appropriate participation of all the \clients throughout the training. The algorithm executed at the server to make scheduling decisions is described in~\autoref{sec:centralized-problem-2}.
%\end{inparaenum}
	%!TEX root = ../main.tex

\section{System Model}\label{sec:setup}

In this section,
we present the setup considered throughout the article.
In~\cref{sec:system-framework,sec:background-fl}, we introduce an \ac{FL} task solved by \clients that move within an urban environment served by BSs. In~\autoref{sec:rem}, we describe the channel model that \clients use to assess whether they should join a round and optimize their next transmissions to the server. % how convenient it is for them to take part in .
			%!TEX root = ../main.tex

\subsection{Vehicular federated learning}
\label{sec:system-framework}

We consider a set $\vset \doteq \{1, \dots, N\}$ of $N$ \clients that move within an urban environment.
The mobility area is served by $N_\text{bs}$ \acs{5G} \acp{BS}. 
\review{As \clients travel, 
	they collect data to improve tasks of interest \review{for assisted or autonomous driving},
	such as semantic segmentation for local navigation,
	pedestrian detection to enhance safety of road users,
	route optimization based on real-time traffic information,
	or allocation of wireless resources used by other tasks or \clients that share the network.}
Each \client $v\in\vset$ collects a local,
private dataset denoted as
%(which may be occasionally updated as the \client travels)
$\mathcal{D}_v$. 

To efficiently learn complex tasks from locally gathered data, 
\clients in $\vset$ are connected to an edge server running an \ac{FL} algorithm.
This allows the \clients to cooperatively learn a common \ac{ML} model without uploading their collected data to the server, 
which may be impractical through high data volume or undesirable because of privacy concerns.
%Moreover,
%this allows each \client to retain privacy of the collected data.

%\begin{rem}[Performing FL on traveling \clients]
\review{In principle,
the \clients may learn a model while they are \textit{static}, \eg parked.
This would ease resource management for \acs{FL},
such as scheduling of updates that could be performed under constant channel conditions.
However,
recent work~\cite{siemens2021avdata,1GBPerSecIntelligent} remarked that the sensing capabilities of autonomous vehicles may generate \si{\giga\byte}s of sensory data per second. 
The cost for storing such a massive amount of data
is high in terms of both energy consumption and storage capacity,
and it would be impractical and expensive for the vehicles to run \acs{FL} tasks after one or multiple trips.
\review[1]{This issue urges to run \acs{FL} tasks \emph{on traveling \clients}~\cite{VehFL_bennis, ObjectDetection6G}},
%This %, besides the possible need to actually learn a model as they move, % for time-critical tasks, 
%allows the \clients to %train local models while driving
%	--thus while they are collecting the data--
%	and 
which could
destroy training data right after using them and save on storage and energy consumption.
However,
including mobility into resource management is nontrivial.
The aggregator may not know routes of \clients,
which affect the experienced channel quality.
On the other hand,
the \clients may not know channel conditions across the whole environment (\eg a city) and share limited computation resources with other driving-related jobs.}
%Conversely, running the FL task on idle \clients after they have collected all data imposes high, if not impractical, storage requirements and energy consumption.
%\end{rem}
			%!TEX root = ../main.tex

\subsection{Preliminaries on federated learning}
\label{sec:background-fl}

An \acs{FL} task is described by the optimization problem
\begin{mini}
	{\param\in\paramset}
	{\mathcal{L}(\param) \doteq \loss\lr\param;\lb\mathcal{D}_v\rb_{v\in\vset}\rr + \lambda\norm{\param}^2}
	{\label{eq:FL-problem}\tag{FL}}
	{}
\end{mini}
where $\param$ is the model parameter,
$\loss$ is the loss function that depends on the dataset,
$\norm{\param}^2$ is a regularization term that, 
in words,
penalizes ``complex'' models,
and $\lambda$ is the regularization weight.
In the following,
%as standard in Machine Learning,
we refer to the total cost $\mathcal{L}(\param)$ as (regularized) loss. % for the sake of simplicity.
The size of the parameter $\param$ amounts to $\bits$ [\si{\bit}].
%The training data for~\eqref{eq:FL-problem} are split across \clients,
%and each \client $v$ has its own dataset $\mathcal{D}_v$ that may not be willing or capable to share.
\review{Problem~\eqref{eq:FL-problem} is tackled in an iterative fashion that alternates between \begin{inparaenum}[(1)]
	\item \clients updating their local models (\textit{local update}) and 
	\item the server aggregating all or some local models and sending the updated global model to all \clients.
\end{inparaenum}}

%Problem~\eqref{eq:FL-problem} has been extensively studied in literature
%and many algorithms have been proposed to solve it.
\review{We highlight up-front that our proposed co-design algorithms can be tailored to any choice of algorithm used to solve~\eqref{eq:FL-problem}.
	For instance,
	\algoname can accommodate both synchronous and asynchronous aggregations as explained in \cref{rem:sync-async}.
	\arxiv{}{This is possible because \clients can upload their local models within a given time window,
	and the global model is updated only after all models are received or the deadline expires.}
	Nonetheless,}
	for the sake of exposition and to ground the discussion,
	in the rest of this article we will assume that the \clients train their local models via \ac{GD} or \ac{SGD} and that the edge server runs \ac{FedAvg}~\cite{mcmahan2017}.
	\review{This choice is motivated by the simplicity and the popularity of this FL algorithm.
	Other learning and aggregation schemes,
	such as FedDrop~\cite{feddrop} or FedLin~\cite{mitra2021linear},
	require to minimally adjust our algorithms as described in \cref{rem:extensions}.}
A high-level snippet of vanilla \ac{FedAvg} (without client scheduling) is provided in~\cref{alg:fedavg},
where all clients perform $H$ \acs{SGD} steps for each local update.
%Each stages of local updates and global aggregation
%are repeated in an iterative fashion
%until a target accuracy is hit or a predefined deadline expires.
We refer to a cycle composed of local updates and global aggregation (\cref{alg:fedavg-server2clients,alg:fedavg-local-update,alg:fedavg-loop,alg:fedavg-client2server,alg:fedavg-aggregation})
as \textit{(learning) iteration} or \textit{(learning) round},
whose duration coincides with the time interval between two consecutive updates of the global model,
and the edge server runs FedAvg for $T$ rounds\remove{ in total}.
We denote by $\param[t]$ the global parameter at iteration $t$, and by $\param[t][t+H][v]$ the locally updated parameter of \client $v$ during round $t$ (before transmission to the aggregator).

\review[1]{As commonly assumed in the literature~\cite{PALORA, SchedulingPoor}}, each learning round has a deadline after which global aggregation is executed,
regardless of the status of local updates,
and a new round begins afterward.
Time is slotted into slots of duration $\tau$~[\si{\second}], which we consider the finest granularity to allocate resources in a time-varying fashion.
%The available time in each learning round has to be used for all operations of that round.
\review{Among the steps of \cref{alg:fedavg}},
our present work addresses local training (\cref{alg:fedavg-local-update}) and transmission of local models from \clients to the server (\cref{alg:fedavg-client2server}).
Hence,
we further assume that at every round the \clients have a deadline of $\latencymax$~[\si{\second}] to update their local models and to upload them to the server, % $t$,
corresponding to $\floor{\latencymax / \tau}$ time slots. % at iteration $t$ to
The set of time slots available during iteration $t$,
\review{when we allocate resources for training and transmission of each \client,}
is denoted by $\slots{t}$. In the following, we will refer to the local training performed by the \clients as \textit{computation} to highlight the allocation of computational resources.

\begin{algorithm}[tbh]
	\caption{Vanilla FedAvg~\cite{mcmahan2017}}
	\label{alg:fedavg}
	\DontPrintSemicolon
	\KwIn{Loss $\mathcal{L}$, rounds $T$, parameter $\param[0]$, local steps $H$.}
	\KwOut{Learned parameter $\param[t]$.}
	\For{$t = 0,1,\dots,T$}{
		server broadcasts global parameter $\param[t]$ to \clients;\;\label{alg:fedavg-server2clients}
		\ForEach{\client $ v\in\vset $}{\label{alg:fedavg-loop}
			train local model: $\param[t][t+H][v]\leftarrow\mathrm{SGD}\lr\param[t], H; \mathcal{D}_v\rr$;\;\label{alg:fedavg-local-update}
			transmit local parameter to server;\;\label{alg:fedavg-client2server}
		}
		server aggregates local models: $\param[t+1] \leftarrow \sum_{v\in\vset} w_v\param[t][t+H][v]$;\;\label{alg:fedavg-aggregation}
	}
\end{algorithm}

\begin{comment}
	For FedAvg,
	choosing the number of local descent steps of GD or SGD 
	for each \client is an important aspect that can enhance or degrade performance of~\eqref{eq:FL-problem}~\cite{fedavg}.
	On the one hand,
	few steps slow down the training convergence;
	on the other hand,
	too many steps pull each local model to the one that fits the local dataset of the corresponding \client,
	which need not coincide with the global minimum of~\eqref{eq:FL-problem}
	and hence leads to a biased global model with poor generalization capability.
\end{comment}
%Carefully choosing the number of local descent steps is one algorithmic contribution of this work,
%which we describe in detail in~\autoref{sec:problem-formulation}.
%a cost function depending on the local datasets.
			%!TEX root = ../main.tex

\subsection{Radio environment and Bitrate}\label{sec:rem}
%We assume that the \clients have knowledge about the (average) channel quality experienced across the traversed environment.
{The area served by the BSs
	features a heterogeneous channel quality depending on the network coverage at different geographical locations.}
\review{To assess channel quality across the urban environment,
	we use an \ac{REM}.}
This is a database
that links a (quantized) geographical location
to the (estimated) value of some channel quality metric. 
In this work, 
we are interested in \remove{knowledge of} the average bitrate associated with a location $x\in\Real{2}$.
\review{As so,}
we consider an REM that contains information about the \ac{SINR} experienced on average \review{at each location}. 
Given a wireless transmission setting and a geographical location $x$, the value of the SINR associated with $x$ contained in the REM can be used to infer the expected average bitrate experienced by a user located at $x$. 
This information is crucially used for resource orchestration by VREM-FL.
\review{In fact,
\clients may experience different channel quality depending on their real-time location.}
%This can be converted to a value of bitrate,
%which we consider as a measure of channel quality.
Formally,
we express the location-to-bitrate map as follows. Given a vehicle $v$ located at $x_t^v$ at time $t$, the corresponding estimated average bitrate $\bitrate_t^v$ is
\begin{equation}\label{eq:rem-map}
	\bitrate_t^v = \beta(\REM(x_t^v), \eta_t^v),
\end{equation}
\review{where $\REM(\cdot)$ is the location-to-SINR map encoded by the REM, $\beta(\cdot)$ is the SINR-to-bitrate map,
and $\eta_t^v$ is the bandwidth used by vehicle $v$ at time $t$.}

In practice,
we assume that an REM $\gamma$ of the environment is computed \textit{a priori} via some estimation technique~\cite{sato2017kriging, sato2021space, chowdappa2018distributed}
and is \remove{deployed} \review{stored} at the \ac{BS}, which broadcasts this information to the \clients \review{taking part in FL}.
%\marginnote{Edge server or BS?}
Because \acp{REM} are approximately constant across time~\cite {bi2019rem, dalfabbro2022rem},
\remove{as previously mentioned,}
\remove{we assume that}
they are sent to the \clients only once during FL training.
%\marginnote{I would remove this: a vehicle may need more than a single BS coverage info along their path in the next 1-2 minutes, I think they should receive an REM of an area wider than a single cell}
\review{When a handover occurs and a \client enters coverage area of a new \acs{BS},
	the latter sends its corresponding \acs{REM} to the \client,
	updating the channel information that the \client uses to allocate computation and transmission resources.}
We assume that the routes traversed by the \clients are planned in advance (at least partially) so that they know their respective trajectories in the near future.
Specifically, at time $t$, \client $v$ knows the locations it is about to traverse over the next $D$ time instants, denoted by $\location{t}{v},\dots,\location{t+D}{v}$, 
for some time horizon $D$.
This information is converted to estimated bitrate values $\bitraterem{t}{v},\dots,\bitraterem{t+D}{v}$ for a \client's trajectory,
where $\bitraterem{k}{v}=\beta\left(\REM(\location{k}{v}), \eta_t^v\right)$ as per~\eqref{eq:rem-map}.
This information is used in our resource-allocation algorithms as detailed in~\autoref{sec:problem-formulation}.

\begin{comment}
	\myParagraph{Computation-communication trade-off}
	With machine learning models of considerable complexity (e.g., neural networks) and in constrained hardware like the vehicles' ones, computation time is not negligible, may be time-varying and needs to be taken into account together with the quality of the local model, the quality of the data and the rate at which data are being collected by that specific device.
\end{comment}
	%!TEX root = ../main.tex

\section{Problem formulation}
\label{sec:problem-formulation}

We now formalize the \codesign problem at the core of our contribution.
We first define the design parameters for the considered decision-making (\autoref{sec:design-params}),
and then formally write the objective function~\eqref{eq:high-level-objective-cost} along with the overall optimization problem (\autoref{sec:optimization-problem}).
			%!TEX root = ../main.tex

\subsection{Design parameters}\label{sec:design-params}
Given problem~\eqref{eq:FL-problem} and an algorithm to solve it (FedAvg), 
\review{our space of intervention is}
\remove{we are concerned with} 
the threefold decision-making associated with the algorithm workflow discussed in~\autoref{sec:high-level-problem}. %that we assume available for design.

\subsubsection{Scheduling}
%Given such a constraint,
%the FL performance may crucially depend on which \clients are selected at each round,
%which calls for a careful \textit{scheduling} strategy.
%Designing a scheduling strategy that picks the best subset of \clients $\vsetsch{t}$ is the other design contribution of this work.
First, we design a \textit{scheduling strategy} to select the \clients that participate in each learning iteration. 
Scheduling is needed because the total amount of \clients involved in an \ac{FL} task is typically large and cannot be handled at once due to the limited communication resources available at the \acp{BS}.
It descends that only a (small) fraction of the \clients can be simultaneously served through the available bandwidth to ensure an acceptable quality of service. 
We denote the subset of \clients that participate in round~$t$ by $\vsetsch{t}\subset\vset$ and the maximum number of \clients that can be scheduled in round $t$ by $\vsetsccard{t}< N$, with $\lvert\vsetsch{t}\rvert\le\vsetsccard{t}$.
%In light of what discussed in the previous section,
%Given that FL tasks require both to learn an accurate final model and to end within a reasonable time,
%the scheduling relies also on the channel quality to favor those \clients that can send updates quickly.
%We use the REM in conjunction with the vehicular mobility to estimate the channel quality
%experienced by the \clients at each iteration.

\subsubsection{Computation}
%Choosing how to allocate the slots in $\slots{t}$ for computation and transmission at each \client is one design contribution of this work.
For each \client scheduled in a learning round, we consider two aspects for co-design.
First,  we allocate the amount of \textit{computation at the \client}, that is, the number of descent steps (of GD or SGD) that the \client performs during that round to train the local model. Through a careful choice of the number of local steps, we can effectively trade convergence speed of \ac{FedAvg} for the quality of the final global model. For each time slot $k\in\slots{t}$ available during \review{learning iteration} $t$, we denote by $\compslot{k}{v}\in\{0,1\}$ the \textit{computation decision} of vehicle $v$ for slot $k$: if $\compslot{k}{v}=1$, it means that $v$ performs a batch of local steps during slot $k$, otherwise no computation is carried out in that slot. %In particular, we assume %that \client $v$ can where is the , that $v$ runs $\bs{v}$ local steps (\textit{computation batch size}) during one time slot.
\review{The total number of time} slots used by \client $v$ for local \review{model update} during \review{learning round} $t$ is denoted by $\compslots{t}{v} \doteq \sum_{k\in\slots{t}}\compslot{k}{v}$. 
To make the training meaningful, we allocate at least $\compslotsmin\ge1$ slots for computation at every round.
%To this aim,
%we rely on a proxy of the convergence rate
%that can be used to approximately find the number of local steps that maximize the convergence speed.

\subsubsection{Communication}
After a \client has updated its local model, we optimize the \textit{transmission from \client to server}. Although the training time for \ac{FL} is trivially minimized if local models are immediately transmitted, %after they have been updated,
in this work we are additionally interested in an efficient allocation of network resources. 
This allows us \review{to reduce both channel bandwidth occupancy} 
\remove{the number of slots during which the  is reserved for}
\review{and transmission energy needed for \ac{FL}}.
\remove{In particular, w}
We assume that all \clients have constant transmission power, so that optimizing for communication resources is equivalent to transmitting where the \ac{SINR} is high.
We denote by $\txslot{k}{v}\in\{0,1\}$ the \textit{transmission decision} of vehicle $v$ in slot $k$.
If $\txslot{k}{v}=1$, 
then $v$ uses time slot $k$ to transmit its local model to the server;
otherwise,
no transmission occurs during slot $k$. 
The total time \client $v$ \review{takes} to \review{upload} its local model during round $t$ is denoted by $\txslots{t}{v} \doteq \sum_{k\in\slots{t}}\txslot{k}{v}$. 
Further, 
we denote by $\latency{t}{v}$ the total time elapsed from the beginning of local \review{training} \remove{model update} to \review{reception of the updated local model} \remove{the end of its transmission to} at the server,
which we name \textit{round latency} of \client $v$ \review{in learning round $t$}.
			%!TEX root = ../main.tex

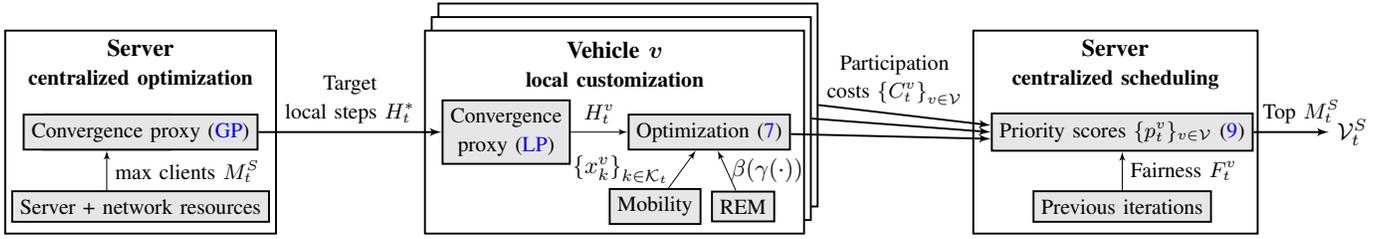
\begin{figure*}[t]
	\centering
	%!TEX root = ../main.tex

\tikzstyle{int}=[draw, fill=black!10, minimum size=1em, thick]
\newcommand{\topboxheight}{1.5}
\newcommand{\bottomboxheight}{.4}
\scalebox{.9}{
	\begin{tikzpicture}[node distance=1.5cm,auto,>=latex',every text node part/.style={align=center}]
		\draw [thick] (0,0) rectangle (4,3);
		\node at (2,2.5) (ph1) {\textbf{Server} \\ {\small \textbf{\phaseone}}};
		\node [int] at (1.5,\bottomboxheight) (resghost) {server};
		\node [int] at (2,\bottomboxheight) (res) {{\small Server + network resources}};
		\node [int] at (1.5,\topboxheight) (gpghost) {{\small Convergence}};
		\node [int] at (2,\topboxheight) (gp) {{\small Convergence proxy~\eqref{eq:global-proxy}}};
		\path[->] (resghost) edge node[right]{{\small max clients $\vsetsccard{t}$}} (gpghost);
		
		\draw [thick,fill=white] (6.4,.4) rectangle (12,3.4) [add reference=v1];
		\draw [thick,fill=white] (6.3,.2) rectangle (11.9,3.2) [add reference=v2];
		\draw [thick,fill=white] (6.2,0) rectangle (11.8,3) [add reference=v3];
		\node at (9.005,2.5) (ph2) {\textbf{Vehicle $\boldsymbol{v}$} \\ {\small \textbf{\phasetwo}}};
		\node [int] at (7.4,\topboxheight) (lp) {{\small Convergence} \\ {\small proxy~\eqref{eq:local-proxy}}};
		\node [int] at (10.4,\topboxheight) (opt) {{\small Optimization~\eqref{eq:local-communication-optimization}}};
		\path[->] (lp) edge node[above]{{\small $H_t^v$}} (opt);
		\node [int] at (9.6,\bottomboxheight) (traj) {{\small Mobility}};
		\node [int] at (10.9,\bottomboxheight) (rem) {{\small REM}};
		\path[->] (traj) edge node[left] {$\lb x_k^v\rb_{k\in\slots{t}}$\hspace{0.75mm}} (opt);
		\path[->] (rem) edge node[right] {$\!\beta(\REM(\cdot))$} (opt);
		
		\path[->,thick] (gp) edge node[above]{{\small Target} \\ {\small local steps $H_t^*$}} (lp);
		
		\draw [thick] (14.3,0) rectangle (18.5,3);
		\node at (16.4,2.5) (ph3) {\textbf{Server} \\ {\small \textbf{\phasethree}}};
		\node [int] at (16.5,\topboxheight) (priority) {{\small Priority scores $\{\priority{t}{v}\}_{v\in\vset}$~\eqref{eq:priority-score}}};
		\node [int] at (16.5,\bottomboxheight) (server) {{\small Previous iterations}};
		\path [->] (server) edge node[right]{{\small Fairness $\fair{t}{v}$}} (priority);
		
		\path[->,thick] (opt) edge ([yshift=-1mm]priority.west); 
		\path[->,thick] (v2 east) edge (priority.west);
		\path[->,thick] (v1 east) edge node[above]{\shortstack{\small Participation \\ {\small costs $\lb C_t^v\rb_{v\in\vset}$}}\hspace{3mm}} ([yshift=1mm]priority.west);
		
		\node at (19.9, \topboxheight) (schedule) {$\vsetsch{t}$};
		\path [->,thick] (priority) edge node[above]{{\small \hspace{3mm}Top $\vsetsccard{t}$}} (schedule);
	\end{tikzpicture}
}
	\caption{\textbf{Workflow of \algoname.}
		At the beginning of learning iteration $t$,
		the server runs \textit{centralized optimization} and transmits the globally optimal local steps to the clients.
		Then,
		each vehicle locally runs \textit{local customization} to fine-tune computation and communication resources to be used in the round.
		Finally,
		the server performs \textit{centralized scheduling} taking into account both feedback from the \clients and global participation information.
		}
	\label{fig:vrem-diagram}
\end{figure*}
			%!TEX root = ../main.tex

\subsection{Optimization problem}\label{sec:optimization-problem}
\begin{comment}
	\begin{equation}
		f(\pi,a_t, b_t) = \sum_{v\in\mathcal{V}}\sum_{t=1}^T \alpha E^{(v)}_t + \beta F_t^{(v)} + \gamma (T^{\text{tx},v}_{t}+T^{\text{cpu},v}_{t})
	\end{equation}
\end{comment}
We aim to jointly optimize the three costs in~\eqref{eq:high-level-objective-cost}. 
As discussed above, optimizing computation and scheduling resources reduces training loss $\review[1]{\mathrm{cost}_\text{loss}}$ and latency $\review[1]{\mathrm{cost}_\text{latency}}$ of the \ac{FL} task, 
while optimizing communication resources translates into efficient channel usage $\review[1]{\mathrm{cost}_\text{channel}}$ during training.

With a slight abuse of notation, we express $\review[1]{\mathrm{cost}_\text{loss}}$ as a function of both computation and scheduling (note that it also depends on the final parameters learned by FedAvg, $\param[T]$):
\begin{equation}\label{eq:number-of-rounds}
	\review[1]{\mathrm{cost}_\text{loss}} = \cost\lr\lb\vsetsch{t},\{\compslot{t}{v}\}_{v\in\vsetsch{t}}\rb_{t\in\rounds}; \param[T] \rr,\tag{TL}
\end{equation}
%To evaluate the former objective (training performance),
%we consider the value of the objective cost in~\eqref{eq:FL-problem}
%that is reached after a pre-defined maximum training time.
%Note that,
%even with a fixed training time,
%the total number of learning rounds $T$ varies depending on how resources are allocated.
%the total number of learning iterations needed to make the cost in~\eqref{eq:FL-problem} reach value $\epsilon$,
%which is a pre-defined parameter that we refer to as \textit{target accuracy}.
%Denoting by $\mathcal{L}^*$ the final value of the objective function of~\eqref{eq:FL-problem} that is reached within the allowed training time,
%we consider the following cost:
where $\compslot{t}{v}\doteq\lb\compslot{k}{v}\rb_{k\in\slots{t}}$ denotes all computation decisions of \client $v$ throughout round $t$ and $\rounds \doteq \{1,\dots,T\}$ gathers all learning rounds.
%Note that,
%while the design of transmissions is not tailored to the learning performance,
%the actual number of learning rounds $T$ depends on how both computation and communication are allocated.
%The expectation is taken w.r.t. random elements such as data and SGD descent steps and the bitrate values.
The term $\review[1]{\mathrm{cost}_\text{latency}}$ is upper bounded by $T\latencymax$ but varies depending on 
\begin{inparaenum}
	\item allocation of computation and communication resources, and
	\item the channel quality experienced by the scheduled \clients.
\end{inparaenum}
We formalize this as
\begin{equation}\label{eq:overall-latency}
	\review[1]{\mathrm{cost}_\text{latency}} = K\lr\lb\lb\compslot{t}{v},\txslot{t}{v};\bitraterem{t}{v}\rb_{v\in\vsetsch{t}}\rb_{t\in\rounds}\rr,\tag{OL}
\end{equation}
where $\txslot{t}{v}\doteq\lb\txslot{k}{v}\rb_{k\in\slots{t}}$ and $\bitraterem{t}{v}\doteq\lb\bitraterem{k}{v}\rb_{k\in\slots{t}}$ denote all transmission decisions of \client $v$ and %of \client $v$ throughout round $t$ and
bitrate values experienced by $v$ during round $t$, respectively.
Finally, 
we quantify the channel usage as the \review{total} time (number of \review{time} slots) the \clients reserve channel bandwidth to upload their local models to the server.
\review{This quantity amounts to summing transmission times $\txslots{t}{v}$ across all scheduled \clients and all learning rounds. 
	Because the transmission time $\txslots{t}{v}$ is defined by transmission decisions $\txslot{t}{t}$,
    which in turn depend on the bitrate $\bitraterem{t}{v}$ experienced by \client $v$ at round $t$,
	we express $\review[1]{\mathrm{cost}_\text{channel}}$ as}
\begin{equation}\label{eq:tx-slots}
	\review[1]{\mathrm{cost}_\text{channel}} = \sum_{t\in\rounds} \sum_{v\in\mathcal{V}_t^S}\txslots{t}{v}\lr \txslot{t}{v};\bitraterem{t}{v}\rr.\tag{CU}
\end{equation}
The total cost~\eqref{eq:high-level-objective-cost} addressed in our co-design amounts to
\begin{align}\label{eq:cost-function}
	\begin{split}
		\mathrm{cost}\lr\vsetsch{},\compslot{}{},\txslot{}{}; h, \param[T]\rr %&= \mathcal{L}^* + c_\text{tx}\sum_{t\in\rounds} \sum_{v\in\mathcal{V}_t^S} \txslots{t}{v}\\
		&=\begin{aligned}[t]
			&\cost\lr\vsetsch{},\compslot{}{}; \param[T] \rr\\
			&+ (1-\wtx) K\lr\compslot{}{},\txslot{}{};\bitraterem{}{}\rr\\
			&+ \wtx \sum_{t\in\rounds} \sum_{v\in\mathcal{V}_t^S} \txslots{t}{v}\lr \txslot{t}{v};\bitraterem{t}{v}\rr
		\end{aligned}
	\end{split}
\end{align}
where  $\vsetsch{}\doteq\{\vsetsch{t}\}_{t\in\rounds}$ denotes the full client schedule, %across all learning rounds,
$\compslot{}{}\doteq\{\compslot{t}{v}\}_{v\in\vsetsch{t},t\in\rounds}$, $\txslot{}{}\doteq\{\txslot{t}{v}\}_{v\in\vsetsch{t}.t\in\rounds}$, and $\bitraterem{}{}\doteq\lb\bitraterem{k}{v}\rb_{k\in\slots{t},t\in\rounds}$ gather all computation and transmission decisions, and bitrate values associated with scheduled \clients across rounds.
The weight $\wtx\in[0,1]$ trades $\review[1]{\mathrm{cost}_\text{latency}}$~\eqref{eq:overall-latency} for $\review[1]{\mathrm{cost}_\text{channel}}$~\eqref{eq:tx-slots}.
If $\wtx=0$,
only the training time is penalized in~\eqref{eq:cost-function};
if $\wtx=1$,
latency is neglected and usage of network resources is discouraged.
The learning cost~\eqref{eq:number-of-rounds} and latency cost~\eqref{eq:overall-latency} jointly depend on all scheduled clients,
while the resource cost~\eqref{eq:tx-slots} decomposes linearly across those.

Equipped with the mathematical definition~\eqref{eq:cost-function} of~\eqref{eq:high-level-objective-cost},
we are now ready to formalize the \codesign problem tackled in the rest of this work.

\begin{table}
	\caption{List of symbols used in this article.}
	\label{table-params}
	\begin{center}
		\footnotesize
		\begin{tabular}{lp{6.5cm}}
			\toprule
			$\vset$ & set of \clients\\
			\midrule
			$\bits~[\si{\bit}]$ & size of model parameter $\theta$\\
			\midrule
			$\tau~[\si{\second}]$ & duration of one time slot\\
			\midrule
			%	\item $T_t = K_t\tau$ maximum time duration of iteration $t$;
			$\latencymax~[\si{\second}]$ & maximum latency of every learning iteration\\
%			\midrule
%			$\slotsmax$ & maximum number of slots available for any iteration\\ % for iteration $t$\\
			\midrule
			$\compslotsmin$ & minimum number of computation slots at every iteration\\
			\midrule
%			$\mathcal{K}_t = \{1, ..., D_t\}$
			$\slots{t}$ & set of time slots available for iteration $t$\\
			\midrule
			$\bs{v}$ & local steps that vehicle $v$ runs in one time slot\\
			\midrule
			$\compslot{k}{v}$ & computation decision of vehicle $v$ for slot $k$ \\
			\midrule
			$\txslot{k}{v}$ & transmission decision of vehicle $v$ for slot $k$\\
			\midrule
			%    \item $k_t^{\text{tx}, v} \in \mathcal{K}_t$ slot in which transmission of vehicle $v$'s model starts, i.e., $b_{t, k_t^{\text{tx}, v}} = 1$ and $b_{t,k}^{(v)} = 0 \ \ \forall k \in \mathcal{K}_t : k < k_t^{\text{tx}, v}$.
%			$T^{\text{cpu},v}_{t} = \sum_{k \in\mathcal{K}_t} a_{t, k}^{(v)}$
			$\compslots{t}{v}$ & number of computation slots of vehicle $v$ in iteration $t$\\
			\midrule
%			$T^{\text{tx},v}_{t} = \sum_{k \in\mathcal{K}_t} b_{t,k}^{(v)}$
			$\txslots{t}{v}$ & number of transmission slots of vehicle $v$ in iteration $t$\\
			\midrule
			$\bitraterem{k}{v}~[\si{\bit}/\si{\second}]$ & bitrate experienced by vehicle $v$ in slot $k$\\
			\midrule
			%    The bitrate is directly related to the radio map and to the interference level.
%			$B_{t}^{(v)} = \tau \sum_{k \in\mathcal{K}_t} b_{t,k}^{(v)}h_{t,k}^{(v)}$
			$\txbits{t}{v}~[\si{\bit}]$ & bits that vehicle $v$ transmits during iteration $t$\\
			\midrule
			%    \item $B_{t, \text{max}}^{(v)} = \tau \sum_{k = T^{\text{cpu}}_\text{min}}^{K_t}h_{t,k}^{(v)}$ maximum number of bits that can be transmitted by vehicle $v$ in iteration $t$; %when starting the transmission at slot $k_t^{\text{tx}}$;
			%    \item $\tau_t^{\text{tx}, v} = \tau T^{\text{tx},v}_{t}$ transmission time of vehicle $v$ at iteration $t$
%			$K_t^{(v)} = \max \{k : b_{t,k}^{(v)} = 1\}$
			$\latency{t}{v}~[\si{\second}]$ & latency of vehicle $v$ in iteration $t$\\
			\midrule
%			$\vset$ & set of vehicles available for iteration $t$\\
%			\midrule
			$\vsetsch{t}$ & set of \clients scheduled for iteration $t$\\
			\midrule
			$\vsetsccard{t}$ & maximum number of \clients scheduled for iteration $t$\\
			\bottomrule
		\end{tabular}
	\end{center}
\end{table}

\begin{prob}[Optimal \codesign for vehicular FL]\label{problem:co-design}
	Given
	\begin{inparaenum}[(i)]
		\item a set of \clients $\vset$,
		%		\item an algorithm that solves~\eqref{eq:FL-problem};
		\item an REM of the environment $\REM$,
		\item an FL algorithm,
		\item model parameters $B, \latencymax, \compslotsmin$,
	\end{inparaenum}
	find
	\begin{inparaenum}[(1)]
		\item a \client schedule $\vsetsch{}$, %\doteq \{\vsetsch{t}\}_{t\in\rounds}$ with $\vsetsch{t}\in\vset$;
		\item computation decisions $a$, %$\compslot{}{} \doteq \{\{\compslot{k}{v}\}_{k\in\slots{t},v\in\vsetsch{t}}\}_{t\in\rounds}$,
%		with $\compslot{k}{v} \in\{0,1\}$;
		\item transmission decisions $b$, %$\txslot{}{} \doteq \{\{\txslot{k}{v}\}_{k\in\slots{t},v\in\vsetsch{t}}\}_{t\in\rounds}$
%		with $\txslot{k}{v} \in\{0,1\}$;
	\end{inparaenum}
	so as to optimize FL training and transmission resources:
	\begin{argmini!}
		{\vsetsch{},\compslot{}{},\txslot{}{}}
		{\mathrm{cost}\lr\vsetsch{},\compslot{}{},\txslot{}{};h,\param[T]\rr\protect\label{eq:co-design-objective}\tag{Pa}}
		{\protect\label{eq:co-design}}
		{\text{(P)}\quad \ }
		\addConstraint{\latency{t}{v}}{\le\latencymax\quad}{\forall v\in\vsetsch{t}, \forall t\in\rounds\protect\label{eq:co-design-constraint-max-latency}\tag{Pb}}
		%		\addConstraint{\compslots{t}{v}}{\ge\compslotsmin}{\forall v\in\vsetsch{t},\forall t=1\dots,T\protect\label{eq:co-design-constraint-min-comp}}
		%		\addConstraint{a_{t,k}^{(v)} +\frac{1}{k-1} \sum_{l = 1}^{k-1}b_{t, l}^{(v)}}{\le 1}{\quad \forall k\in\slots{t}, \forall t=1\dots,T
			%		\protect\label{eq:co-design-contraint-comp-tx-order}}
		%\addConstraint{\txbits{t}{v}}{\ge B\quad}{\forall v\in\vsetsch{t}, t\in\rounds \protect\label{eq:co-design-constraint-min-bits}\tag{Pc}}
		\addConstraint{\compslots{t}{v}}{\ge \compslotsmin \quad}
		{\forall v\in\vsetsch{t}, \forall t\in\rounds.\protect\label{eq:co-design-contraint-min-comp}\tag{Pc}}
	\end{argmini!}
\end{prob}

In words,
constraint~\eqref{eq:co-design-constraint-max-latency} ensures that each round ends within the pre-assigned deadline,
%constraint~\eqref{eq:co-design-constraint-min-bits} forces the scheduled \clients to transmit the model in time,
and constraint~\eqref{eq:co-design-contraint-min-comp} requires the scheduled \clients to perform a minimal number of descent steps.
The meaning of all symbols is provided in~\autoref{table-params}.
In the next section,
we propose co-design algorithms to solve~\cref{problem:co-design} that crucially rely %the proxies~\eqref{eq:global-proxy} and~\eqref{eq:local-proxy} of the convergence (performance metric)
%the that is estimated by \clients based on the REM and 
on vehicular mobility and the REM to estimate the channel quality experienced by \clients.

\section{Algorithms for co-design}
\label{sec:algorithms}

The \codesign problem (P) requires designing both local operations performed by \clients (computation and transmission) and global scheduling decisions the edge server makes. 
To efficiently tackle it, 
we propose a cascade procedure executed at the beginning of each round $t$,
involving three phases (split between edge server and \clients).
%that is,
%\textit{(i)} a \textit{centralized client scheduling},
%run by the global scheduler,
%and \textit{(ii)} a \textit{local computation-communication optimization},
%run by each vehicle.
%The full pipeline and is composed of three phases:
\begin{description}
	\item[Centralized optimization:] the edge server computes the optimal number of local steps $ H_t^* $ to be performed by the \clients in round $t$ and broadcasts them this value.
	\item[Local customization:] each \client refines its local steps and transmission (time slots to send the local model update),
	and sends its cost $ \vcost{v}{t} $ to participate in round $t$ to the server.
	\item[Centralized scheduling:] the server receives all the costs $\vcost{v}{t}$ and uses both such local information and global knowledge about the fairness of updates to select the subset of vehicles that will actually take part in the model update at round $t$.
%	which are subsequently triggered and actually start learning round $ t $.
\end{description}
\Cref{fig:vrem-diagram} provides a schematic representation of the workflow of \algoname with the three phases summarized above. 
In the following, 
the operations of \algoname are described in detail.
%First,
%we introduce a proxy that we use to estimate convergence speed and accuracy of FedAvg.
%Then,
%we provide the details of each of the phases listed above.
			%!TEX ROOT = ../main.tex

\subsection{Centralized optimization}\label{sec:centralized-problem-1}

In this phase, the edge server first sets the maximum number $\vsetsccard{t}$ of vehicles that are allowed to participate in round $t$. %according to available resources
%based on its available local resources (e.g., transmission channel resources and computational capability for model aggregation). %, such as computational capability for model aggregation and available bandwidth. 
Then, the server computes an approximate number of local steps to be performed by the scheduled \clients in order to speed up the training.
To evaluate at runtime the %performance-related term $\cost$ and the training time $K$ in the objective cost of~\eqref{eq:co-design},
relation between the number of local steps at the \clients and the training time,
%we need to evaluate the convergence of FedAvg online.
%To this aim,
%the expected number of rounds~\eqref{eq:number-of-rounds} online,
%we use two proxies that account respectively 
%for the global accuracy when all \clients performs the same computation (\textit{global proxy})
%and for the accuracy of the local model at a \client (\textit{local proxy}).
%As for the ,
we use the proxy for \textit{global FL convergence} proposed in~\cite{li2019convergence}. 
This proxy assumes that $M$ clients are scheduled at every round, and that each of them performs $H$ local steps at every local update.
The proxy is expressed \review{in~\cite[Eq.~(1)]{li2019convergence}} as
\begin{equation}\label{eq:global-proxy}
	T\epsilon \approxeq \gproxy{H}{M}\doteq\frac{C}{H}+\left(1+\frac{1}{M}\right)H\tag{GP}
\end{equation}  %$$\mathbb{E}\left[F(\bm w_t) - F^*)\right] \sim \frac{A+(1+\frac{1}{M})E^2}{t}$$
where $\epsilon$ is the estimated accuracy in $T$ rounds,
%$T$ is the number of iterations to reach accuracy ,
\ie $\cost(\param[T])\le\epsilon$,
and $C$ is a constant that depends on the data distribution.

The server homogeneously chooses $H_t^*$ as the minimizer of the proxy~\eqref{eq:global-proxy} with respect to $H$,
setting $M=\vsetsccard{t}$
and assuming that all \clients run $H_t^*$ local steps at all iterations:
\begin{argmini}
	{H}
	{\gproxy{\vsetsccard{t}}{H}.}
	{\label{eq:nominal-local-steps}}
	{H_t^*=}
\end{argmini}
This first subproblem is unconstrained and convex. % because the objective cost~\eqref{eq:global-proxy} 
%is convex in $H$. %is the only variable and $M$ is a parameter. 
Its solution is
\begin{equation}
    \label{eq:optimal-H}
    H_t^* = \sqrt{C\left(1+\frac{1}{\vsetsccard{t}}\right)^{-1}}.
\end{equation}
This reveals how the optimal number of local steps $H_t^*$ depends on the scheduled clients.
It is a strictly increasing and concave function of $M$ ($M \ge 1$) that saturates to $\sqrt{C}$ for $M \to +\infty$\arxiv{}{ (horizontal asymptote)}{}.

After computing $H_t^*$, the server broadcasts this value to all the \clients for the second phase.
%A snippet of this subroutine is given in~\cref{alg:centralized-algo-step-1}.

%\begin{itemize}
%	%	\item $C_1$ depends on gradient boundedness only, $C_2$ on the sum of i) gradient boundedness, ii) local SGD variance, iii) amount of non-iidness.
%	\item $H$ is the number of local steps performed by the \clients (assumed constant across rounds);
%	\item $M$ is the number of \clients scheduled in each iteration;
%	\item $T$ is the number of iterations to reach accuracy $\epsilon$,
%	\ie $\mathcal{L}^*\le\epsilon$ after $T$ rounds;
%	\item $C_1$ and $C_2$ are constants that depend on the overall dataset.
%	%	\item $\epsilon$ is the user-defined target precision,
%	%	that is,
%	%	the final value of the objective function of~\eqref{eq:FL-problem}.
%\end{itemize}
%As explained later,
%we use the proxy~\eqref{eq:global-proxy} in the phase \textit{\phaseone} to obtain a guess $H_t^*$
%on the optimal number of local steps that the scheduled \clients should perform at round $t$.

%To exploit vehicle decentralized computation,
%preserve privacy about some user data,
%and to make the scheduling algorithm amenable to large-scale scenarios,
%the edge server first
%%uses available information to decide
%sets the maximum number of vehicles that can participate in the round (based on available resources)
%and then
%estimates the optimal number of local computation slots at each vehicle.
			%!TEX ROOT = ../main.tex

\subsection{Local customization}\label{sec:decentralized-problem}

In this phase, each \client independently executes a local subroutine to allocate slots for computation and communication. This allocation attempts to optimize (i) convergence of local training and (ii) channel utilization. The resulting number of local steps is temporarily stored by the \clients, which use it later to perform the local model update in case they are actually scheduled.
If a \client has a means to (efficiently) estimate the loss gradient, it first refines the number of local steps $H_t^*$ communicated by the server (\textit{computation refinement}) based on a proxy for local convergence, obtaining a new number of local steps $H_t^v$, and then it optimizes for transmission (\textit{communication optimization}). Instead, if evaluating the gradient is expensive, the \client skips the computation refinement at this stage and sets the number of local steps as $H_t^v=\max\lb H_t^*, \bs{v}\compslotsmin\rb$. %optimize for transmission.

%To prioritize optimization of learning convergence,
%computation and communication slots are sequentially chosen based on two mutually constrained optimization problems.

%While the computation could be theoretically allocated through a sparse pattern $\compslot{v}{t}$,
%the optimal choice to optimize vehicle resources requires knowledge of other computational tasks
%that need to be executed close in time to the federated learning.
%However,
%assuming knowledge of other tasks \textit{a priori} is unrealistic and makes the allocation choice hard.
%Hence,
%we allocate all computation slots at the beginning of the round,
%namely,
%the computation decisions are $ a_{t,k}^{(v)} = 1 $ for $k = 1,\dots,\bar{k}$
%and $ a_{t,k}^{(v)} = 0 $ for $k = \bar{k},\dots,D_t$,
%for some $\bar{k}$.

%\begin{description}%[leftmargin=*]
%	\item[Computation refinement.] 
\subsubsection{Computation refinement}
\begin{figure}[t]
	\centering
	\unitlength=1cm
	\begin{picture}(8,1.2)(0,0)
		\put(0,0){\includegraphics[trim=5mm 0 0 0, clip, width=0.95\columnwidth]{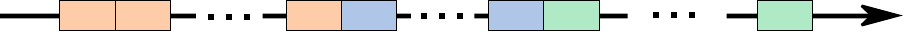}}
		\put(.2,-0.3){\scriptsize $0$}
		\put(.22,0.05){$\hm{\big\vert}$}
		\put(4,-0.3){\scriptsize round $t$}
		\put(0.46,0.08){\footnotesize $\tau$}
		\put(0.93,0.08){\footnotesize $2\tau$}
		\put(7.04,0.08){\footnotesize $N\tau$}
		\put(0.55,0.4){$\overbrace{\hspace{2.2cm}}$}
		\put(3.15,0.4){$\overbrace{\hspace{1.6cm}}$}
		\put(5.1,0.4){$\overbrace{\hspace{2.2cm}}$}
		\put(0.85,0.8){\scriptsize computation $\compslots{t}{v}$}
		\put(3.75,0.8){\scriptsize idle}
		\put(5.45,0.8){\scriptsize transmission $\txslots{t}{v}$}
		\put(7.47,0.05){$\hm{\big\vert}$}
		\put(7.4,-0.3){\scriptsize $\latency{t}{v}$}
		\put(7.88,0.05){$\hm{\big\vert}$}
		\put(7.8,-0.3){\scriptsize $\latencymax$}
		%		\put(7.3,-0.3){\scriptsize round $t\!+\!1$}
	\end{picture}
	\vspace{3mm}
	\caption{\textbf{Behavior of a selected client at each round.} If a client $v$ is selected by the server, it goes through the following steps.  
	 Each slot has duration $\tau$~[\si{\second}]. At first, the client performs some local (S)GD steps (\textit{computation}) according to its computation allocation. Eventually, it transmits its updated local model (\textit{transmission}) after time $\latency{t}{v}$ and within the maximum allowed latency $\latencymax$, possibly waiting for some time slots (\textit{idle}) to enjoy a better channel quality.
	 }
	\label{fig:access_node}
\end{figure}
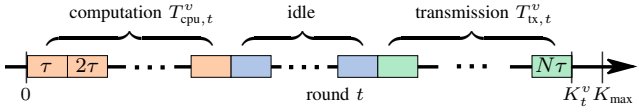

To optimize the local updates of the \client,
we consider a local proxy that jointly keeps into account the individual client convergence properties and the global recommendation $H_t^*$ indicated by the server. As such, the proxy is obtained as the sum of multiple terms. First, we consider a "convergence" proxy $\Theta^v_t\lr H_t^v\rr$ related to the local optimality gap, which bounds the client deterministic gradient norm after $H_t^v$ local steps of gradient descent %at round $t$,
starting from the global parameter $\param[t]$:
 % that bounds the expected distance from of the model locally trained by \client $v$ at iteration $t$:
\begin{equation}\label{eq:local-proxy}
	%\mathbb{E}\left[\norm{\nabla\loss^v(\param_{t}^v)}\right] 
  \norm{\nabla\loss^v\!\lr\param[t][t+H_{t}^v][v]\rr} \le \Theta^v_t\lr H_t^v\rr \doteq
	\norm{\nabla\loss^v\!\lr\param[t]\rr}\!\lr1-\kappa_v^{-1}\rr^{H_t^v-1}.\tag{LP}
\end{equation}
The constant $\kappa_v>0$ is the condition number associated with the local loss $\loss^v$. 
The proxy \eqref{eq:local-proxy} is based on quadratic cost functions \review{and we derive it explicitly in \arxiv{the technical report~\cite[Appendix~A]{arxiv}, using results from \cite{dal2022shed}}{Appendix~\ref{app:proxy}}}. 
The proxy~\eqref{eq:local-proxy} is motivated by the fact that the optimality gap $\|\theta - \theta^*\|$ is in general proportional to the gradient norm, but we have only access to the gradient, while $\theta^*$ is unknown.
It can be computed by each client based on their local cost only and does not take into account the distributed nature of the FL problem.
Hence, during this step, the \client $v$ refines its computation by solving the optimization problem
%\begin{itemize}
%	\item $\loss^v$ is the loss that accounts only for data of $v$
%	and $\loss^{v,*}$ is its global minimum;
%	\item $L_1^v$ and $L_2^v$ are constants that depend on $\loss^v$;
%%		\item $\param_{t-1}$ is the global model parameter after $t-1$ iterations (before running GD or SGD at the current round);
%%		\item $\param^v_t$ is the local parameter after $v$ has run GD or SGD starting from $\param_{t-1}$.
%\end{itemize}
\begin{argmini!}
	{H\in\mathbb{N}}
	{\Theta^v_t\lr H\rr + \dfrac{\rho_1H}{\norm{\nabla\loss^{v}\lr\param[t]\rr)}} + \rho_2 \lr H - H_t^*\rr^2\label{eq:local-convergence-optimization-cost}}
	{\label{eq:local-convergence-optimization}}
	{H_t^{v}=}
	\addConstraint{H}{\ge \bs{v}\compslotsmin.\label{eq:local-convergence-optimization-constr}}
\end{argmini!}
The second addend in~\eqref{eq:local-convergence-optimization-cost} accounts for how close the \client is to a local minimum,
forcing a few steps if the \client has (locally) almost converged and many steps if it is still far from convergence.
The third addend encourages the chosen local steps $H_t^v$ to be close to the target $H_t^*$ computed by the server.
The cost~\eqref{eq:local-convergence-optimization-cost} is convex and grows unbounded as $H\rightarrow+\infty$,
so that problem~\eqref{eq:local-convergence-optimization} can be efficiently solved by a linear search. 
The number of computation slots is then set as $\compslots{t}{v}=\ceil{\nicefrac{H_t^v}{\bs{v}}}$.
For the sake of simplicity,
we require that the \client allocates all slots for computation at the beginning of the round.
Formally,
this means $\compslot{k}{v} = 1$ for $k=1,\dots,\compslots{t}{v}$ and $\compslot{k}{v} = 0$ for $k>\compslots{t}{v}$
for all time slots $k\in\slots{t}$.
The slots with $\compslot{k}{v} = 1$ are the leftmost ``computation'' slots in~\autoref{fig:access_node}.
%	Vehicle $ v $ finds the optimal local computation by solving
%	\red{}
%	where $L_0 = \left(\ell(x_0)-\ell^*-L_1\right)$,
%	$ \rho_1 $ and $\rho_2$ are user-defined weights,
%	$ \ell^{(v)} $ is the local loss of vehicle $ v $,
%	and $ x_{t}^{(v)} $ is the local parameter of $ v $ at the beginning of round $ t $.
%	In words,
%	$ \nabla\ell^{(v)}(x_{t}^{(v)}) $ quantifies the local error
%	incurred by applying the current value of the local parameter (\ie before running the $ t $th learning round):
%	if such error is small,
%	it means that vehicle $ v $ is close to its local minimizer,
%	hence
%	the penalty term multiplied by $ \rho $ refrains $ v $ from running many descent steps (see notes below).
%	Conversely,
%	the third addend acts as a regularizer with respect to the global computation target $H_t^*$.
%	Note that any norms (not necessarily equal) may be used for the latter two addends.
%	\red{and $ \rho(\cdot) $ is a convex nonnegative function that penalizes deviation of
	%		the actual number of computation steps from the centralized indication $ H_t^* $.
	%		Standard choices are $ \rho(\cdot) = |\cdot| $ or $ \rho(\cdot) = (\cdot)^2 $.}
%	\item[Communication optimization.] %Given the design assumption above (computations without interruptions),

\subsubsection{Communication optimization}
\begin{figure}
	\centering
	\input{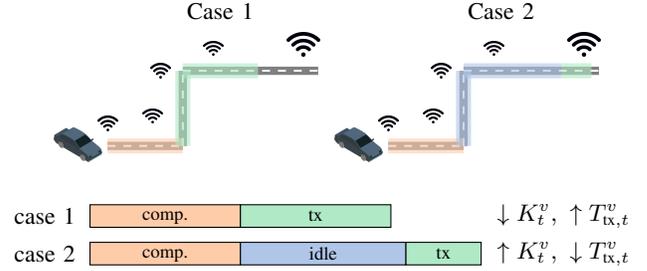}\\
	\vspace{3mm}
	%!TEX root = ../main.tex

\scalebox{1}{
	\begin{tikzpicture}
		% patterns and costs
		\node at (-3.6, -2.15) (case1) {{\small case 1}}; 
		\draw[fill=comp, opacity=1] (-3,-2.3) rectangle (-1,-2);
		\node (comp1) at (-2,-2.17) {{\scriptsize comp.}};
		\draw[fill=comm, opacity=1] (-1,-2.3) rectangle (1,-2);
		\node (tx1) at (0,-2.15) {{\scriptsize tx}};
		
		\node at (3.3,-2.15) (lat1) {{\small $\downarrow\latency{t}{v}, \ \uparrow\txslots{t}{v}$}};
		
		\node at (-3.6, -2.65) (case2) {{\small case 2}}; 
		\draw[fill=comp, opacity=1] (-3,-2.8) rectangle (-1,-2.5);
		\node (comp2) at (-2,-2.67) {{\scriptsize comp.}};
		\draw[fill=idle, opacity=1] (-1,-2.8) rectangle (1.2,-2.5);
		\node (idle2) at (.1,-2.65) {{\scriptsize idle}};
		\draw[fill=comm, opacity=1] (1.2,-2.8) rectangle (2.2,-2.5);
		\node (tx2) at (1.7,-2.65) {{\scriptsize tx}};
		
		\node at (3.3,-2.65) (lat1) {{\small $\uparrow\latency{t}{v}, \ \downarrow\txslots{t}{v}$}};
	\end{tikzpicture}
}
	\caption{\textbf{\ac{REM}-based solution for communication optimization.}
		Two evaluations of the objective cost~\eqref{eq:local-communication-optimization-obj}. In Case 1, transmission occurs under poor channel conditions (low bitrate) and takes long time. On the contrary,
		in Case 2, idle slots delay communication until when the vehicle travels through a well served area (high bitrate), so that transmission time is shorter.
	}
	\label{fig:mobility-rem}
\end{figure}

In this step, the \client chooses the time slots to transmit its local model to the server, adjusting the computation slots if needed.
For simplicity, we require the \client to allocate a batch of consecutive slots also for transmission.
Formally, 
the communication decisions for round $t$ are $ \txslot{k}{v} = 0 $ for $k = 1,\dots,\bar{k}_1$,
and $ \txslot{k}{v} = 1$ for $k = \bar{k}_1+1,\dots,\bar{k}_2$,
for some $\bar{k}_1\ge\compslots{t}{v}$ and $\bar{k}_2 > \bar{k}_1$.
We denote the set of such transmission patterns %with this structure
%		that are compatible with the maximum round latency $ \latencymax $ and minimum computation $T^{\text{cpu}}_\text{min}$ 
by $ \mathcal{B}_t^v(\compslots{t}{v}) $.
The allocation of slots for computation and communication is depicted in~\autoref{fig:access_node},
where the slots with $\txslot{k}{v}=1$ are the rightmost ``transmission'' slots.
\review[1]{The \client considers an allocation feasible if it estimates that its local model ($\bits$~[\si{\bit}]) can be uploaded before the deadline $\latencymax$ using the transmission slots.}
%also considering varying channel quality due to mobility.
To choose the communication pattern, given $\compslots{t}{v}$ slots of computation, the \client attempts to solve the following optimization problem:
\begin{argmini!}
	{b\in\mathcal{B}_t^v(\compslots{t}{v})}
	{C_t^v \doteq (1-\wtx) \latency{t}{v} + \wtx \txslots{t}{v}\protect\label{eq:local-communication-optimization-obj}}
	{\label{eq:local-communication-optimization}}
	{\txslot{t}{v}=}
	\addConstraint{\latency{t}{v}}{\le \latencymax}
	\addConstraint{\txbits{t}{v}}{\ge \bits.}
\end{argmini!}
The cost $C_t^v$ in~\eqref{eq:local-communication-optimization-obj} is designed to trade round latency $\latency{t}{v}$ for channel occupancy $\txslots{t}{v}$,
according to the overall objective cost~\eqref{eq:cost-function}.
To solve~\eqref{eq:local-communication-optimization},
vehicular mobility in conjunction with the REM plays a crucial role.
\review[1]{Each \client $v$ inspects the REM to estimate the available bitrate $\bitrate_k^v$ in each slot $k\in\slots{t}$ of round $t$ according to~\eqref{eq:rem-map} and, in turn, the number of time slots needed to upload the model in round $t$.}
For example, suppose the \client is about to travel close to a well-served area (\eg a main urban road). In that case, it will likely experience a high channel quality and thus take a short time for the upload, whereas, if it approaches a weakly served area (\eg a tunnel), it will predict poor channel conditions. It might even declare~\eqref{eq:local-communication-optimization} infeasible, giving up on joining the learning round. 
\Cref{fig:mobility-rem} pictorially represents the mobility-aware \ac{REM}-based evaluation of the cost~\eqref{eq:local-communication-optimization-obj}. In this figure, for case 1, transmission slots are allocated in a greedy fashion, whereas for case 2, the \client defers the transmission of the local model, waiting for the channel quality to improve. This trades some extra delay \review[1]{(higher round latency $\latency{t}{v}$)} for a better usage of channel resources \review[1]{(shorter upload time $\txslots{t}{v}$)}. 

Computation and communication decisions for the round are set by first solving~\eqref{eq:local-convergence-optimization} -- if possible -- to fine-tune the number of computation slots, and then~\eqref{eq:local-communication-optimization} until a feasible computation-communication pattern is found or the whole allocation is declared infeasible. % and the cost $C_t^v$ is set to infinity.
\cref{alg:local-step} summarizes the workflow of this phase. Eventually,
each \client $v$ transmits its estimated cost $C_t^v$ for participation in round $t$ to the server, to inform the latter on the expected benefit of scheduling $v$ for the present round. We use the convention that, if \client $v$ cannot find a feasible allocation for round $t$ (\ie problem~\eqref{eq:local-communication-optimization} turns out to be infeasible), it will communicate an infinite cost $C_t^v$, see~\cref{alg:local-step-initialize-cost}.

\begin{algorithm}
	\caption{Subroutine \texttt{customize\_local}}
	\label{alg:local-step}
	\KwIn{Local proxy $\Theta^v$, minimum computation $\compslotsmin$, maximum round latency $\latencymax$,
		target steps $H_t^*$.}
	\KwOut{Cost $C_t^v$, computation and communication decisions $\compslot{t}{v}$ and $\txslot{t}{v}$ for round $t$.}
	\If{can efficiently estimate $\nabla\loss_t^v$}{
		compute $ H_t^v $ as the solution to~\eqref{eq:local-convergence-optimization}\;	
	}
	\Else{
		set $H_t^v \leftarrow \max\lb H_t^*, \bs{v}\compslotsmin\rb$\;
	}
	set $ \compslots{t}{v} \leftarrow \ceil{\nicefrac{H_t^{v}}{\bs{v}}} $\;
	set $C_t^v\leftarrow+\infty$\;\label{alg:local-step-initialize-cost}
	\Repeat{$\compslots{t}{v}<\compslotsmin$}{
		compute $\txslot{t}{v}$ as the solution to~\eqref{eq:local-communication-optimization}\;
		\If{problem~\eqref{eq:local-communication-optimization} is feasible}{
			set $C_t^v\leftarrow C_t^v(\txslot{t}{v})$\;
			store computation $\compslots{t}{v}$ and communication $\txslot{t}{v}$\;
			\textbf{break}\;
		}
		\Else{
			set $ \compslots{t}{v} \leftarrow \compslots{t}{v} - 1$\;
		}
	}
	%			transmit $C_t^v$ to the server.
\end{algorithm}

\begin{rem}[Computation refinement for scheduled \client]\label{rem:computation-refinement}
	If a \client cannot evaluate the cost~\eqref{eq:local-convergence-optimization-cost} in reasonable time and skips the refinement~\eqref{eq:local-convergence-optimization},
	scheduled \clients may refine their local steps \textit{after} they have been scheduled.
	To this aim,
    they can compute or approximate the gradient $\nabla\loss^v(\param[t])$ at the beginning of the round,
	\eg by running a few descent steps,
	and then solve for $H_t^v$ according to~\eqref{eq:local-convergence-optimization} with the (approximate) gradient just computed.
	Then,
	they complete the local update by running \acs{SGD} until they reach $H_t^v$ total steps.
\end{rem}

\begin{rem}[Reducing latency \vs network resources]\label{rem:local-communication-optimization}
	Optimization~\eqref{eq:local-communication-optimization} makes both the \clients reduce their respective latency $\latency{t}{v}$ for round $t$ concerning the deadline $\latencymax$ and the server aware of their expected latency.
	This favors those \clients that are likely to send their local models in short time	and in turn speeds up the whole training.
	We experimentally demonstrate this via ablation studies and comparisons against scheduling benchmarks in \autoref{sec:results}.
	In particular,
	by tuning the weight $\wtx$ in~\eqref{eq:local-communication-optimization-obj},
	a system designer can encourage a short training (small $\wtx$)
	or a frugal usage of network resources (large $\wtx$).
\end{rem}

%		Given the initialization $ H_t^{v,*} $ for computational slots 
%		and the criterion stated above for communication slots,
%		all resources are adjusted according to the following cases.
%		\LB{If we don't have a smarter idea to allocate slots in Case 2,
	%		the next two cases can just be written as a $\min$}
%		\begin{description}
	%			\item[Initialization:] :
	%			\item[Case 1:] if~\eqref{eq:local-communication-optimization} is feasible,
	%			solve~\eqref{eq:local-communication-optimization} and \textbf{return};
	%			\item[Case 2:] if~\eqref{eq:local-communication-optimization} is not feasible:
	%			\begin{description}
		%				\item[Case 2.1:] if $T^{\text{cpu},v}_{t}>T^\text{cpu}_\text{min}$,
		%				set $T^{\text{cpu},v}_{t} \leftarrow T^{\text{cpu},v}_{t} - 1$
		%				and go to \textbf{Case 1};
		%				\item[Case 2.2:] else,
		%				set $ C_t^{v,*} = \infty $ and \textbf{return}.
		%			\end{description}
	%			
	%		\end{description}
%and allocated computational slots $ H_t^{v,*} $ for round $ t $.
%\end{description}
			%!TEX ROOT = ../main.tex

\subsection{Centralized scheduling}\label{sec:centralized-problem-2}

After the server receives information from all \clients about their (predicted) cost for the round,
the \clients are scheduled based on both this cost,
that measures the training performance,
and on fairness metrics such as the \ac{AoI} and scheduling frequency,
accounting for the learning accuracy of the global model.
In particular,
drawing inspiration from~\cite{GunduzComputationTime},
we define the fairness $\fair{t}{v}$ for \client $v$ at round $t$ as
\begin{equation}\label{eq:fairness}
	\fair{t}{v} \doteq  \dfrac{1}{\freq{t}{v}} + \aoi{t}{v}
\end{equation}
where $\freq{t}{v}$ is the scheduling frequency of \client $v$ before round $t$ and $\aoi{t}{v}$ is its AoI at the server.
%\begin{itemize}
%	\item fairness, given by AoI $ A $ and scheduling frequency $ f $:\\
%	$ F_t^{(v)} = $;
%	\item weighed sum of latency and transmission time: \\
%	$C_t^{(v)} = \gamma K_t^{(v)} + \delta T^{\text{tx},v}_{t}$.
%\end{itemize}
%
%This allow us to write the schedule in~\eqref{eq:cost-function} as a function of fairness and latency cost:
%
%\begin{equation}\label{eq:schedule-function}
%	\mathcal{V}_t^S = \mathcal{V}_t^S\left(\{F_t^{(v)},C_t^{(v)}\}_{v\in\mathcal{V}}\right).
%\end{equation}
%
%The scheduler receives by all vehicles their estimated cost for participating in round $ t $.
%and the amount of computation they intend to perform.
To schedule the participating vehicles,
the server 
%uses information for each vehicle:
%latency in the round $ K_t^{v,*} $ (which decreases scheduling priority),
%%amount of computation $ H_t^{v,*} $ (which decreases scheduling priority),
%and fairness as quantified by Age of Information $ A_t^{(v)} $ (which increases scheduling priority)
%and scheduling frequency $ f_t^{(v)} $ (which decreases scheduling priority).
%
%To choose which vehicles are scheduled,
%the scheduler can then 
assigns a \textit{priority score} $ \priority{t}{v} $ to each vehicle $v$ for round $t$:
\begin{equation}\label{eq:priority-score}
	\priority{t}{v} \doteq \begin{cases}
		\dfrac{1}{C_t^{v}} + w_{\aoi{}{}}\fair{t}{v} & \text{if } C_t^{v} <+\infty\\
		-1 & \text{if } C_t^{v} =+\infty.
	\end{cases}
	                                %+ \dfrac{\beta}{H_t^{v,*}}
%									+ \gamma A_t^{(v)} + \dfrac{\delta}{f_t^{(v)}}
\end{equation}
The weight $w_{\aoi{}{}}$ in~\eqref{eq:priority-score} should be chosen so as to strike a balance between high-performing \clients,
which may significantly reduce the objective cost~\eqref{eq:cost-function} in the short run,
and overall training in the long run that needs to gather information from all \clients to eventually learn an accurate global model.

Formally, the vehicles with the highest priority scores are scheduled,
according to the following optimization problem:
%\begin{equation}\label{eq:scheduled}
%	\vsetsch{t} = \lb v\in\vset : |\vsetsch{t}|\le\vsetsccard{t}, \ \priority{t}{v} \ge \priority{t}{v'} \ \forall v'\notin\vsetsch{t}, \priority{t}{v} > 0\rb.
%\end{equation}
\begin{argmaxi!}
	{\vsetsch{}\subseteq\vset}
	{\sum_{v\in\vsetsch{}}\priority{t}{v}\protect\label{eq:scheduled-obj}}
	{\label{eq:scheduled}}
	{\vsetsch{t} =}
	\addConstraint{\left|\vsetsch{}\right|}{\le\vsetsccard{t}.}
	%\addConstraint{\priority{t}{v}}{>0 \; \forall v \in \vsetsch{}.\label{eq:scheduled-constr-positive-priority}}
\end{argmaxi!}
According to our convention described in~\autoref{sec:decentralized-problem},
the clients that communicate infeasible participation are assigned a negative priority score as per~\eqref{eq:priority-score}, which automatically excludes them from the round according to maximization of~\eqref{eq:scheduled-obj}.

The set $\vsetsch{t}$ contains the \clients that the server schedules for transmitting their local updates in the current round $t$.
The workflow of \algoname is provided in~\cref{alg:codesign}.

\begin{algorithm}
	\caption{\algoname}
	\label{alg:codesign}
	\DontPrintSemicolon
	\KwIn{Global proxy $\Gamma$, local proxy $\Theta^v$, maximal round latency $\latencymax$, minimal computation slots $\compslotsmin$, maximal number of scheduled \clients $\vsetsccard{t}$ .}
	\KwOut{Scheduled \clients $\vsetsch{t}$.}
	\nonl\textbf{centralized optimization (at the edge server):}\;
	%compute $\vsetsccard{t}$ based on resources available for round $t$;\;
	compute $ H_t^* $ as the solution to~\eqref{eq:nominal-local-steps};\;
	broadcast $ H_t^* $ and maximal latency $ \latencymax $ to \clients;\;
	\nonl\textbf{local customization (at the \clients):}\;
	\ForEach{\client $v\in\vset$}{
		$C_t^v \leftarrow$ \texttt{customize\_local}$(\Theta^v, \compslotsmin, \latencymax, H_t^*)$;\;
		send $C_t^v$ to server;\;
	}
	\nonl\textbf{centralized scheduling (at the edge server):}\;
	\ForEach{\client $v\in\vset$}{
		compute $\priority{t}{v}$ as per~\eqref{eq:priority-score};\;
	}
	%$\vsetsch{t} \leftarrow\emptyset$;\;
	%\While(\tcp*[f]{{\small arbitrarily break ties}}){$|\vsetsch{t}|<\vsetsccard{t}$ \textbf{and} $\max_{v\notin\vsetsch{t}}\priority{t}{v}>0$}{
		%	$\vsetsch{t} \leftarrow \vsetsch{t} \cup \argmax_{v\notin\vsetsch{t}}\priority{t}{v}$;\;
		%}
	populate $\vsetsch{t}$ according to~\eqref{eq:scheduled};\;
	\textbf{return} $\vsetsch{t}$.
\end{algorithm}

\review{\begin{rem}[\algoname supports asynchronous aggregation]\label{rem:sync-async}
	\algoname assumes that the aggregator updates the global model after all scheduled \clients have transmitted their updates or the deadline $\latencymax$ expires.
	This means that the aggregation scheme can be \emph{asynchronous}.
	In fact,
	if \clients train and upload local models with the settings output by \texttt{customize\_local} (respectively $\compslot{t}{v}$ and $\txslot{t}{v}$),
	the timing of training and transmission will be in general different for each \client,
	yielding asynchronous reception of local models at the aggregator.
	For this,
	using knowledge of the \acs{REM} and mobility pattern at the \client,
	as we propose,
	is essential;
	broadly speaking,
	transmitting when the channel is good is advantageous regardless of aggregation schemes.
	This behavior is observed in our experiments,
	where moreover not all scheduled \clients transmit their local models in time.
	However,
	while this happens rarely when using \algoname,
	which leverages knowledge of the channel quality experienced by \clients,
	other scheduling strategies incur several missed updates that contribute to degrading learning and wasting resources.
	See~\autoref{fig:client-fraction} for our experimental comparison.
\end{rem}}

\review{\begin{rem}[Extension to other FL algorithms]\label{rem:extensions}
The convergence proxies~\eqref{eq:global-proxy} and~\eqref{eq:local-proxy} are the only elements that depend on the algorithms used to solve~\eqref{eq:FL-problem} and they can be adjusted to use \algoname with other local training or aggregation schemes.
\end{rem}}
			\review{%!TEX root = ../main.tex

\subsection{Complexity Analysis of VREM-FL}\label{sec:complexity-analysis}

A strength of \algoname is its light computational and communication requirements,
which can accommodate a large number of \clients with modest computational power onboard.

\subsubsection{Centralized optimization}
During this phase, the edge server computes the target global local steps $H_t^*$ as~\eqref{eq:optimal-H}, which has complexity $O(1)$. The required communication corresponds to broadcasting this value once to all \clients.

\subsubsection{Local customization}
\paragraph{Computation refinement}

The first part of the second phase, if executed, requires each \client to solve problem~\eqref{eq:local-convergence-optimization}, which is the minimization of a scalar submodular function on $\mathbb{N}$ and can be solved via linear search.

\paragraph{Communication optimization}

In the second part of the second phase, each \client solves problem~\eqref{eq:local-communication-optimization} that requires the evaluation of the cost function~\eqref{eq:local-communication-optimization-obj} at least $|\slots{t}|-\compslots{t}{v}$ times, where $\compslots{t}{v}$ is the number of slots allocated for local training from either problem~\eqref{eq:local-convergence-optimization} or the received $H_t^*$.
In the worst case, when~\eqref{eq:local-communication-optimization} is infeasible, 
the \client performs about $(\compslots{t}{v}+1)|\slots{t}| - \nicefrac{\compslots{t}{v}}{2}$ evaluations of~\eqref{eq:local-communication-optimization-obj}, 
which is linear with $|\slots{t}|$.
In our realistic experiments, the vehicles almost always solved~\eqref{eq:local-communication-optimization} in a few attempts, resulting in a low computational requirement.
Each \client transmits to the server its cost $C_t^v$.

\subsubsection{Centralized scheduling}

The edge server computes one priority score per \client and selects the participating ones via problem~\eqref{eq:scheduled}. Both operations are linear with the number of \clients.
A fast implementation computes the priority~\eqref{eq:priority-score} of each \client $v$ as soon as its cost $C_t^v$ is received, and keeps the \clients ordered by priority with \texttt{InsertionSort}.
In this case, the computational complexity to solve~\eqref{eq:scheduled} is $O(1)$.}
	%!TEX root = ../main.tex

\section{Numerical results}
\label{sec:results}

%To validate the effectiveness of \algoname,
\review{We perform \acs{FL} experiments with both synthetic and real-world data.
We address vehicular mobility by both generating trajectories of vehicles with a realistic simulator and using real-world mobility data.
Without loss of generality,
we set a constant bandwidth $\eta_t^v \equiv \eta$,
identical for all vehicles and learning iterations.}
\remove{Nonetheless,
we remark that VREM-FL can handle the case in which the bandwidth is different across vehicles and iterations at no additional cost. We defer experiments and design studies with heterogeneous bandwidths to future work.}
\review{In~\autoref{sec:simulation-mobility-radio-generation},
we describe the urban environment, mobility data,
and \acs{REM} generation.
In \autoref{sec:exp-benchmarks},
we present the benchmarks compared with \algoname.}
\review{In~\autoref{sec:exp-ls},
we showcase results for a linear regression model on a least-squares problem with synthetic data.
This allows us to conduct ablation studies that isolate the effects of several features of \algoname and highlight their benefits.}
\review{In~\autoref{sec:exp-nn},
we use \algoname to train a deep neural network model for semantic segmentation,
a task of interest for assisted and autonomous driving.
For this experiment,
we use the real-world dataset \texttt{ApolloScape}~\cite{huang2018apolloscape} and both simulated and real-world mobility data.\footnote{
    Code for simulations available at \url{https://github.com/lucaballotta/vrem-fl}.
}
}
\remove{This results section is divided into two parts. In the first one, we use a synthetically generated dataset with non-iid data, used to train a least-squares regression problem to show in detail the operation of VREM-FL. In the second part, VREM-FL is applied to the real-world semantic segmentation dataset \texttt{ApolloScape}~\cite{huang2018apolloscape}, to test the algorithm on a task of specific interest for vehicular applications. The state-of-the-art \texttt{deeplabv3}~\cite{chen2017rethinking} model is trained, choosing \texttt{mobilenetv3-large}~\cite{howard2019searching} as a backbone.}
			%!TEX root = ../main.tex

\subsection{Mobility and urban radio environment generation}
\label{sec:simulation-mobility-radio-generation}

We implement the simulations in Python.
We use the map of Padova, Italy,
from \texttt{OpenStreetMap}~\cite{osm} 
\review{and use \ac{SUMO}~\cite{sumo}} to simulate $1,000$ vehicles that move across the city for one hour, discarding the first ten minutes of simulation to let the road map populate with a sufficiently large number of vehicles.
\review{For the second experiment, we use a real-world mobility dataset that we describe in \autoref{sec:exp-nn-taxi}.}

On top of the city map, BSs are deployed with an inter-site distance of $600$ \si{\meter}, according to typical 5G deployment criteria. Average SINR values and the corresponding bitrates have been obtained through the Matlab 5G NR link-level simulation tool~\cite{mathworks}. SINR values are calculated considering (i) the transmission power of vehicles, (ii) physical settings of the 5G NR, (iii) propagation models, and (iv) interference and noise power. We set the transmission power of vehicles to $23$~\si{\dBm}. Without loss of generality, we assume to allocate (in the frequency domain) a fixed number of $10$ resource blocks for data transmission.
The carrier frequency is set to 3.5~\si{\giga\hertz}. The sub-carrier space and the resulting size (in the frequency domain) of each single radio resource block are respectively set to $30$~\si{\kilo\hertz} and $360$~\si{\kilo\hertz}. Hence, the per-client bandwidth for $10$ resource blocks is $\eta = 3.6$~\si{\mega\hertz} (see~\eqref{eq:rem-map}).
%Carrier frequency and sub-carrier space (that is the size of each single radio resource block in the frequency domain) are set to $3.5$~\si{\giga\hertz} and $30$~\si{\kilo\hertz}, respectively. Hence, the per-client bandwidth for $10$ resource blocks is $\eta = 300$kHz (see~\eqref{eq:rem-map}). 
The antenna height of BSs and vehicles is set to $25$~\si{\meter} and $1.5$~\si{\meter}, respectively. Path loss parameters are set according to the urban microcell scenario, as defined in the TR 38.901 specification of 3GPP. The %fast-fading effect is modeled through the Tapped Delay Line fading channel (specifically, TPL-A for taking care of NLOS links), as defined in the TR 38.901 specification of 3GPP. Moreover, 
slow-fading (shadowing) is added to the coverage area of each BS following the 3GPP guidelines~\cite{3GPP_Shadowing}, with a de-correlation distance of $25$~\si{\meter} and standard deviation of $6$~\si{\deci\bel}, which are common values for urban environments~\cite{ITU_PL}. Finally, the noise figure, used to derive the noise power, is set to $6$~\si{\deci\bel}. 
The bitrate corresponding to a given SINR average value -- \ie the map $\beta(\cdot)$ in~\eqref{eq:rem-map} --  is obtained according to the physical uplink shared channel (namely 5G NR PUSCH) throughput experienced in a 5G New Radio link~\cite{3GPP_PUSCH_211, 3GPP_PUSCH_212, 3GPP_PUSCH_214} for that value of SINR. 
The average throughput is calculated through the Matlab 5G NR link-level simulation tool implementing the 3GPP NR standard~\cite{mathworks}, 
assuming that vehicles experience a given average SINR value during the time slot $\tau$ of $1$ \si{\second}. 
%The street map of Padova together with the BS locations and shaded SNR obtained is depicted in Fig.~\textcolor{blue}{fig}.

\begin{table}
	\caption{\review{Parameters used in the two experiments on linear regression (\autoref{sec:exp-ls}) and deep learning (\autoref{sec:exp-nn}).
			Curly brackets indicate multiple trials\remove{ for the same parameter}.
			Bold font indicates the best choice w.r.t.~\eqref{eq:cost-function}.
		}
	}
	\label{table-simulation-params}
	\begin{center}
		\footnotesize
		\begin{tabular}{cccc}
			\toprule
			& \review{Linear regression}	& \multicolumn{2}{c}{\review{Deep learning}}\\
			&													& \acs{SUMO}	& roma/taxi\\
			\midrule
			$\bits$     & $400$~\si{\byte}        & \multicolumn{2}{c}{$42.3$~\si{\mega\byte}}\\
			\midrule
			learning horizon 		& $30$ rounds & \multicolumn{2}{c}{$1$~\si{\hour}}	\\
			\midrule
			$\tau$ 		& $1~\si{\second}$ & \multicolumn{2}{c}{$1~\si{\second}$}\\
			\midrule
			$\latencymax$ 		& $100~\si{\second}$ & \multicolumn{2}{c}{$2~\si{\minute}$}\\
			\midrule
			$\compslotsmin$ 		& $1$ & \multicolumn{2}{c}{$1$}\\
			\midrule
			$\bs{v}$	& $1$ (GD)	& \multicolumn{2}{c}{$3$ (SGD with $\text{bz}=32$)} \\
			\midrule
			$|\vset|$	& $1000$ & $1000$	& $92$\\
			\midrule
			$\vsetsccard{t}$	&$30$ & $\{5,10,\textbf{15},20,25\}$ & $10$\\
			\midrule
			$C$ in~\eqref{eq:global-proxy} & $200$ 		& \multicolumn{2}{c}{$1000$} \\
			\midrule
			$(\rho_1, \rho_2)$ in~\eqref{eq:local-convergence-optimization-cost} & $(0.001,1)$ & \multicolumn{2}{c}{$(1, 0.02)$}\\
			\midrule
			$\wtx$ in~\eqref{eq:local-communication-optimization-obj}	& $\{0, \textbf{0.5}, 1\}$	& \multicolumn{2}{c}{$0.9$}\\
			\midrule
			$w_A$ in~\eqref{eq:priority-score} & $0$ 	& $0.01$	& \{0.01, \textbf{0}\}\\
			\bottomrule
		\end{tabular}
		\remove{\begin{tabular}{ll}
			\toprule
			$T=50$~\si{\minute} & simulation horizon\\
			\midrule
			$\tau=1~\si{\second}$ & duration of one time slot\\
			\midrule
			$\latencymax=100~\si{\second}$ & maximum latency of every iteration\\
			\midrule
			$\compslotsmin=1$ & minimum computation slots\\
			\midrule
			$\vsetsccard{t}=30$ & maximum number of \clients scheduled\\
			\midrule
			$\rho_1=0.001$, $\rho_2=1$ & weight coefficients of Eq.~\eqref{eq:local-convergence-optimization}\\
			\midrule
			$C=200$ & weight coefficient of Eq.~\eqref{eq:global-proxy}\\
			\midrule
			$w_C=1, w_A=0$ & weight coefficients of Eq.~\eqref{eq:priority-score}\\
			\bottomrule
		\end{tabular}}
	\end{center}
\end{table}
			%!TEX root = ../main.tex

\subsection{Scheduling Benchmarks}\label{sec:exp-benchmarks}
We compare VREM-FL against \review{four} scheduling benchmarks:
\begin{description}
    \item[Round robin:] Vehicles are chosen by the scheduler based on a round robin policy, \ie in a cyclic order.
	\item[\protect\review{FedAvg~\cite{mcmahan2017}}:] Vehicles are chosen by sampling them randomly with equal probability (uniform random variable).
	\item[\protect\review{Fairness~\cite{GunduzComputationTime}}:] This scheme optimizes the metric proposed in~\cite{GunduzComputationTime} %Besides, other references use AoI-based metrics (\textcolor{red}{refs}). 
	and is equivalent to \remove{VREM-FL when setting $w_C=0$ in~\eqref{eq:priority-score},
	\ie} computing scheduling priority as $\priority{t}{v}=\fair{t}{v}$.
	A crucial difference between this method and \algoname is that the cost $C_t^v$ is not used and thus the two cases in~\eqref{eq:priority-score} are indistinguishable.
    \review{That is,
    even if a \client could communicate that its participation is infeasible}, ``Fairness'' neglects this information and uses only fairness-related information $\fair{t}{v}$ \review{known to the scheduler}.
    \item[\protect\review{Centr-SNR~\cite{chen2024efficient}:}]\review{This algorithm is adapted from~\cite{chen2024efficient}, where clients are scheduled in a centralized way based on the uplink channel gain (SNR) reported to the server.
    	To draw a fair comparison, we select the \clients with the best estimated bitrate at the beginning of each learning round. 
    	We use this scheme for the second experiment in \autoref{sec:exp-nn}.}
\end{description}

\subsection{First Experiment: Linear Regression on Synthetic Data}
\label{sec:exp-ls}

\review{We first address a least squares problem with a linear regression model to illustrate the functioning of our proposed algorithm.
In~\autoref{sec:exp-ls-dataset},
we describe the synthetic dataset used for this experiment.
In~\autoref{sec:results_estimated_REM},
we study how performance of \algoname varies with the precision of the \acs{REM},
considering both perfect channel knowledge and estimated \acp{REM} that may differ from actual channel conditions.
In \autoref{sec:exp-ls-comparison},
we compare \algoname against literature benchmarks and demonstrate its superiority in reducing training time while more sparingly allocating channel resources.
In \autoref{sec:exp-ls-ablation-algo},
we perform several ablations to inspect how the metrics included in~\eqref{eq:cost-function} are affected by different design choices of \algoname parameters.}
					%!TEX root = ../main.tex

\subsubsection{Synthetic dataset generation}
\label{sec:exp-ls-dataset}
We generate $S$ synthetic data samples ${x}_1, ..., {x}_S$, each obtained as a random linear combination of the form ${x} = \sum_{j = 1}^{n}z_j\Tilde{{u}}_j$, where $z_1, ..., z_n$ are i.i.d. Gaussian random variables, and $\{\Tilde{{u}}_1, ..., \Tilde{{u}}_n\} = \{\sigma_1{u}_1, ..., \sigma_n{u}_n\}$, with $\sigma_1, ..., \sigma_n$ constants between $10^{-2}$ and $1$, while $\{{u}_1, ..., {u}_n\}$ are $n$ linearly independent basis vectors. Response values $y_1, ..., y_S$ are generated from the data samples as $y_i = {x}_i^T{\theta}^*$, $i = 1, ..., S$, where the generating parameter $\theta^*$ is also obtained as a random linear combination of $\{\Tilde{{u}}_1, ..., \Tilde{{u}}_n\}$.

%We consider a linear data generator in which data points are vectors in $\mathbb{R}^n$. To build the generator, we consider a set $\mathcal{U}$ of $n$ linearly independent basis vectors, $\mathcal{U} = \{{u}_1, ..., {u}_n\}$. 
%We scale these vectors by constants $\sigma_1, ..., \sigma_n$ that take values between $10^{-2}$ and $1$. The resulting generating set of vectors is $\widetilde{\mathcal{U}} = \{\Tilde{{u}}_1, ..., \Tilde{{u}}_n\} = \{\sigma_1{u}_1, ..., \sigma_n{u}_n\}$. Each synthetic data sample $x$ is generated by a random linear combination of the form ${x} = \sum_{j = 1}^{n}z_j\Tilde{{u}}_j$, where $z_1, ..., z_n$ are i.i.d. Gaussian random variables. We generate $S$ data samples. Accordingly, $S$ response values $y_1, ..., y_S$ are generated from the data samples ${x}_1, ..., {x}_S$ as $y_i = {x}_i^T{\theta}^*$, where the generating parameter $\theta^*$ is also obtained as a random linear combination of $\{\Tilde{{u}}_1, ..., \Tilde{{u}}_n\}$. 
%For this synthetic dataset, we study regularized least squares, \ie the instantiation of~\eqref{eq:FL-problem} with loss function $\loss(\theta) = \sum_{i = 1}^S(\theta^\top x_i - y_i)^2$ and regularization term $\lambda\|\theta\|^2$, 
The least squares loss in~\eqref{eq:FL-problem} is $\loss(\theta) = \sum_{i = 1}^S(\theta^\top x_i - y_i)^2$.
\review{We show here the results obtained with regularization parameter $\lambda = 10^{-4}$.
We obtained similar results with $\lambda \in \{10^{-3}, 10^{-5}\}$ that are illustrated in \arxiv{the technical report~\cite[Appendix~B]{arxiv}.}{Appendix~\ref{app:lambda}.}}
For the purpose of the ablation study with synthetic data, 
we set a small parameter dimension $n = 25$.
Accordingly, 
to get meaningful results with respect to the scheduling design, 
we rescale the bitrate values by a factor $2\times 10^{-5}$. %\review{Here, we show results obtained with $\lambda = 10^{-4}$, but we obtained equivalent results with $\lambda = 10^{-3}, 10^{-5}$ as we illustrate in the extended ArXiv version of the paper.} 
					%!TEX root = ../main.tex

\subsubsection{Performance using estimated REMs}
\label{sec:results_estimated_REM}

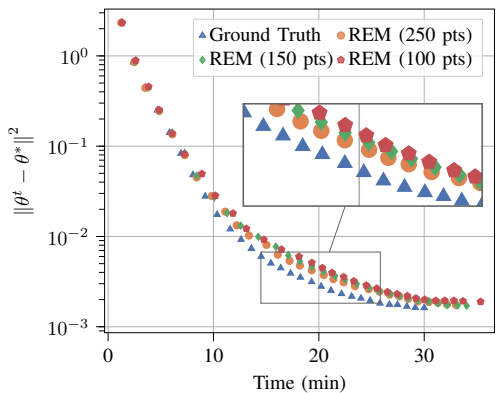
\begin{figure}
	\centering
	\resizebox{.75\columnwidth}{!}{%!TEX root = ../main.tex

\begin{tikzpicture}[spy using outlines={magnification=2, rectangle, width=4.2cm, height=1.8cm, white!40!black, connect spies, }]
	
	\definecolor{color0}{rgb}{0.298039215686275,0.447058823529412,0.690196078431373}
	\definecolor{color1}{rgb}{0.866666666666667,0.517647058823529,0.32156862745098}
	\definecolor{color2}{rgb}{0.333333333333333,0.658823529411765,0.407843137254902}
	\definecolor{color3}{rgb}{0.768627450980392,0.305882352941176,0.32156862745098}
	
	\begin{axis}[
		legend cell align={left},
		legend style={fill opacity=0.8, draw opacity=1, text opacity=1, draw=white!80!black, font=\normalfont},
		legend columns=2,
		legend pos=north east,
		log basis y={10},
		tick align=outside,
		tick pos=left,
		x grid style={white!69.0196078431373!black},
		xlabel={Time (min)},
		xmajorgrids,
		xmin=580, xmax=2800,
		xtick style={color=black},
		y grid style={white!69.0196078431373!black},
		ylabel={$\norm{\param[t] - \param[*]}^2$},
		ymajorgrids,
		ymin=0.000859176560317704, ymax=3.30138766297069,
		ymode=log,
		ytick style={color=black},
		xtick={600, 1200, 1800, 2400},
		xticklabels={0, 10, 20, 30}
		]
		\addplot [semithick, color0, mark=triangle*, mark size=1.75, mark options={solid}, only marks]
		table {%
			675 2.33333586142001
			746 0.8514212266794
			812 0.449806713281362
			880 0.25032570945003
			950 0.141498960772782
			1015 0.0830970386150636
			1083 0.0480850568610122
			1153 0.0278797832547482
			1221 0.017490182655642
			1295 0.0120615827962728
			1357 0.00925033393685673
			1416 0.00735212338949192
			1472 0.00600338558023308
			1527 0.00504797765191402
			1583 0.00446929675711829
			1639 0.00390639087518032
			1695 0.00352449917444298
			1759 0.00314589902378307
			1813 0.00280111667015795
			1869 0.00251918864514811
			1929 0.00232059585848131
			1987 0.00216962663851192
			2040 0.002065786870015
			2092 0.00194942747097282
			2148 0.00182026992538023
			2202 0.00176352618584475
			2248 0.00175836420750116
			2301 0.00168049988766852
			2349 0.00162948614400753
			2400 0.00162100593963466
		};
		\addlegendentry{Ground Truth}
		\addplot [semithick, color1, mark=*, mark size=1.75, mark options={solid}, only marks]
		table {%
			675 2.33333586142001
			749 0.853092556115803
			812 0.442897779584161
			890 0.246244072337061
			966 0.135595436880131
			1037 0.0797145943456762
			1105 0.0450507766903584
			1191 0.0280341648266049
			1267 0.0188273522552796
			1332 0.0132761445635579
			1402 0.0102451363657111
			1502 0.00805376403700091
			1566 0.00628383621524744
			1633 0.00537195068928746
			1692 0.00475710734941072
			1759 0.00421943709318484
			1828 0.00373865903102936
			1883 0.00337273126411465
			1941 0.00310717364105591
			2006 0.00281351388854218
			2084 0.00260737232718643
			2145 0.00244746377330389
			2203 0.00226750196152121
			2258 0.00217036768792033
			2308 0.00203804645977425
			2364 0.00191090751809603
			2426 0.00187665669230959
			2478 0.00184043215432866
			2530 0.00176780459580434
			2582 0.00173935576997355
		};
		\addlegendentry{REM (250 pts)}
		\addplot [semithick, color2, mark=diamond*, mark size=1.75, mark options={solid}, only marks]
		table {%
			679 2.33333586142001
			754 0.871670884270222
			831 0.446190201410155
			893 0.246070586880993
			965 0.140857131441631
			1037 0.0839868842361182
			1105 0.0470163353543503
			1205 0.0272059655401561
			1278 0.0181522677660178
			1355 0.0131548050528954
			1455 0.00991340960957829
			1555 0.00769470145022519
			1626 0.00617042087318048
			1692 0.00531092695498359
			1760 0.00463524034388272
			1827 0.0040954101313818
			1892 0.00365987746171128
			1949 0.00333057486906564
			2016 0.00299227966720677
			2086 0.00270957209883918
			2145 0.00247696441885566
			2207 0.00231632651638831
			2259 0.00218676142208906
			2309 0.00207844660000886
			2363 0.00195919450832297
			2427 0.00191310292758036
			2479 0.00181750577128029
			2531 0.00174698615499813
			2590 0.00172398907728549
			2642 0.00171004126413973
		};
		\addlegendentry{REM (150 pts)}
		\addplot [semithick, color3, mark=pentagon*, mark size=1.75, mark options={solid}, only marks]
		table {%
			680 2.33333586142001
			757 0.891034334848887
			831 0.456143880367214
			891 0.251431698084957
			964 0.138555415325583
			1034 0.0817696342251805
			1134 0.0494833152154145
			1211 0.0283356228264463
			1311 0.01804421698844
			1387 0.0122109601139056
			1487 0.00921231898741419
			1587 0.00717105418142819
			1687 0.00598507649994105
			1760 0.00511323290092182
			1820 0.00448369945808975
			1875 0.0039639788944426
			1941 0.00355966054353269
			1999 0.00319817867899162
			2070 0.00287339871602872
			2130 0.00265322984553361
			2190 0.00242081452120854
			2246 0.00230424623517016
			2304 0.00219300493438473
			2353 0.00206722684912112
			2403 0.0020048233525809
			2465 0.00194628750621685
			2517 0.00195053724686327
			2569 0.00194828454868992
			2621 0.00192288325972618
			2721 0.00189459894854986
		};
		\addlegendentry{REM (100 pts)}
		\coordinate (c1) at (axis cs: 1810, 3.5e-3);
		\coordinate (c2) at (axis cs: 2050, 8e-2);
		\spy on (c1) in node[fill=white] at (c2);
	\end{axis}
	
\end{tikzpicture}}
	\caption{\review{Model convergence with} VREM-FL and varying \ac{REM} quality \review{for linear regression experiment.}
		\remove{with the synthetic dataset (see \autoref{subsec:synthGen})}
		\review{The curves correspond to different estimates} \remove{considering estimations} of the \ac{REM} obtained by varying granularity of available measures.}
	\label{fig:estimation}
\end{figure}

We show the performance of \algoname on the synthetic dataset described in~\autoref{sec:exp-ls-dataset} when the REM is estimated via Gaussian Process Regression (GPR)~\cite{dalfabbro2022rem, muppirisetty2015spatial} performed on a limited number of measurements. 
We consider the cases where $100$, $150$, or $250$ measurements are available in each BS cellular sector. 
For each sector, we select the measuring locations uniformly at random. 
To perform GPR, we assume that the standard deviation and the de-correlation distance of the shadowing process are known \textit{a priori}.\footnote{If this is not the case, they can be estimated via standard techniques~\cite{dalfabbro2022rem}.}
\review{The parameters \remove{in the simulations} for \acs{FL} and \algoname \remove{settings} are
%unless differently specified, 
reported in the first column of~\autoref{table-simulation-params}.}

In~\autoref{fig:estimation}, we show the distance from the optimal parameter $\theta^*$ of the least squares problem as a function of the simulation time, where different lines denote a different number of measures available to generate the REM maps through GPR. Markers denote the simulation slot where the server has received all the updates from the clients and computes the average, \ie the end of a learning round. Triangular blue markers correspond to the ground truth, namely, the real REM is available at the scheduler. We can see that the more points are available to generate the REM estimation, the more likely it is to avoid stragglers. Stragglers are those nodes that slow down the learning process because they do not have sufficient computational or communication resources. %To visualize the presence of strugglers, let us focus on the zoomed area in Fig.~\ref{fig:estimation}: The estimation method using 100 points shows three big latency gaps. This is because some clients have been chosen for which the estimation was good but significantly far from the ground truth. Hence, these nodes took much more time than expected to transmit their local model. It is worth noting that the specific moment where the estimation fails can have a high impact: For the green line relative to the REM generated with 150 measures, there are only two significant gaps caused by strugglers, but, due to an unlucky case, they both come at an early stage of the learning process, being thus more harmful.
The presence of scheduled nodes that had an estimated channel quality superior to the real one makes the learning round last longer, hence the total latency increases. 
The estimated map obtained using $100$ points produces a total learning duration approximately $5$ minutes longer than the ground truth and $2$ minutes longer than the estimated map with $250$ points. For the rest of the experiments, we used the REM estimated with $250$ points.
					%!TEX root = ../main.tex

\subsubsection{Comparison with other scheduling policies}
\label{sec:exp-ls-comparison}
\begin{figure}
	\centering
	\resizebox{.75\columnwidth}{!}{%!TEX root = ../main.tex

% This file was created by tikzplotlib v0.9.9.
\begin{tikzpicture}
	
	\definecolor{color0}{RGB}{76,114,176}
	\definecolor{color1}{RGB}{221,132,82}
	\definecolor{color2}{RGB}{85,168,104}
	\definecolor{color3}{RGB}{196,78,82}
	
	\begin{axis}[
		legend cell align={left},
		legend style={fill opacity=0.8, draw opacity=1, text opacity=1, draw=white!80!black},
		log basis y={10},
		tick align=outside,
		tick pos=left,
		x grid style={white!69.0196078431373!black},
		xlabel={Time (min)},
		xmajorgrids,
		xmin=580, xmax=3726,
		xtick style={color=black},
		y grid style={white!69.0196078431373!black},
		ylabel={$\norm{\param[t] - \param[*]}^2$},
		ymajorgrids,
		ymin=0.00072961617297943, ymax=3.32670278045643,
		ymode=log,
		ytick style={color=black},
		xtick={600, 1200, 1800, 2400, 3000, 3600},
		xticklabels={0,10,20,30,40,50}
		]
		\addplot [semithick, color0, mark=triangle*, mark size=1.75, mark options={solid}, only marks]
		table {%
			675 2.33333586142001
			749 0.853092556115803
			812 0.442897779584161
			890 0.246244072337061
			966 0.135595436880131
			1037 0.0797145943456762
			1105 0.0450507766903584
			1191 0.0280341648266049
			1267 0.0188273522552796
			1332 0.0132761445635579
			1402 0.0102451363657111
			1502 0.00805376403700091
			1566 0.00628383621524744
			1633 0.00537195068928746
			1692 0.00475710734941072
			1759 0.00421943709318484
			1828 0.00373865903102936
			1883 0.00337273126411465
			1941 0.00310717364105591
			2006 0.00281351388854218
			2084 0.00260737232718643
			2145 0.00244746377330389
			2203 0.00226750196152121
			2258 0.00217036768792033
			2308 0.00203804645977425
			2364 0.00191090751809603
			2426 0.00187665669230959
			2478 0.00184043215432866
			2530 0.00176780459580434
			2582 0.00173935576997355
		};
		\addlegendentry{\textbf{VREM-FL} (ours)}
		\addplot [semithick, color1, mark=*, mark size=1.75, mark options={solid}, only marks]
		table {%
			699 2.33333586142001
			799 0.726945537355726
			899 0.288899001980842
			996 0.146050714071503
			1096 0.0836145755065814
			1196 0.0368259698699993
			1296 0.022289982773136
			1397 0.0165096625276609
			1497 0.0129173062876742
			1597 0.0104637595491899
			1689 0.00852151609746017
			1789 0.00744049533834833
			1885 0.00642134491498948
			1985 0.00566125219804927
			2085 0.00497976070253144
			2185 0.00433509302162672
			2285 0.00399412438740049
			2385 0.00367738953472874
			2484 0.00334179888761564
			2584 0.00319278282573915
			2684 0.00293687344265747
			2785 0.00285887492293827
			2885 0.00264022250532367
			2981 0.00248008321615337
			3081 0.00243307342095823
			3181 0.00216313583095321
			3281 0.00212590480849489
			3381 0.00210097803746592
			3481 0.00214189964982943
			3581 0.00205431303916492
		};
		\addlegendentry{Fairness\review{~\cite{GunduzComputationTime}}}
		\addplot [semithick, color2, mark=diamond*, mark size=1.75, mark options={solid}, only marks]
		table {%
			700 2.33333586142001
			800 0.908099210359196
			900 0.401300138701901
			1000 0.185152648635043
			1100 0.0987993389463055
			1200 0.047285913313019
			1300 0.0278495404250482
			1400 0.0207905487143592
			1500 0.0157344373094457
			1600 0.0129868443213985
			1700 0.0105637347862774
			1800 0.0090123095328713
			1900 0.00767630717863796
			2000 0.00688276175580384
			2100 0.00598955942123269
			2200 0.00527191672607396
			2300 0.00463717577620376
			2400 0.00407552631048042
			2500 0.00362254167622566
			2600 0.0032592029619292
			2700 0.00306261057445072
			2800 0.0028432674717651
			2900 0.00267383469478831
			3000 0.00255851442676609
			3100 0.00237437759613857
			3201 0.00223515916275939
			3301 0.00210334268540908
			3401 0.00193572437647017
			3501 0.00188825326207389
			3601 0.0018647844137811
		};
		\addlegendentry{\review{FedAvg~\cite{mcmahan2017}}}
		\addplot [semithick, color3, mark=pentagon*, mark size=1.75, mark options={solid}, only marks]
		table {%
			700 2.33333586142001
			801 0.714529514670697
			901 0.28851993487167
			1001 0.157509427002554
			1101 0.087354206131316
			1201 0.0521829958425579
			1301 0.0310450935638149
			1401 0.0208949898976962
			1501 0.0159375688210431
			1601 0.0127988877404029
			1701 0.0101770159878817
			1801 0.00862939458276253
			1901 0.00738006851941269
			2001 0.0064646852838215
			2101 0.00569403385235481
			2201 0.00501450559282826
			2301 0.00439152022642423
			2401 0.00403960688438366
			2501 0.00363793735484017
			2601 0.00337338277856817
			2701 0.003114736344167
			2801 0.00298156142248396
			2901 0.00284109327894249
			3001 0.00268622069658731
			3101 0.00245730098092312
			3201 0.00217908319465244
			3301 0.00211885158759981
			3401 0.00207751535487729
			3501 0.00205861677920064
			3601 0.00202666533824634
		};
		\addlegendentry{Round robin}
	\end{axis}
	
\end{tikzpicture}}
	\caption{\remove{Model convergence as a function of time.}
		\review{Model convergence with} VREM-FL \vs \remove{the considered} \review{scheduling benchmarks} \remove{in terms of client selection algorithm} \review{for linear regression experiment}.
		\review{All algorithms achieve comparable accuracy,
			but \algoname significantly reduces training latency ($33~\si{\minute}$ \vs $50~\si{\minute}$).}
		\remove{with the synthetic dataset (see \autoref{subsec:synthGen})}}
	\label{fig:scheduling}
\end{figure}
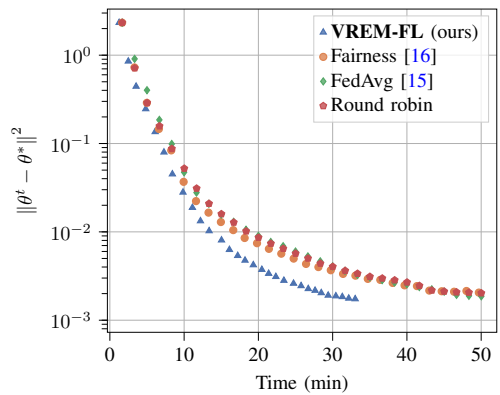

We show the comparison of VREM-FL against scheduling benchmarks in~\autoref{fig:scheduling}. 
``FedAvg'' and ``Round robin'' perform similarly.
The learning accuracy converges, 
but it takes longer than the other two scheduling algorithms.
The target of $30$ rounds is not even reached after the simulation horizon of $50$ minutes. 
Also,
``Fairness'' uses the whole simulation horizon. 
Nonetheless, it yields a slight advantage in learning performance at early stages because it integrates information in a smarter way,
ensuring that all \clients contribute evenly. 
Under the given settings, 
\algoname reduces the total latency by at least $28\%$ while providing the same model accuracy. 
This is achieved by wisely using the network resources through estimated channel conditions.

					%!TEX root = ../main.tex

\subsubsection{Ablation study}
\label{sec:exp-ls-ablation-algo}

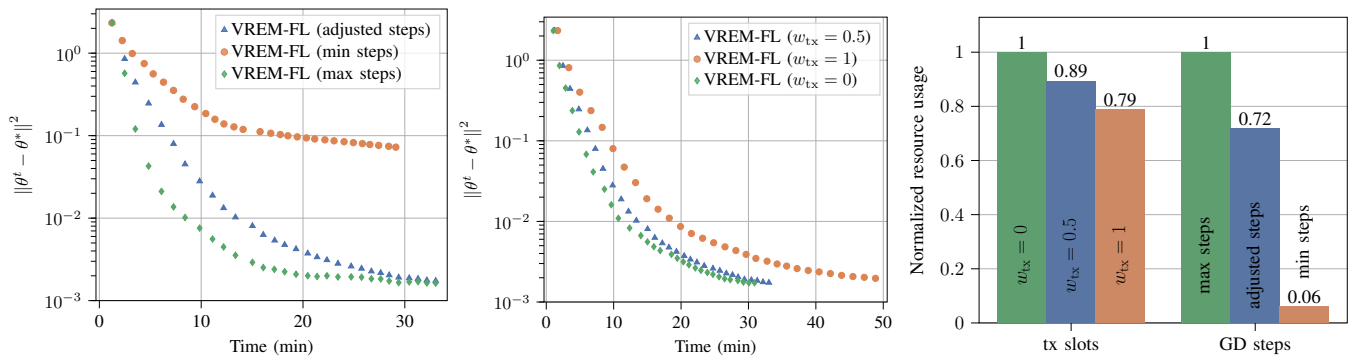
\begin{figure*}
	\centering
	\subfloat[\review{Model convergence varying computation policy}.
	\label{fig:cpu}]{\resizebox{.33\textwidth}{!}{% This file was created by tikzplotlib v0.9.9.
\begin{tikzpicture}

	\definecolor{color0}{RGB}{76,114,176}
	\definecolor{color1}{RGB}{221,132,82}
	\definecolor{color2}{RGB}{85,168,104}
	
	\begin{axis}[
		legend cell align={left},
		legend style={fill opacity=0.8, draw opacity=1, text opacity=1, draw=white!80!black, font=\normalfont},
		log basis y={10},
		tick align=outside,
		tick pos=left,
		x grid style={white!69.0196078431373!black},
		xlabel={Time (min)},
		xmajorgrids,
		xmin=580, xmax=2650,
		xtick style={color=black},
		y grid style={white!69.0196078431373!black},
		ylabel={$\norm{\param[t] - \param[*]}^2$},
		ymajorgrids,
		ymin=0.001, ymax=3.43023837057512,
		ymode=log,
		ytick style={color=black},
		xtick={600,1200,1800,2400},
		xticklabels={0, 10, 20, 30},
		]
		\addplot [semithick, color0, mark=triangle*, mark size=1.75, mark options={solid}, only marks]
		table {%
			675 2.33333586142001
			749 0.853092556115803
			812 0.442897779584161
			890 0.246244072337061
			966 0.135595436880131
			1037 0.0797145943456762
			1105 0.0450507766903584
			1191 0.0280341648266049
			1267 0.0188273522552796
			1332 0.0132761445635579
			1402 0.0102451363657111
			1502 0.00805376403700091
			1566 0.00628383621524744
			1633 0.00537195068928746
			1692 0.00475710734941072
			1759 0.00421943709318484
			1828 0.00373865903102936
			1883 0.00337273126411465
			1941 0.00310717364105591
			2006 0.00281351388854218
			2084 0.00260737232718643
			2145 0.00244746377330389
			2203 0.00226750196152121
			2258 0.00217036768792033
			2308 0.00203804645977425
			2364 0.00191090751809603
			2426 0.00187665669230959
			2478 0.00184043215432866
			2530 0.00176780459580434
			2582 0.00173935576997355
		};
		\addlegendentry{VREM-FL (adjusted steps)}
		\addplot [semithick, color1, mark=*, mark size=1.75, mark options={solid}, only marks]
		table {%
			674 2.33333586142001
			735 1.42195483482975
			794 0.990731071984778
			864 0.748055015499839
			917 0.560781304700199
			978 0.444880543559006
			1037 0.353552093048721
			1094 0.276505226867783
			1162 0.224503738301285
			1227 0.185493961506558
			1281 0.158047014122008
			1334 0.13872445045031
			1393 0.127734341232522
			1447 0.119093758717996
			1547 0.111822224982869
			1610 0.107042414214847
			1666 0.103189940425908
			1711 0.0995989439458539
			1766 0.0966714696285318
			1820 0.0938587222826449
			1869 0.0912416779337241
			1929 0.088951198676083
			1983 0.0866226230765037
			2042 0.0843667239842829
			2097 0.0822594825976238
			2150 0.0801393609416328
			2194 0.0781620778308839
			2246 0.0762445831792462
			2305 0.0743486181960971
			2347 0.0725408022095851
		};
		\addlegendentry{VREM-FL (min steps)}
		\addplot [semithick, color2, mark=diamond*, mark size=1.75, mark options={solid}, only marks]
		table {%
			675 2.33333586142001
			749 0.570795690395965
			812 0.120570130221757
			890 0.0427667296464793
			966 0.0211028117414148
			1037 0.0137000218670984
			1105 0.0101731493457605
			1191 0.00757446847173546
			1267 0.00558405898116279
			1332 0.00447400195166156
			1402 0.00354486023536542
			1502 0.00290987549356871
			1566 0.00251627083714635
			1631 0.0023958973677077
			1692 0.00224167577992413
			1759 0.00208988913865007
			1828 0.00200549986016008
			1883 0.00196377519741194
			1941 0.00200144908117935
			2006 0.00193660347641649
			2084 0.00194088378391313
			2145 0.00192245955787187
			2203 0.0018395613801075
			2258 0.00183293831139496
			2308 0.00173143150119014
			2364 0.0016539515903132
			2426 0.00168670706182245
			2478 0.00170046660242446
			2530 0.00163816408612371
			2582 0.00163111521785991
		};
		\addlegendentry{VREM-FL (max steps)}
	\end{axis}
	
\end{tikzpicture}}}
	\subfloat[\review{Model convergence varying transmission policy}.
	\label{fig:tx}]{\resizebox{.33\textwidth}{!}{% This file was created by tikzplotlib v0.9.9.
\begin{tikzpicture}
	
	\definecolor{color0}{RGB}{76,114,176}
	\definecolor{color1}{RGB}{221,132,82}
	\definecolor{color2}{RGB}{85,168,104}
	
	\begin{axis}[
		legend cell align={left},
		legend style={fill opacity=0.8, draw opacity=1, text opacity=1, draw=white!80!black, font=\normalfont},
		log basis y={10},
		tick align=outside,
		tick pos=left,
		x grid style={white!69.0196078431373!black},
		xlabel={Time (min)},
		xmajorgrids,
		xmin=580, xmax=3635,
		xtick style={color=black},
		y grid style={white!69.0196078431373!black},
		ylabel={$\norm{\param[t] - \param[*]}^2$},
		ymajorgrids,
		ymin=0.001, ymax=3.42670278045643,
		ymode=log,
		ytick style={color=black},
		xtick={600, 1200, 1800, 2400, 3000, 3600},
		xticklabels={0,10,20,30,40,50}
		]
		\addplot [semithick, color0, mark=triangle*, mark size=1.75, mark options={solid}, only marks]
		table {%
			675 2.33333586142001
			749 0.853092556115803
			812 0.442897779584161
			890 0.246244072337061
			966 0.135595436880131
			1037 0.0797145943456762
			1105 0.0450507766903584
			1191 0.0280341648266049
			1267 0.0188273522552796
			1332 0.0132761445635579
			1402 0.0102451363657111
			1502 0.00805376403700091
			1566 0.00628383621524744
			1633 0.00537195068928746
			1692 0.00475710734941072
			1759 0.00421943709318484
			1828 0.00373865903102936
			1883 0.00337273126411465
			1941 0.00310717364105591
			2006 0.00281351388854218
			2084 0.00260737232718643
			2145 0.00244746377330389
			2203 0.00226750196152121
			2258 0.00217036768792033
			2308 0.00203804645977425
			2364 0.00191090751809603
			2426 0.00187665669230959
			2478 0.00184043215432866
			2530 0.00176780459580434
			2582 0.00173935576997355
		};
		\addlegendentry{VREM-FL ($\wtx = 0.5$)}
		\addplot [semithick, color1, mark=*, mark size=1.75, mark options={solid}, only marks]
		table {%
			701 2.33333586142001
			801 0.80812949332428
			898 0.401344706951644
			998 0.23701110153088
			1098 0.146956763088126
			1194 0.0800479732296692
			1294 0.0471997933562695
			1395 0.0303136307227965
			1495 0.0191223091122654
			1595 0.014169734621299
			1695 0.0109908718121477
			1795 0.00864158699424043
			1892 0.00708237535515974
			1992 0.00618394605525423
			2090 0.00543165723965524
			2191 0.00483525541636699
			2288 0.0043535895442204
			2379 0.00389902047959521
			2466 0.00347404874037111
			2566 0.00320975907731772
			2655 0.00296409632937498
			2740 0.00279067722828134
			2833 0.00259862323976005
			2932 0.00246068707683878
			3029 0.00236516717909864
			3129 0.00225474752310631
			3229 0.00214126888327629
			3330 0.00207029211662157
			3431 0.00202435159441779
			3531 0.0019667335179257
		};
		\addlegendentry{VREM-FL ($\wtx = 1$)}
		\addplot [semithick, color2, mark=diamond*, mark size=1.75, mark options={solid}, only marks]
		table {%
			663 2.33333586142001
			718 0.858405314726996
			772 0.452249237478739
			832 0.236855241230514
			892 0.129324875210032
			956 0.0680867002140461
			1018 0.041130889460813
			1118 0.0251093853389025
			1179 0.0160684351112556
			1242 0.0109457448266865
			1342 0.00827842382361575
			1442 0.00667374773513075
			1501 0.00557093518951498
			1558 0.00485885644920086
			1614 0.00434169853690987
			1714 0.00390497968204803
			1766 0.00349241660986959
			1820 0.00315357443176943
			1870 0.00289623896201417
			1927 0.00264546726199518
			1977 0.00247117452908136
			2034 0.00232242657405142
			2084 0.00220314957647238
			2138 0.00206035314295487
			2188 0.00194278930272546
			2238 0.00189478369715876
			2306 0.00184587556054318
			2355 0.00176463143705506
			2404 0.00173258089329449
			2454 0.00173228343725736
		};
		\addlegendentry{VREM-FL ($\wtx = 0$)}
	\end{axis}
	
\end{tikzpicture}}}
	\subfloat[\review{Resource usage under different policies.}
	\label{fig:resources}]{\resizebox{.33\textwidth}{!}{% This file was created by tikzplotlib v0.9.9.
\begin{tikzpicture}

\definecolor{color0}{rgb}{0.347058823529412,0.458823529411765,0.641176470588235}
\definecolor{color1}{rgb}{0.798529411764706,0.536764705882353,0.389705882352941}
\definecolor{color2}{rgb}{0.374019607843137,0.618137254901961,0.429901960784314}

\begin{axis}[
legend cell align={left},
legend style={fill opacity=0.8, draw opacity=1, text opacity=1, draw=white!80!black},
tick align=outside,
tick pos=left,
x grid style={white!69.0196078431373!black},
xlabel={},
xmin=-0.5, xmax=1.5,
xtick style={color=black},
xtick={0,1},
xticklabels={tx slots, GD steps},
y grid style={white!69.0196078431373!black},
ylabel={Normalized resource usage},
ymin=0, ymax=1.12896585410068,
ytick style={color=black},
ymajorgrids,
nodes near coords align=below,
]
\draw[draw=none,fill=color2] (axis cs:-0.4,0) rectangle (axis cs:-0.133333333333333,1);
%\addlegendimage{ybar,ybar legend,draw=none,fill=color2}
%\addlegendentry{max}

\draw[draw=none,fill=color2] (axis cs:0.6,0) rectangle (axis cs:0.866666666666667,1);
\draw[draw=none,fill=color0] (axis cs:-0.133333333333333,0) rectangle (axis cs:0.133333333333333,0.893047);
%\addlegendimage{ybar,ybar legend,draw=none,fill=color0}
%\addlegendentry{opt}

\draw[draw=none,fill=color0] (axis cs:0.866666666666667,0) rectangle (axis cs:1.13333333333333,0.716606);
\draw[draw=none,fill=color1] (axis cs:0.133333333333333,0) rectangle (axis cs:0.4,0.787105);
%\addlegendimage{ybar,ybar legend,draw=none,fill=color1}
%\addlegendentry{min}

\draw[draw=none,fill=color1] (axis cs:1.13333333333333,0) rectangle (axis cs:1.4,0.0600196);
\draw (axis cs:-0.266666666666667,1) ++(0pt,0pt) node[
  scale=0.5,
  anchor=south,
  text=black,
  rotate=0.0
]{\huge{1}};
\draw (axis cs:-0.266666666666667,0.25) ++(0pt,0pt) node[
  scale=0.5,
  anchor=center,
  text=black,
  rotate=90.0
]{\huge{$\wtx = 0$}};
\draw (axis cs:0.733333333333333,1) ++(0pt,0pt) node[
  scale=0.5,
  anchor=south,
  text=black,
  rotate=0.0
]{\huge{1}};
\draw (axis cs:0.733333333333333,0.25) ++(0pt,0pt) node[
  scale=0.5,
  anchor=center,
  text=black,
  rotate=90.0
]{\huge{max steps}};
\draw (axis cs:0,0.893047) ++(0pt,0pt) node[
  scale=0.5,
  anchor=south,
  text=black,
  rotate=0.0
]{\huge{0.89}};
\draw (axis cs:0,0.25) ++(0pt,0pt) node[
  scale=0.5,
  anchor=center,
  text=black,
  rotate=90.0
]{\huge{$\wtx = 0.5$}};
\draw (axis cs:1,0.716606) ++(0pt,0pt) node[
  scale=0.5,
  anchor=south,
  text=black,
  rotate=0.0
]{\huge{0.72}};
\draw (axis cs:1,0.25) ++(0pt,0pt) node[
  scale=0.5,
  anchor=center,
  text=black,
  rotate=90.0
]{\huge{adjusted steps}};
\draw (axis cs:0.266666666666667,0.787105) ++(0pt,0pt) node[
  scale=0.5,
  anchor=south,
  text=black,
  rotate=0.0
]{\huge{0.79}};
\draw (axis cs:0.266666666666667,0.25) ++(0pt,0pt) node[
  scale=0.5,
  anchor=center,
  text=black,
  rotate=90.0
]{\huge{$\wtx = 1$}};
\draw (axis cs:1.26666666666667,0.0600196) ++(0pt,0pt) node[
  scale=0.5,
  anchor=south,
  text=black,
  rotate=0.0
]{\huge{0.06}};
\draw (axis cs:1.26666666666667,0.3) ++(0pt,0pt) node[
  scale=0.5,
  anchor=center,
  text=black,
  rotate=90.0
]{\huge{min steps}};
\end{axis}

\end{tikzpicture}}}
	\caption{\review{Ablation study on \algoname with linear regression experiment}.
		\remove{with the synthetic dataset (see \autoref{sec:exp-ls-dataset})}
		\review{For computation policies,
			we vary the number of local \acs{GD} steps with fixed transmission weight $\wtx = 0.5$.
			For transmission policies,
			we vary the number of slots when \clients occupy the channel while using the adjusted steps computation policy.
		}
		\remove{while varying the policies relative to the computation and communication (number of slots where the clients fill the channel) resources.}}
	\label{fig:ablation}
\end{figure*}
We propose an ablation study to isolate the effects of the proposed local steps adaptation strategy (\autoref{fig:cpu}) and transmission policy (\autoref{fig:tx}). 
\Cref{fig:cpu} shows that performing the minimum number of GD steps (label ``min steps'', \ie only one step) locally at each round makes convergence steady but slow.
The policy labeled ``max steps'' is obtained by filling all idle slots between computation and transmission (see~\autoref{fig:access_node}) with additional GD steps. 
In this way,
the total latency is the same of our optimized solution and local models are transmitted during the same time slots, 
but the total number of local iterations is higher than the proposed \algoname (label ``adjusted steps'').
Indeed,
the latter version may limit the number of local steps based on convergence proxies~\eqref{eq:local-proxy} and~\eqref{eq:global-proxy}. 
Although ``max steps'' is initially faster than ``adjusted steps'', 
both solutions converge after $40$ minutes (\ie approximately $25$ rounds). 
\Cref{fig:resources} shows the GD steps normalized with respect to the strategy ``max steps'' \remove{referred to this experiment} on the right. 
Strategy ``min steps'' uses only $6\%$ of the total steps and its convergence is too slow. 
Noteworthy, ``adjusted steps'' reduces the GD steps by $28\%$, 
which directly translates into higher energy efficiency and better usage of computation resources while reaching the same accuracy in a comparable time.  

%Figure~\ref{fig:tx} shows the results relative to varying the coefficients in~\eqref{eq:local-communication-optimization-obj} for communication optimization.
In \autoref{fig:tx}, the results relative to varying the weight $\wtx$ in the optimization problem~\eqref{eq:local-communication-optimization} are shown. 
Specifically, tuning $\wtx$ between $0$ and $1$ makes the solution move along the Pareto front between the two extremes $\wtx=1$ (orange circles),
which corresponds to minimizing the number of slots where the channel is filled with communication, and $\wtx=0$ (green diamonds), which corresponds to minimizing the total latency. 
Setting $\wtx=0$ corresponds to transmitting the local models as soon as the GD steps are completed,
while the policy $\wtx=1$ uses the REM and waits for the available transmission window with the highest bitrate. 
Any coefficients in between correspond to a weighted solution between these two criteria.
For example, we show the results for $\wtx=0.5$ (blue triangles). 
As $\wtx \rightarrow 0$, convergence is faster because this solution minimizes the total latency. 
The scenario with $\wtx=0.5$ is reasonably close and it takes $2$ minute longer to complete the given $30$ rounds.
Conversely, 
setting $\wtx=1$ penalizes only the usage of transmission resources and uses all the available $50$ minutes. 
By looking at the resource usage (\autoref{fig:resources}, leftmost bars), 
we see that,
in this context, 
it is possible to reduce the number of slots used for communications by at most $21\%$ (\ie from $\wtx=0$ to $\wtx=1$). 
Interestingly, by setting $\wtx=0.5$, we reduce the resource usage by $11\%$,
hence significantly improving the efficiency while performing very close to $\wtx=0$ in terms of overall training latency. 
			%!TEX root = ../main.tex

\subsection{Second Experiment: Deep Learning for Real-World Semantic Segmentation}
\label{sec:exp-nn}

\review{In this section,
we address a vision-based semantic segmentation task by training a deep neural network model.
This experiment allows us to validate the effectiveness of \algoname for real-world applications.
In \autoref{sec:exp-nn-apollo},
we describe the real-world 
semantic segmentation dataset Apollo,
while in \autoref{sec:exp-nn-params} we report details on the learning model and \algoname parameters.
Then,
we perform two sets of experiments with different mobility patterns.
For both experiments, 
we use the bitrate obtained through the estimation method of \autoref{sec:results_estimated_REM} with $250$ samples. 
In \autoref{sec:exp-nn-sumo},
we use the vehicular mobility simulated with \acs{SUMO} for the first experiment.
In~\autoref{sec:exp-nn-taxi},
we use a real-world dataset with taxi trips in Rome, Italy.}
%Even though the \clients' routes are just \textit{inputs} to \algoname,
%this experiment indeed corroborates its effectiveness in a real-world mobility scenario.}
					%!TEX root = ../main.tex

\subsubsection{Real-world dataset ApolloScape and learning performance}
\label{sec:exp-nn-apollo}

\review{We use the real-world \texttt{ApolloScape} lane segmentation dataset~\cite{huang2018apolloscape} to show the performance of \algoname on a realistic task. 
	It consists of more than $110$ thousand annotated frames from $73$ street scene videos recorded in China with various weather conditions. 
	We realistically split the dataset by assigning one record to each available vehicle,
    so that data are correlated in time and space within the same client and are non-iid across clients.
	In words,
	we simulate video streams independently acquired by \clients with onboard cameras while traveling.
	We downsampled the frames to make the training compatible with limited computing resources of \clients.
}

To measure the \review{learning} performance,
we use the \ac{mIoU} score, 
a popular evaluation metric for segmentation tasks.
The \ac{mIoU} is obtained by first computing the ratio of the area of the intersection between predicted and ground truth regions to the area of their union, 
and then taking their average. 
Formally, 
let $\mathcal{R}$ denote the set of semantic regions and $\vert\mathcal{R}\vert$ denote its cardinality,
then it holds
\begin{equation}
	\label{eq:miou}
	\text{mIoU} \doteq \frac{1}{\vert\mathcal{R}\vert}\sum_{r\in\mathcal{R}} \frac{r^\text{pred} \cap r^\text{true}}{r^\text{pred} \cup r^\text{true}}.
\end{equation}
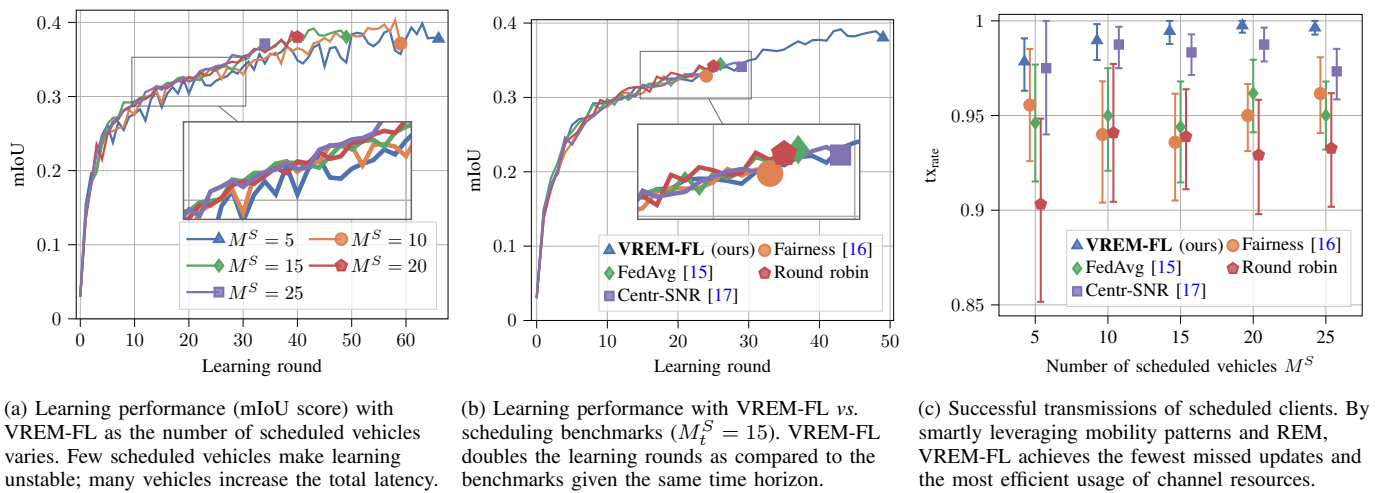
\begin{figure*}
	\centering
	\subfloat[Learning \review{performance (\acs{mIoU} score)} with \algoname \review{as the number of scheduled \clients varies.
		Few scheduled \clients make learning unstable;
		many \clients increase the total latency.
	}
	\label{fig:n_clients}]{\resizebox{.33\textwidth}{!}{%!TEX root = ../main.tex

\begin{tikzpicture}[spy using outlines={magnification=2, rectangle, width=4.2cm, height=1.8cm, white!40!black, connect spies, }]
	
	\definecolor{color0}{rgb}{0.298039215686275,0.447058823529412,0.690196078431373}
	\definecolor{color1}{rgb}{0.866666666666667,0.517647058823529,0.32156862745098}
	\definecolor{color2}{rgb}{0.333333333333333,0.658823529411765,0.407843137254902}
	\definecolor{color3}{rgb}{0.768627450980392,0.305882352941176,0.32156862745098}
	\definecolor{color4}{rgb}{0.505882352941176,0.447058823529412,0.701960784313725}
	\definecolor{lightgray204}{RGB}{204,204,204}
	
	\begin{axis}[
		legend cell align={left},
		legend style={fill opacity=0.8, draw opacity=1, text opacity=1, draw=lightgray204},
		legend pos=south east,
		legend columns=2,
		tick align=outside,
		tick pos=left,
		x grid style={white!69.0196078431373!black},
		xlabel={Learning round},
		xmajorgrids,
		xmin=-0.75, xmax=67.75,
		xtick style={color=black},
		y grid style={white!69.0196078431373!black},
		ylabel={mIoU},
		ymajorgrids,
		ymin=0, ymax=0.416446551680565,
		ytick style={color=black},
		%legend image post style={mark indices={}}
		]
		\addplot+[very thick, color0, mark=triangle*, mark size=3, mark options={solid, mark indices=67}, forget plot]
		table {%
			0 0.03
			1 0.15426766872406
			2 0.196397304534912
			3 0.200714871287346
			4 0.247574403882027
			5 0.261715412139893
			6 0.237930566072464
			7 0.265266388654709
			8 0.254858314990997
			9 0.272626757621765
			10 0.276408761739731
			11 0.282615303993225
			12 0.267503589391708
			13 0.295877009630203
			14 0.306293696165085
			15 0.285524278879166
			16 0.303648561239242
			17 0.312726497650146
			18 0.303985685110092
			19 0.323117285966873
			20 0.30393847823143
			21 0.32114976644516
			22 0.328525871038437
			23 0.315311282873154
			24 0.314583778381348
			25 0.321263700723648
			26 0.32455763220787
			27 0.327127486467361
			28 0.334221333265305
			29 0.330893099308014
			30 0.340272665023804
			31 0.347026973962784
			32 0.336232930421829
			33 0.357008725404739
			34 0.344593614339828
			35 0.357591420412064
			36 0.356251776218414
			37 0.359825730323792
			38 0.367981970310211
			39 0.362394213676453
			40 0.348733156919479
			41 0.344496488571167
			42 0.352526992559433
			43 0.366617411375046
			44 0.370374530553818
			45 0.340676486492157
			46 0.365052938461304
			47 0.366705328226089
			48 0.352354764938354
			49 0.345822811126709
			50 0.383851081132889
			51 0.385833770036697
			52 0.384486794471741
			53 0.372162580490112
			54 0.367785215377808
			55 0.373659193515778
			56 0.386312752962112
			57 0.389315247535706
			58 0.361914545297623
			59 0.35717710852623
			60 0.378054827451706
			61 0.392904460430145
			62 0.384106487035751
			63 0.398454427719116
			64 0.378488153219223
			65 0.379293322563171
			66 0.377886265516281
		};
		\addlegendimage{very thick, color0, mark=triangle*, mark options={mark indices=3}, mark size=3}
		\addlegendentry{$\vsetsccard{}=5$}
		\addplot+[very thick, color1, mark=*, mark size=3, mark options={solid, mark indices=60}, forget plot]
		table {%
			0 0.03
			1 0.143306836485863
			2 0.180149987339973
			3 0.205717116594314
			4 0.233514010906219
			5 0.251493901014328
			6 0.253265500068664
			7 0.257279545068741
			8 0.259317576885223
			9 0.281907051801682
			10 0.275620222091675
			11 0.292076170444489
			12 0.297143965959549
			13 0.301203578710556
			14 0.307414442300797
			15 0.291248232126236
			16 0.316446423530579
			17 0.317703038454056
			18 0.324598342180252
			19 0.326413482427597
			20 0.319074779748917
			21 0.325676411390305
			22 0.325804740190506
			23 0.337100833654404
			24 0.336329489946365
			25 0.336865425109863
			26 0.323534905910492
			27 0.337482184171677
			28 0.337749421596527
			29 0.336194217205048
			30 0.329705268144607
			31 0.342333644628525
			32 0.349029153585434
			33 0.349126756191254
			34 0.351050555706024
			35 0.356295496225357
			36 0.35912275314331
			37 0.344651341438293
			38 0.368147522211075
			39 0.360371798276901
			40 0.365257829427719
			41 0.362484574317932
			42 0.356657236814499
			43 0.371829032897949
			44 0.373937040567398
			45 0.379012435674667
			46 0.373943507671356
			47 0.38594862818718
			48 0.380143076181412
			49 0.3763587474823
			50 0.382613062858581
			51 0.38092365860939
			52 0.381798267364502
			53 0.392410963773727
			54 0.390527367591858
			55 0.391819715499878
			56 0.39716249704361
			57 0.38715186715126
			58 0.402982890605927
			59 0.371337950229645
		};
		\addlegendimage{very thick, color1, mark=*, mark options={mark indices=3}, mark size=3}
		\addlegendentry{$\vsetsccard{} = 10$}
		\addplot+[very thick, color2, mark=diamond*, mark size=3, mark options={solid, mark indices=50}, forget plot]
		table {%
			0 0.03
			1 0.136917069554329
			2 0.184220820665359
			3 0.197270050644874
			4 0.236371919512749
			5 0.262913048267365
			6 0.270995438098907
			7 0.272052496671677
			8 0.292096108198166
			9 0.291834950447083
			10 0.2926344871521
			11 0.298482954502106
			12 0.299544870853424
			13 0.300683736801147
			14 0.313879162073135
			15 0.312760472297668
			16 0.31836262345314
			17 0.31849867105484
			18 0.324597001075745
			19 0.325404345989227
			20 0.32512354850769
			21 0.323833614587784
			22 0.321960091590881
			23 0.329784035682678
			24 0.337255209684372
			25 0.334385871887207
			26 0.339867889881134
			27 0.340722769498825
			28 0.336114495992661
			29 0.345849573612213
			30 0.349622160196304
			31 0.351895391941071
			32 0.364183872938156
			33 0.363447219133377
			34 0.362266659736633
			35 0.364459216594696
			36 0.370034158229828
			37 0.375414967536926
			38 0.371384352445602
			39 0.376884251832962
			40 0.377445638179779
			41 0.376256465911865
			42 0.388925731182098
			43 0.391718596220016
			44 0.382584750652313
			45 0.38218492269516
			46 0.38507154583931
			47 0.387276142835617
			48 0.388964146375656
			49 0.380482196807861
		};
		\addlegendimage{very thick, color2, mark=diamond*, mark options={mark indices=3}, mark size=3}
		\addlegendentry{$\vsetsccard{} = 15$}
		\addplot+[very thick, color3, mark=pentagon*, mark size=3, mark options={solid, mark indices=41}, forget plot]
		table {%
			0 0.03
			1 0.133709669113159
			2 0.18105611205101
			3 0.234858468174934
			4 0.215229988098145
			5 0.248106628656387
			6 0.262856304645538
			7 0.268905848264694
			8 0.281645983457565
			9 0.285945296287537
			10 0.2874935567379
			11 0.296888291835785
			12 0.296523243188858
			13 0.301623463630676
			14 0.310072511434555
			15 0.312769532203674
			16 0.311116009950638
			17 0.316374033689499
			18 0.321750909090042
			19 0.318407744169235
			20 0.326162278652191
			21 0.324947655200958
			22 0.332738667726517
			23 0.32796248793602
			24 0.329629123210907
			25 0.338787257671356
			26 0.339648008346558
			27 0.344343245029449
			28 0.346845746040344
			29 0.34538409113884
			30 0.354157447814941
			31 0.356391280889511
			32 0.363623410463333
			33 0.364161223173141
			34 0.359792143106461
			35 0.364376127719879
			36 0.368342727422714
			37 0.376589506864548
			38 0.367274045944214
			39 0.387277752161026
			40 0.380225211381912
		};
		\addlegendimage{very thick, color3, mark=pentagon*, mark options={mark indices=3}, mark size=3}
		\addlegendentry{$\vsetsccard{} = 20$}
		\addplot+[very thick, color4, mark=square*, mark size=2.25, mark options={solid, mark indices=35}, forget plot]
		table {%
			0 0.03
			1 0.141066834330559
			2 0.182010173797607
			3 0.222498923540115
			4 0.246907234191895
			5 0.252149194478989
			6 0.266292661428451
			7 0.271349281072617
			8 0.280638217926025
			9 0.283688426017761
			10 0.293873518705368
			11 0.289517611265182
			12 0.305985361337662
			13 0.309185594320297
			14 0.31341490149498
			15 0.308967351913452
			16 0.314956396818161
			17 0.319019734859466
			18 0.320700585842133
			19 0.32496452331543
			20 0.320928156375885
			21 0.322194695472717
			22 0.333972036838531
			23 0.33675029873848
			24 0.337662220001221
			25 0.340989679098129
			26 0.341958463191986
			27 0.345142900943756
			28 0.356126368045807
			29 0.354535967111588
			30 0.353348314762115
			31 0.355006009340286
			32 0.360868185758591
			33 0.366358667612076
			34 0.371159702539444
		};
		\addlegendimage{very thick, color4, mark=square*, mark options={mark indices=3}, mark size=2.25}
		\addlegendentry{$\vsetsccard{} = 25$}
		%\legend{$M=5$, $M=10$, $M=15$, $M=20$, $M=25$};
		\coordinate (c1) at (axis cs: 20, 0.32);
		\coordinate (c2) at (axis cs: 40, 0.2);
		\spy on (c1) in node[fill=white] at (c2);
	\end{axis}
	
\end{tikzpicture}}}
	\hfill
	\subfloat[\review{Learning performance with \algoname \vs scheduling benchmarks ($\vsetsccard{t}=15$).
	\algoname doubles the learning rounds as compared to the benchmarks given the same time horizon.}
	\label{fig:scheduling-apollo}]{\resizebox{.33\textwidth}{!}{%!TEX root = ../main.tex

% This file was created by tikzplotlib v0.9.9.
\begin{tikzpicture}[spy using outlines={magnification=2, rectangle, width=4.2cm, height=1.8cm, white!40!black, connect spies, }]
	
	\definecolor{color0}{rgb}{0.298039215686275,0.447058823529412,0.690196078431373}
	\definecolor{color1}{rgb}{0.866666666666667,0.517647058823529,0.32156862745098}
	\definecolor{color2}{rgb}{0.333333333333333,0.658823529411765,0.407843137254902}
	\definecolor{color3}{rgb}{0.768627450980392,0.305882352941176,0.32156862745098}
    \definecolor{color4}{rgb}{0.505882352941176,0.447058823529412,0.701960784313725}
	\definecolor{lightgray204}{RGB}{204,204,204}
	
	\begin{axis}[
		legend cell align={left},
		legend style={fill opacity=0.8, draw opacity=1, text opacity=1, draw=lightgray204},
		legend pos=south east,
		legend columns=2,
		tick align=outside,
		tick pos=left,
		x grid style={white!69.0196078431373!black},
		xlabel={Learning round},
		xmajorgrids,
		xmin=-0.75, xmax=50.5,
		xtick style={color=black},
		y grid style={white!69.0196078431373!black},
		ylabel={mIoU},
		ymajorgrids,
		ymin=0, ymax=0.404458672553301,
		ytick style={color=black}
		]
		\addplot+[very thick, color0, mark=triangle*, mark size=3, mark options={solid, mark indices=50}, forget plot]
		table {%
			0 0.03
			1 0.136917069554329
			2 0.184220820665359
			3 0.197270050644874
			4 0.236371919512749
			5 0.262913048267365
			6 0.270995438098907
			7 0.272052496671677
			8 0.292096108198166
			9 0.291834950447083
			10 0.2926344871521
			11 0.298482954502106
			12 0.299544870853424
			13 0.300683736801147
			14 0.313879162073135
			15 0.312760472297668
			16 0.31836262345314
			17 0.31849867105484
			18 0.324597001075745
			19 0.325404345989227
			20 0.32512354850769
			21 0.323833614587784
			22 0.321960091590881
			23 0.329784035682678
			24 0.337255209684372
			25 0.334385871887207
			26 0.339867889881134
			27 0.340722769498825
			28 0.336114495992661
			29 0.345849573612213
			30 0.349622160196304
			31 0.351895391941071
			32 0.364183872938156
			33 0.363447219133377
			34 0.362266659736633
			35 0.364459216594696
			36 0.370034158229828
			37 0.375414967536926
			38 0.371384352445602
			39 0.376884251832962
			40 0.377445638179779
			41 0.376256465911865
			42 0.388925731182098
			43 0.391718596220016
			44 0.382584750652313
			45 0.38218492269516
			46 0.38507154583931
			47 0.387276142835617
			48 0.388964146375656
			49 0.380482196807861
		};
		\addlegendimage{only marks, very thick, color0, mark=triangle*, mark options={mark indices=3}, mark size=3}
		\addlegendentry{\textbf{VREM-FL} (ours)}
		\addplot [very thick, color1, mark=*, mark size=3, mark options={solid, mark indices=25}, forget plot]
		table {%
			0 0.03
			1 0.144283324480057
			2 0.178732976317406
			3 0.200921967625618
			4 0.225503921508789
			5 0.238948345184326
			6 0.258273482322693
			7 0.272080063819885
			8 0.274417638778687
			9 0.284610837697983
			10 0.287422835826874
			11 0.295020937919617
			12 0.305659592151642
			13 0.306031435728073
			14 0.303107589483261
			15 0.305523633956909
			16 0.316770821809769
			17 0.319173067808151
			18 0.3186896443367
			19 0.319089949131012
			20 0.323007017374039
			21 0.331868708133698
			22 0.327499985694885
			23 0.33142751455307
			24 0.328839242458344
		};
		\addlegendimage{only marks, very thick, color1, mark=*, mark options={mark indices=3}, mark size=3}
		\addlegendentry{Fairness\review{~\cite{GunduzComputationTime}}}
		\addplot [very thick, color2, mark=diamond*, mark size=3, mark options={solid, mark indices=27}, forget plot]
		table {%
			0 0.03
			1 0.148537427186966
			2 0.190939247608185
			3 0.212860986590385
			4 0.243344813585281
			5 0.253902494907379
			6 0.266270846128464
			7 0.27303472161293
			8 0.290686130523682
			9 0.284579247236252
			10 0.297641962766647
			11 0.29016700387001
			12 0.303792387247086
			13 0.296774238348007
			14 0.304936408996582
			15 0.314841270446777
			16 0.318204015493393
			17 0.315589904785156
			18 0.328173696994781
			19 0.316070646047592
			20 0.329036295413971
			21 0.325723439455032
			22 0.330133587121964
			23 0.339832901954651
			24 0.336940973997116
			25 0.34617018699646
			26 0.344876080751419
		};
		\addlegendimage{only marks, very thick, color2, mark=diamond*, mark options={mark indices=3}, mark size=3}
		\addlegendentry{\review{FedAvg~\cite{mcmahan2017}}}
		\addplot [very thick, color3, mark=pentagon*, mark size=3, mark options={solid, mark indices=26}, forget plot]
		table {%
			0 0.03
			1 0.138769179582596
			2 0.175301715731621
			3 0.203264340758324
			4 0.238184794783592
			5 0.247472032904625
			6 0.254564553499222
			7 0.273452967405319
			8 0.276583135128021
			9 0.296668440103531
			10 0.292583525180817
			11 0.297805935144424
			12 0.307431548833847
			13 0.312425225973129
			14 0.313977926969528
			15 0.313537806272507
			16 0.30768272280693
			17 0.327921211719513
			18 0.324173696994781
			19 0.333070646047592
			20 0.331036295413971
			21 0.327723439455032
			22 0.329133587121964
			23 0.342832901954651
			24 0.337940973997116
			25 0.34217018699646
		};
		\addlegendimage{only marks, very thick, color3, mark=pentagon*, mark options={mark indices=3}, mark size=3}
		\addlegendentry{Round robin}
  \addplot+[very thick, color4, mark=square*, mark size=2.25, mark options={solid, mark indices=30}, forget plot]
  table {%
0 0.03
1 0.149585321545601
2 0.18137389421463
3 0.202846020460129
4 0.245930150151253
5 0.236531600356102
6 0.252171725034714
7 0.267236709594727
8 0.274443507194519
9 0.282312124967575
10 0.290414303541184
11 0.296765089035034
12 0.297072619199753
13 0.303363531827927
14 0.301505416631699
15 0.31730905175209
16 0.313160836696625
17 0.31491830945015
18 0.316444158554077
19 0.323038578033447
20 0.326779782772064
21 0.330115377902985
22 0.331881999969482
23 0.330132961273193
24 0.336367100477219
25 0.34016677737236
26 0.338850200176239
27 0.343916267156601
28 0.346666902303696
29 0.341241627931595
};
\addlegendimage{only marks, very thick, color4, mark=square*, mark options={mark indices=3}, mark size=2.25}
\addlegendentry{\review{Centr-SNR~\cite{chen2024efficient}}}
		\coordinate (c1) at (axis cs: 22.5, 0.33);
		\coordinate (c2) at (axis cs: 30, 0.2);
		\spy on (c1) in node[fill=white] at (c2);
	\end{axis}
	
\end{tikzpicture}}}
	\hfill
	\subfloat[\review{Successful transmissions of scheduled clients.
		By smartly leveraging mobility patterns and \acs{REM},
		\algoname achieves the fewest missed updates and the most efficient usage of channel resources.
	}
	\remove{Scheduled clients delivering the weights in time.}
	\label{fig:client-fraction}]{\resizebox{.33\textwidth}{!}{%!TEX root = ../main.tex

% This file was created by tikzplotlib v0.9.9.
\begin{tikzpicture}
	
	\definecolor{color0}{rgb}{0.298039215686275,0.447058823529412,0.690196078431373}
	\definecolor{color1}{rgb}{0.866666666666667,0.517647058823529,0.32156862745098}
	\definecolor{color2}{rgb}{0.333333333333333,0.658823529411765,0.407843137254902}
	\definecolor{color3}{rgb}{0.768627450980392,0.305882352941176,0.32156862745098}
	\definecolor{color4}{rgb}{0.505882352941176,0.447058823529412,0.701960784313725}
	
	\begin{axis}[
		legend cell align={left},
		legend style={
			fill opacity=0.8,
			draw opacity=1,
			text opacity=1,
			at={(0.97,0.03)},
			anchor=south east,
			draw=white!80!black
		},
		legend columns=2
		tick align=outside,
		tick pos=left,
		unbounded coords=jump,
		x grid style={white!69.0196078431373!black},
		xlabel={Number of scheduled \clients $\vsetsccard{}$},
		xmajorgrids,
		xmin=-0.5, xmax=4.5,
		xtick style={color=black},
		xticklabels={0, 5, 10, 15, 20, 25},
		y grid style={white!69.0196078431373!black},
		ylabel={\review{$\txrate$}}, %{Fraction of active \clients},
		ymajorgrids,
		ymin=0.844193557673885, ymax=1.00341935439648,
		ytick style={color=black},
		ytick align=outside,
		xtick align=outside
		]
		\addplot [line width=1.08pt, color0, mark=triangle*, mark size=3, mark options={solid}, only marks]
		table {%
			-0.15 0.978461539745331
			0.85 0.98965517097506
			1.85 0.994557823453631
			2.85 0.997499999403954
			3.85 0.996363634412939
		};
		\addlegendentry{\textbf{VREM-FL} (ours)}
		\addplot [line width=1.08pt, color0, forget plot]
		table {%
			-0.2 0.96307692527771
			-0.1 0.96307692527771
			nan nan
			-0.15 0.96307692527771
			-0.15 0.990769231319428
			nan nan
			-0.2 0.990769231319428
			-0.1 0.990769231319428
		};
		\addplot [line width=1.08pt, color0, forget plot]
		table {%
			0.8 0.979310342977787
			0.9 0.979310342977787
			nan nan
			0.85 0.979310342977787
			0.85 0.998275861657899
			nan nan
			0.8 0.998275861657899
			0.9 0.998275861657899
		};
		\addplot [line width=1.08pt, color0, forget plot]
		table {%
			1.8 0.987755102770669
			1.9 0.987755102770669
			nan nan
			1.85 0.987755102770669
			1.85 1
			nan nan
			1.8 1
			1.9 1
		};
		\addplot [line width=1.08pt, color0, forget plot]
		table {%
			2.8 0.993749998509884
			2.9 0.993749998509884
			nan nan
			2.85 0.993749998509884
			2.85 1
			nan nan
			2.8 1
			2.9 1
		};
		\addplot [line width=1.08pt, color0, forget plot]
		table {%
			3.8 0.992727268825878
			3.9 0.992727268825878
			nan nan
			3.85 0.992727268825878
			3.85 1
			nan nan
			3.8 1
			3.9 1
		};
		\addplot [line width=1.08pt, color1, mark=*, mark size=3, mark options={solid}, only marks]
		table {%
			-0.075 0.955555558204651
			0.925 0.939999995231628
			1.925 0.935897439718246
			2.925 0.949999993046125
			3.925 0.961599996089935
		};
		\addlegendentry{Fairness\review{~\cite{GunduzComputationTime}}}
		\addplot [line width=1.08pt, color1, forget plot]
		table {%
			-0.125 0.925925930341085
			-0.025 0.925925930341085
			nan nan
			-0.075 0.925925930341085
			-0.075 0.985185186068217
			nan nan
			-0.125 0.985185186068217
			-0.025 0.985185186068217
		};
		\addplot [line width=1.08pt, color1, forget plot]
		table {%
			0.875 0.903999996185303
			0.975 0.903999996185303
			nan nan
			0.925 0.903999996185303
			0.925 0.968099999427795
			nan nan
			0.875 0.968099999427795
			0.975 0.968099999427795
		};
		\addplot [line width=1.08pt, color1, forget plot]
		table {%
			1.975 0.905128210783005
			1.875 0.905128210783005
			nan nan
			1.925 0.905128210783005
			1.925 0.961538463830948
			nan nan
			1.875 0.961538463830948
			1.975 0.961538463830948
		};
		\addplot [line width=1.08pt, color1, forget plot]
		table {%
			2.875 0.931249998447796
			2.975 0.931249998447796
			nan nan
			2.925 0.931249998447796
			2.925 0.966666661202908
			nan nan
			2.875 0.966666661202908
			2.975 0.966666661202908
		};
		\addplot [line width=1.08pt, color1, forget plot]
		table {%
			3.875 0.940759989917278
			3.975 0.940759989917278
			nan nan
			3.925 0.940759989917278
			3.925 0.980799996852875
			nan nan
			3.875 0.980799996852875
			3.975 0.980799996852875
		};
		\addplot [line width=1.08pt, color2, mark=diamond*, mark size=3, mark options={solid}, only marks]
		table {%
			0 0.946153849363327
			1 0.949999995529652
			2 0.94400000333786
			3 0.961764700272504
			4 0.949999991059303
		};
		\addlegendentry{\review{FedAvg~\cite{mcmahan2017}}}
		\addplot [line width=1.08pt, color2, forget plot]
		table {%
			-0.05 0.91519231274724
			0.05 0.91519231274724
			nan nan
			0 0.91519231274724
			0 0.976923078298569
			nan nan
			-0.05 0.976923078298569
			0.05 0.976923078298569
		};
		\addplot [line width=1.08pt, color2, forget plot]
		table {%
			0.95 0.920833329297602
			1.05 0.920833329297602
			nan nan
			1 0.920833329297602
			1 0.974999996523062
			nan nan
			0.95 0.974999996523062
			1.05 0.974999996523062
		};
		\addplot [line width=1.08pt, color2, forget plot]
		table {%
			1.95 0.91466667175293
			2.05 0.91466667175293
			nan nan
			2 0.91466667175293
			2 0.968000001907349
			nan nan
			1.95 0.968000001907349
			2.05 0.968000001907349
		};
		\addplot [line width=1.08pt, color2, forget plot]
		table {%
			2.95 0.941176459982115
			3.05 0.941176459982115
			nan nan
			3 0.941176459982115
			3 0.979485292645062
			nan nan
			2.95 0.979485292645062
			3.05 0.979485292645062
		};
		\addplot [line width=1.08pt, color2, forget plot]
		table {%
			3.95 0.931999990344048
			4.05 0.931999990344048
			nan nan
			4 0.931999990344048
			4 0.967999988868833
			nan nan
			3.95 0.967999988868833
			4.05 0.967999988868833
		};
		\addplot [line width=1.08pt, color3, mark=pentagon*, mark size=3, mark options={solid}, only marks]
		table {%
			0.075 0.903225812219804
			1.075 0.940909087657928
			2.075 0.938888892531395
			3.075 0.929166662196318
			4.075 0.93272726373239
		};
		\addlegendentry{Round robin}
		\addplot [line width=1.08pt, color3, forget plot]
		table {%
			0.025 0.851612912070367
			0.125 0.851612912070367
			nan nan
			0.075 0.851612912070367
			0.075 0.948387099850562
			nan nan
			0.025 0.948387099850562
			0.125 0.948387099850562
		};
		\addplot [line width=1.08pt, color3, forget plot]
		table {%
			1.025 0.904431811787865
			1.125 0.904431811787865
			nan nan
			1.075 0.904431811787865
			1.075 0.977272724563425
			nan nan
			1.025 0.977272724563425
			1.125 0.977272724563425
		};
		\addplot [line width=1.08pt, color3, forget plot]
		table {%
			2.025 0.911111116409302
			2.125 0.911111116409302
			nan nan
			2.075 0.911111116409302
			2.075 0.963888891041279
			nan nan
			2.025 0.963888891041279
			2.125 0.963888891041279
		};
		\addplot [line width=1.08pt, color3, forget plot]
		table {%
			3.025 0.897916667163372
			3.125 0.897916667163372
			nan nan
			3.075 0.897916667163372
			3.075 0.958333325944841
			nan nan
			3.025 0.958333325944841
			3.125 0.958333325944841
		};
		\addplot [line width=1.08pt, color3, forget plot]
		table {%
			4.025 0.901818175139752
			4.125 0.901818175139752
			nan nan
			4.075 0.901818175139752
			4.075 0.961818174882369
			nan nan
			4.025 0.961818174882369
			4.125 0.961818174882369
		};
		\addplot [line width=1.08pt, color4, mark=square*, mark size=2.25, mark options={solid}, only marks]
		table{%
			0.15 0.975000023841858
			1.15 0.987500011920929
			2.15 0.98333328962326
			3.15 0.987500071525574
			4.15 0.973333299160004
		};
		\addlegendentry{\review{Centr-SNR~\cite{chen2024efficient}}}
		\addplot [line width=1.08pt, color4, forget plot]
		table {%
			0.15 0.939999938011169
			0.15 1
		};
		\addplot [line width=1.08pt, color4, forget plot]
		table {%
			0.1 0.939999938011169
			0.2 0.939999938011169
		};
		\addplot [line width=1.08pt, color4, forget plot]
		table {%
			0.1 1
			0.2 1
		};
		\addplot [line width=1.08pt, color4, forget plot]
		table {%
			1.15 0.975000023841858
			1.15 0.996874988079071
		};
		\addplot [line width=1.08pt, color4, forget plot]
		table {%
			1.1 0.975000023841858
			1.2 0.975000023841858
		};
		\addplot [line width=1.08pt, color4, forget plot]
		table {%
			1.1 0.996874988079071
			1.2 0.996874988079071
		};
		\addplot [line width=1.08pt, color4, forget plot]
		table {%
			2.15 0.971428513526917
			2.15 0.992857098579407
		};
		\addplot [line width=1.08pt, color4, forget plot]
		table {%
			2.1 0.971428513526917
			2.2 0.971428513526917
		};
		\addplot [line width=1.08pt, color4, forget plot]
		table {%
			2.1 0.992857098579407
			2.2 0.992857098579407
		};
		\addplot [line width=1.08pt, color4, forget plot]
		table {%
			3.15 0.97857141494751
			3.15 0.996428549289703
		};
		\addplot [line width=1.08pt, color4, forget plot]
		table {%
			3.1 0.97857141494751
			3.2 0.97857141494751
		};
		\addplot [line width=1.08pt, color4, forget plot]
		table {%
			3.1 0.996428549289703
			3.2 0.996428549289703
		};
		\addplot [line width=1.08pt, color4, forget plot]
		table {%
			4.15 0.958518505096436
			4.15 0.985222241282463
		};
		\addplot [line width=1.08pt, color4, forget plot]
		table {%
			4.1 0.958518505096436
			4.2 0.958518505096436
		};
		\addplot [line width=1.08pt, color4, forget plot]
		table {%
			4.1 0.985222241282463
			4.2 0.985222241282463
		};
	\end{axis}
	
\end{tikzpicture}}}
	\caption{Performance of \algoname on the real-world \review{semantic segmentation experiment with simulated vehicular mobility.}
		\remove{dataset \texttt{ApolloScape} concerning the benchmarks}}
	\label{fig:apolloscape}
\end{figure*}

\remove{For the experiments,
a resolution-downsampled version of \texttt{ApolloScape} is split assigning to the \clients a single record session among those available to reflect the fact that data collected by \clients are correlated in time and space. }
					%!TEX root = ../main.tex

\subsubsection{Training and \algoname settings}
\label{sec:exp-nn-params}

%\LB{Not sure we want this section, but I wasn't sure where to put learning and param details. I don't think it hurts tho. We may also move this to arxiv.}

We train the deep network \texttt{deeplabv3}~\cite{chen2017rethinking},
a state-of-the-art model for semantic image segmentation,
choosing \texttt{mobilenetv3-large}~\cite{howard2019searching} as a backbone.
The initial local learning rate is set to ${\rm lr} = 2.5\times 10^{-4}$ with a cosine annealing scheduler and the local optimizer is \ac{SGD}.
We train the model with minibatches of size $32$,
and the \clients run $s_v = 3$ local steps of \ac{SGD} per time slot
(measured on an NVIDIA GeForce RTX 2080 GPU).

\review{We 
	\remove{set the fairness regularization coefficient to $w_A=0.01$ and} 
	modify the \algoname parameters to reflect the different nature of the problem;
	see the second column of~\autoref{table-simulation-params}.
	\remove{setting $C=1000$, $\rho_1=1$, $\rho_2=0.02$, and $\wtx=0.9$. }
	These choices \remove{have the effects of} increase the local \ac{SGD} steps and induce \clients to wait for the best transmission window in the given time.}
\remove{The total learning horizon lasts $50$ minutes and each round has a deadline of two minutes during which scheduled vehicles must update their local models and upload them to the server.}
					%!TEX root = ../main.tex

\subsubsection{Simulated mobility}
\label{sec:exp-nn-sumo}

\begin{table}
	\caption{\review{Performance of \algoname \vs the benchmarks in the real-world semantic segmentation experiment with simulated mobility w.r.t. the performance metrics addressed through the cost function~\eqref{eq:cost-function}.
			For the transmission slots we report the mean value with a $95\%$ confidence interval.}}
	\label{table:performance-metrics-apollo-sumo}
	\begin{center}
		\footnotesize
		\review{\begin{tabular}{lccc}
			\toprule
			&\ac{mIoU} & Learning rounds & Tx slots\\
			\midrule
			\textbf{\algoname}	&	$\bm{0.392}$ & $\bm{49}$  &	$\bm{6.74 \pm 0.05}$ \\
			Fairness~\cite{GunduzComputationTime} &	$0.332$	& $24$	&	$15.0 \pm 1.5$ \\
            FedAvg~\cite{mcmahan2017} &	$0.346$ & $26$	&	$15.2 \pm 1.8$ \\
            Round robin & $0.343$	& $25$	&	$17.5 \pm 2.0$ \\
            Centr-SNR~\cite{chen2024efficient} &	$0.347$  &	$29$	&	$9.04 \pm 0.65$ \\
			\bottomrule
		\end{tabular}}
	\end{center}
\end{table}

In \autoref{fig:n_clients}, the \ac{mIoU} score of \algoname is plotted against the learning round as the number of scheduled clients $\vsetsccard{}$ varies. 
Selecting a few clients, 
\eg $\vsetsccard{}=5$, 
the system performs many rounds within the simulation horizon (one hour) because each round is completed in short time.
However, the learning quality is poor as witnessed by frequent spikes because a low number of clients is not representative of the full dataset,
which may vary a lot across rounds. 
On the other hand, increasing the number of clients results in smoother learning curves. \remove{, the learning quality stabilizes as seen by.}
However,
for \eg $\vsetsccard{}=25$,
we trade enhanced learning stability for a longer round duration because scheduling more clients means increasing the chance that at least one of them \remove{is as a straggler,} 
has a poor channel and takes a long time to transmit the parameters. \remove{, and this results in a reduced latency efficiency.}
The zoomed area in~\autoref{fig:n_clients} reveals that the learning curves stabilize for $\vsetsccard{} \ge 15$. \remove{, which provides a better latency behavior if compared to using $\vsetsccard{}>15$.} 
For the following results, we set $\vsetsccard{}=15$ as the best tradeoff between learning quality and latency. 

\Cref{fig:scheduling-apollo} compares \algoname with the benchmark scheduling algorithms.
There is no significant difference w.r.t. the learning quality 
and even ``Fairness'' does not provide advantages concerning ``FedAvg''. \remove{ in this setting}
However, using \algoname,
the server aggregates the parameters at a double rate than the benchmarks. 
Leveraging the \ac{REM},
the scheduler chooses the fastest clients to transmit their local model weights, and the system performs $49$ learning rounds instead of the $24$-$26$ performed by the benchmarks within the same time horizon. 
This directly translates to a higher \ac{mIoU} for the same training time -- about $9.3\%$ higher than the best benchmark strategy.
\review{The detailed comparison with respect to the three performance metrics addressed in \cref{problem:co-design} is provided in \autoref{table:performance-metrics-apollo-sumo}.}

Clients that \review{overestimate the channel quality may fail to} send their model parameters update within the learning round deadline,
wasting computation and communication resources. 
To assess how \algoname prevents resources from being wasted, 
we evaluate the fraction of \review{scheduled} clients uploading the model to the server within the deadline.
\review{Formally,
we compute
\begin{equation}
	\txrate \doteq \dfrac{1}{T}\sum_{t\in\rounds}\dfrac{\left|\{v\in\vsetsch{t} : v \ \mbox{\small uploads update in round }t\}\right|}{\vsetsccard{t}}.
\end{equation}
\Cref{fig:client-fraction} shows $\txrate$ as the number of scheduled clients $\vsetsccard{}$ varies.}
Compared to \review{other algorithms}, the channel quality maps allow \review{\algoname and ``Centr-SNR''} to select only those clients who send their weights in time. This result is consistent since the variance is relatively small. \review{\algoname performs slightly better than ``Centr-SNR'' as the channel quality is evaluated locally at the vehicles estimating the location at transmission time.}
On the other hand, a drop in the share of clients transmitting their weights is observed without REM availability, with a worst-case scenario of $85\%$ in the simulated settings. 
The average value of the benchmark algorithms is around $94\%$, while for \algoname is close to $100\%$.
This demonstrates that not only \algoname outperforms the benchmark strategies in terms of learning performance, 
but it also does so with a more economical and efficient usage of computing resources at \clients and communication resources at the network edge.
					\review{%!TEX root = ../main.tex

\subsubsection{Real-world mobility}
\label{sec:exp-nn-taxi}

We now apply \algoname to the dataset roma/taxi~\cite{crawdad-roma-taxi} that gathers traces of taxi trips recorded in Rome, Italy, 
between February and March 2014.
For training,
we selected the recordings from 7:30 pm to 8:30 pm on February 15,
comprising $92$ different taxis in total.
In the original recordings,
each taxi independently and asynchronously transmitted its real-time location every $15~\si{\second}$.
We linearly interpolate the recorded traces and obtain synchronous trajectories at $1~\si{\hertz}$ to allocate resources in a granular fashion. % and thus more effective.
The \acs{REM} varies a little at a few meter distance,
hence we expect the bitrate at interpolated locations to be very similar to the one the taxis would experience along the true routes.

In \autoref{fig:scheduling-miou-taxi},
the mIoU of \algoname with two values of the fairness weight $w_A$ is compared with the benchmarks. 
Setting $w_A=10^{-2}$ allows us to combine the \ac{REM} with fairness information; cf.~\eqref{eq:priority-score}. 
However, since at least $11\%$ of the vehicles are scheduled in each round,
fairness has a marginal impact because each \client is frequently selected in any case.
\remove{all vehicles will likely be scheduled at some point}
The effect on learning is thus negligible,
and \algoname with $w_A=10^{-2}$ performs only $1$ to $3$ learning rounds more than the benchmarks within the given time horizon.
On the other hand, 
setting $w_A=0$ means scheduling the vehicles based on the sole \ac{REM} and mobility information,
and \algoname performs about $40$ learning rounds more than the benchmarks, reaching a mIoU score of $0.43$ as opposed to $0.37$ achieved by ``Centr-SNR'' and \algoname with $w_A=10^{-2}$.
\Cref{fig:bandwidth-taxi} shows the number of communications slots used by the vehicles to upload the local models (left) and the fraction of successful uploads $\txrate$ (right). 
\algoname with $w_A=0$ significantly outperforms the benchmarks, 
reducing the channel usage by $45\%$ concerning the best benchmark ``Centr-SNR''.
This highlights the need for online, distributed, and predictive scheduling in vehicular contexts as the channel quality experienced both across learning rounds and during each round can vary a lot depending on the vehicle speed and the environment geometry.
\algoname with $w_A=10^{-2}$ is still better than ``FedAvg'' and ``Fairness'' because it incorporates channel quality information to schedule vehicles. 
The successful update rate $\txrate$ confirms the trends just discussed.
 \algoname with $w_A=0$ ensures almost total participation of the scheduled clients ($99.9\%$), although the centralized approximation ``Centr-SNR'' performs close ($99\%$),
 suggesting that in this case the bitrate at the beginning of a round is a good proxy for that experienced during the round.

\begin{figure}
	\centering
	\resizebox{0.9\columnwidth}{!}{%!TEX root = ../main.tex

\begin{tikzpicture}[spy using outlines={magnification=2, rectangle, width=3.6cm, height=2cm, white!40!black, connect spies, }]
	
	\definecolor{color0}{rgb}{0.298039215686275,0.447058823529412,0.690196078431373}
	\definecolor{color1}{rgb}{0.866666666666667,0.517647058823529,0.32156862745098}
	\definecolor{color2}{rgb}{0.333333333333333,0.658823529411765,0.407843137254902}
	\definecolor{color3}{rgb}{0.505882352941176,0.447058823529412,0.701960784313725}
	\definecolor{lightgray204}{RGB}{204,204,204}
	
	\begin{axis}[
		legend cell align={left},
		legend pos = south east,
		legend columns = 2,
		legend style={fill opacity=0.8, draw opacity=1, text opacity=1, draw=lightgray204, font=\footnotesize, /tikz/every even column/.append style={column sep=4pt}},
		tick align=outside,
		tick pos=left,
		x grid style={white!69.0196078431373!black},
		xlabel={Learning round},
		xmajorgrids,
		xmin=-0.7, xmax=76.7,
		xtick style={color=black},
		y grid style={white!69.0196078431373!black},
		ylabel={\ac{mIoU}},
		ymajorgrids,
		ymin=0, ymax=0.44,
		ytick style={color=black}
		]
		\addplot [very thick, color0, mark=triangle*, mark size=3, mark options={solid, mark indices=36}, forget plot]
		table {%
			0 0.03
			1 0.14327897131443
			2 0.17454332113266
			3 0.227302446961403
			4 0.237107500433922
			5 0.245382875204086
			6 0.263785809278488
			7 0.266655832529068
			8 0.268303126096725
			9 0.27655303478241
			10 0.288361757993698
			11 0.294461607933044
			12 0.298503637313843
			13 0.308114439249039
			14 0.30730876326561
			15 0.310398012399673
			16 0.314249604940414
			17 0.316494017839432
			18 0.321610748767853
			19 0.323528468608856
			20 0.309841364622116
			21 0.328306317329407
			22 0.331828981637955
			23 0.324638426303864
			24 0.332220286130905
			25 0.331444323062897
			26 0.347709268331528
			27 0.340571880340576
			28 0.342779338359833
			29 0.348667085170746
			30 0.350449174642563
			31 0.358395010232925
			32 0.348719924688339
			33 0.347127288579941
			34 0.368598878383636
			35 0.371305882930756
		};
		\addlegendimage{only marks, very thick, color0, mark=triangle*, mark options={mark indices=3}, mark size=3}
		\addlegendentry{\textbf{\algoname} \tiny{$\left(w_A=10^{-2}\right)$}}
		\addplot [very thick, color0, mark=triangle, mark size=3, mark options={solid, mark indices=76}, forget plot, dashed]
		table {%
            0 0.03
			1 0.141349270939827
            2 0.176610425114632
            3 0.18967704474926
            4 0.227370008826256
            5 0.253175377845764
            6 0.272819012403488
            7 0.269366234540939
            8 0.290253639221191
            9 0.294719308614731
            10 0.301822453737259
            11 0.310615956783295
            12 0.313522666692734
            13 0.323908865451813
            14 0.324721723794937
            15 0.319760739803314
            16 0.318309724330902
            17 0.313535362482071
            18 0.325064659118652
            19 0.336371034383774
            20 0.328361570835114
            21 0.334523379802704
            22 0.332773804664612
            23 0.329517275094986
            24 0.328268378973007
            25 0.317648738622665
            26 0.332189083099365
            27 0.35318785905838
            28 0.358740955591202
            29 0.357396334409714
            30 0.358019888401031
            31 0.360065519809723
            32 0.363054662942886
            33 0.358558177947998
            34 0.366617739200592
            35 0.35286283493042
            36 0.364190280437469
            37 0.359382420778274
            38 0.372804820537567
            39 0.369237273931503
            40 0.378029733896255
            41 0.377679824829102
            42 0.382879704236984
            43 0.375162601470947
            44 0.365785866975784
            45 0.377929538488388
            46 0.365324914455414
            47 0.387371003627777
            48 0.397583186626434
            49 0.389923274517059
            50 0.394752353429794
            51 0.390387862920761
            52 0.408660113811493
            53 0.389760613441467
            54 0.400239706039429
            55 0.380074232816696
            56 0.402066290378571
            57 0.406554698944092
            58 0.416132271289825
            59 0.405661791563034
            60 0.407948821783066
            61 0.406927287578583
            62 0.404235035181045
            63 0.39434227347374
            64 0.400612592697144
            65 0.411434084177017
            66 0.402245372533798
            67 0.399100244045258
            68 0.401731789112091
            69 0.406154155731201
            70 0.413809269666672
            71 0.41141203045845
            72 0.407216668128967
            73 0.422716706991196
            74 0.407693713903427
            75 0.393300116062164
		};
		\addlegendimage{only marks, very thick, color0, mark=triangle, mark options={mark indices=3}, mark size=3}
		\addlegendentry{\textbf{\algoname} \tiny{$\left(w_A=0\right)$}}
		\addplot [very thick, color1, mark=*, mark size=3, mark options={solid, mark indices=33}, forget plot]
		table {%
			0 0.03
			1 0.134173139929771
			2 0.183690547943115
			3 0.215337410569191
			4 0.219377920031548
			5 0.251424938440323
			6 0.256155580282211
			7 0.269906252622604
			8 0.276302814483643
			9 0.28869754076004
			10 0.284239441156387
			11 0.304697424173355
			12 0.310061991214752
			13 0.301615536212921
			14 0.303818315267563
			15 0.305402785539627
			16 0.317423850297928
			17 0.304470479488373
			18 0.321174532175064
			19 0.320813119411469
			20 0.334650039672852
			21 0.334320813417435
			22 0.334358811378479
			23 0.326058834791184
			24 0.333071619272232
			25 0.338207304477692
			26 0.327449858188629
			27 0.340949714183807
			28 0.349459379911423
			29 0.353758275508881
			30 0.34489244222641
			31 0.35328134894371
			32 0.343952178955078
		};
		\addlegendimage{only marks, very thick, color1, mark=*, mark options={mark indices=3}, mark size=3}
		\addlegendentry{Fairness\review{~\cite{GunduzComputationTime}}}
		\addplot [very thick, color2, mark=diamond*, mark size=3, mark options={solid, mark indices=34}, forget plot]
		table {%
			0 0.03
			1 0.129039838910103
			2 0.157823234796524
			3 0.211724430322647
			4 0.22698001563549
			5 0.250028342008591
			6 0.243565231561661
			7 0.255727529525757
			8 0.267326086759567
			9 0.264894127845764
			10 0.269560158252716
			11 0.291276544332504
			12 0.289685934782028
			13 0.302337646484375
			14 0.296379178762436
			15 0.30966779589653
			16 0.323823690414429
			17 0.313963234424591
			18 0.318092077970505
			19 0.326196700334549
			20 0.326225429773331
			21 0.323608726263046
			22 0.326476156711578
			23 0.338484138250351
			24 0.337130039930344
			25 0.347513824701309
			26 0.34120574593544
			27 0.343838751316071
			28 0.340461105108261
			29 0.356540113687515
			30 0.368121981620789
			31 0.346606910228729
			32 0.363142758607864
			33 0.354771107435226
		};
		\addlegendimage{only marks, very thick, color2, mark=diamond*, mark options={mark indices=3}, mark size=3}
		\addlegendentry{\review{FedAvg~\cite{mcmahan2017}}}
		\addplot [very thick, color3, mark=square*, mark size=2.25, mark options={solid, mark indices=35}, forget plot]
		table {%
			0 0.03
			1 0.144910916686058
			2 0.194422796368599
			3 0.240432158112526
			4 0.25055381655693
			5 0.241214632987976
			6 0.271172612905502
			7 0.284424990415573
			8 0.276473969221115
			9 0.292396366596222
			10 0.297811716794968
			11 0.301283061504364
			12 0.311779409646988
			13 0.309644311666489
			14 0.303609758615494
			15 0.324799597263336
			16 0.317683905363083
			17 0.309526771306992
			18 0.316799014806747
			19 0.319767206907272
			20 0.325682073831558
			21 0.335876196622849
			22 0.325720012187958
			23 0.327781051397324
			24 0.333392024040222
			25 0.328228414058685
			26 0.334472596645355
			27 0.338474124670029
			28 0.348538428544998
			29 0.350011825561523
			30 0.349505364894867
			31 0.351121574640274
			32 0.362425237894058
			33 0.355826556682587
			34 0.369874447584152
		};
		\addlegendimage{only marks, very thick, color3, mark=square*, mark options={mark indices=3}, mark size=2.25}
		\addlegendentry{\review{Centr-SNR}~\cite{chen2024efficient}}
		\coordinate (c1) at (axis cs: 28, 0.35);
		\coordinate (c2) at (axis cs: 40, 0.22);
		\spy on (c1) in node[fill=white] at (c2);
	\end{axis}
	
\end{tikzpicture}}
	\caption{\review{Learning performance (\ac{mIoU} score) with VREM-FL \vs
scheduling benchmarks on the real-world semantic segmentation experiment with real-world vehicular mobility.
        \remove{Performance comparison of \algoname with the benchmarks as per mIoU.} \algoname ($w_A=0$) outperforms the benchmarks as it uses the \ac{REM} information to effectively schedule fastest transmitting vehicles. In this way, it performs more learning rounds in the same time horizon.}
	}
	\label{fig:scheduling-miou-taxi}
\end{figure}
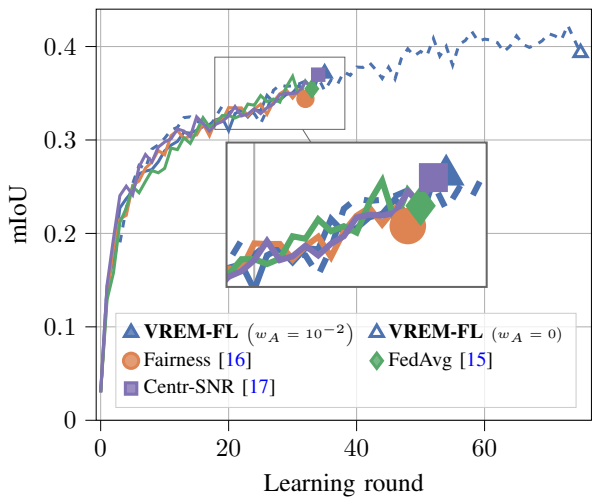

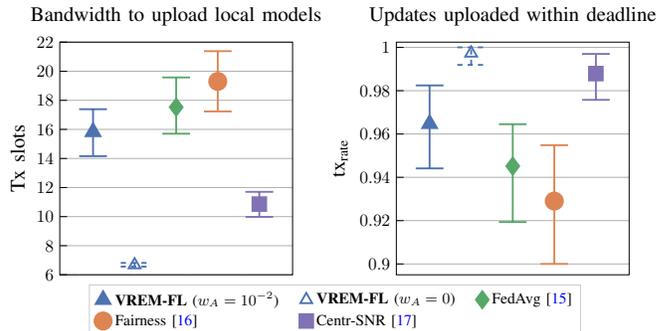
\begin{figure}
	\centering
	\resizebox{\columnwidth}{!}{%!TEX root = ../main.tex

\begin{tikzpicture}
	
	\definecolor{color0}{rgb}{0.298039215686275,0.447058823529412,0.690196078431373}
	\definecolor{color1}{rgb}{0.866666666666667,0.517647058823529,0.32156862745098}
	\definecolor{color2}{rgb}{0.333333333333333,0.658823529411765,0.407843137254902}
    \definecolor{color3}{rgb}{0.505882352941176,0.447058823529412,0.701960784313725}
	\newlength\figureheight
	\newlength\figurewidth
	\setlength\figureheight{.35\textwidth}
	\setlength\figurewidth{.35\textwidth}
	
	\begin{axis}[%
		name=bandwidth,
		width=\figurewidth,
		height=\figureheight,
		title={Bandwidth to upload local models},
		title style={font=\large},
		tick align=inside,
		tick pos=left,
		unbounded coords=jump,
		y grid style={white!69.0196078431373!black},
		ylabel={Tx slots},
		ylabel style={font=\large},
		ymajorgrids,
		ytick style={color=black},
		xtick style={color=white},
		xtick={},
		xticklabels={},
        ymin=6,
        ymax=22,
        ytick distance=2
		]
		\addplot [line width=1.08pt, color0, mark=triangle*, mark size=5, mark options={solid}, only marks]
		table {%
			-0.15 15.8260869565217
		};
		\addplot [line width=1.08pt, color0, forget plot]
		table {%
			-0.2 14.1592753623188
			-0.1 14.1592753623188	
			nan nan
			-0.15 14.1592753623188
			-0.15 17.388768115942
			nan nan
			-0.2 17.388768115942
			-0.1 17.388768115942
		};
		\addplot [line width=1.08pt, color0, mark=triangle, mark options={solid}, mark size=4, only marks]
		table {%
			0 6.67919445037842
		};
		\addplot [line width=1.08pt, color0, forget plot, dashed]
		table {%
			-0.05 6.56372481584549
			0.05 6.56372481584549
			nan nan
			0 6.56372481584549
			0 6.81748312711716
			nan nan
			-0.05 6.81748312711716
			0.05 6.81748312711716
		};
		\addplot [line width=1.08pt, color2, mark=diamond*, mark size=5, mark options={solid}, only marks]
		table {%
			0.15 17.5273311897106
		};
		\addplot [line width=1.08pt, color2, forget plot]
		table {%
			0.1 15.7036977491961
			0.2 15.7036977491961
			nan nan
			0.15 15.7036977491961
			0.15 19.5725080385852
			nan nan
			0.1 19.5725080385852
			0.2 19.5725080385852
		};
        \addplot [line width=1.08pt, color1, mark=*, mark size=5, mark options={solid}, only marks]
		table {%
			0.3 19.3006535947712
		};
		\addplot [line width=1.08pt, color1, forget plot]
		table {%
			0.25 17.2352941176471
			0.35 17.2352941176471
			nan nan
			0.3 17.2352941176471
			0.3 21.3791666666667
			nan nan
			0.25 21.3791666666667
			0.35 21.3791666666667
		};
		\addplot [line width=1.08pt, color3, mark=square*, mark size=4, mark options={solid}, only marks]
		table {%
			0.45 10.8626861572266
		};
		\addplot [line width=1.08pt, color3, forget plot]
		table {%
			0.4 9.98186609745026
			0.5 9.98186609745026
			nan nan
			0.45 9.98186609745026
			0.45 11.7044773101807
			nan nan
			0.4 11.7044773101807
			0.5 11.7044773101807
		};
		\coordinate (left) at (rel axis cs:2.6,-.15);
	\end{axis}%
	
	\hfill
	
	\begin{axis}[%
		name=clients,
		at=(bandwidth.right of south east), anchor=left of south west,
		width=\figurewidth,
		height=\figureheight,
		title={Updates uploaded within deadline},
		title style={font=\large},
		legend to name=leg,
		legend cell align={left},
		legend style={fill opacity=1, draw opacity=1, text opacity=1, draw=white!80!black, legend columns=4,
			/tikz/every even column/.append style={column sep=10pt}, font=\small},
        legend columns=3,
		tick align=inside,
		tick pos=left,
		unbounded coords=jump,
		y grid style={white!69.0196078431373!black},
		ymajorgrids,
		ylabel={$\txrate$},
		ylabel style={font=\large},
		ytick style={color=black},
		xtick style={color=white},
		xtick={},
		xticklabels={},
        ymax=1.0024,
        ymin=0.895,
        ytick distance=0.02
		]
		\addplot [line width=1.08pt, color0, mark=triangle*, mark size=5, mark options={solid}, only marks]
		table {%
			0.85 0.964705877444323
		};
		\addlegendentry{\textbf{\algoname} ($w_A=10^{-2}$)}
		\addplot [line width=1.08pt, color0, forget plot]
		table {%
			0.8 0.944117637241588
			0.9 0.944117637241588
			nan nan
			0.85 0.944117637241588
			0.85 0.982426469816881
			nan nan
			0.8 0.982426469816881
			0.9 0.982426469816881
		};
		\addplot [line width=1.08pt, color0, mark=triangle, mark options={solid}, mark size=4, only marks]
		table {%
			1 0.997297346591949
		};
		\addlegendentry{\textbf{\algoname} ($w_A=0$)}
		\addplot [line width=1.08pt, color0, forget plot, dashed]
		table {%
			.95 0.991891801357269
			1.05 0.991891801357269
			nan nan
			1 0.991891801357269
			1 1
			nan nan
			.95 1
			1.05 1
		};
		\addplot [line width=1.08pt, color2, mark=diamond*, mark size=5, mark options={solid}, only marks]
		table {%
			1.15 0.945161283016205
		};
        \addlegendentry{FedAvg~\cite{mcmahan2017}}
		\addplot [line width=1.08pt, color2, forget plot]
		table {%
			1.1 0.919354831018755
			1.2 0.919354831018755
			nan nan		
			1.15 0.919354831018755
			1.15 0.964516126340435
			nan nan
			1.1 0.964516126340435
			1.2 0.964516126340435
		};
        \addplot [line width=1.08pt, color1, mark=*, mark size=5, mark options={solid}, only marks]
		table {%
			1.3 0.92903225075814
		};\addlegendentry{Fairness~\cite{GunduzComputationTime}}
		\addplot [line width=1.08pt, color1, forget plot]
		table {%
			1.25 0.899999995385447
			1.35 0.899999995385447
			nan nan
			1.3 0.899999995385447
			1.3 0.954838706601051
			nan nan
			1.25 0.954838706601051
			1.35 0.954838706601051
		};
		\addplot [line width=1.08pt, color3, mark=square*, mark size=4, mark options={solid}, only marks]
		table {%
			1.45 0.987878739833832
		};
		\addlegendentry{Centr-SNR~\cite{chen2024efficient}}
		\addplot [line width=1.08pt, color3, forget plot]
		table {%
			1.4 0.975757598876953
			1.5 0.975757598876953
			nan nan		
			1.45 0.975757598876953
			1.45 0.996969759464264
			nan nan
			1.4 0.996969759464264
			1.5 0.996969759464264
		};
		\coordinate (right) at (rel axis cs:0,-.15);
	\end{axis}
	
	\path (left)--(right) coordinate[midway] (center);
	\node[centered] at(center -| current bounding box) {\pgfplotslegendfromname{leg}};
	
\end{tikzpicture}}
	\caption{\review{Resource usage comparison between \algoname and scheduling benchmarks on the semantic segmentation experiment with real-world mobility.
			\algoname ($w_A=0$) reduces the bandwidth for transmissions of updates (left) and increases the fraction of scheduled \clients that upload the updates within the deadline (right), significantly improving resource efficiency.
		}
	}
	\label{fig:bandwidth-taxi}
\end{figure}}
	%!TEX root = ../main.tex

\section{Conclusion and future work}
\label{sec:conclusion}

Motivated by the need of efficient solutions for \acs{FL} tasks in vehicular networks,
we have proposed \algoname, a \codesign algorithm that jointly optimizes a learning-related performance metric and network-related communication resources.
Specifically, \algoname orchestrates local computations at the vehicles, 
transmission of their local models to the edge server, and schedules clients at each learning iteration to strike a good balance between learning accuracy,
training time, and wireless channel usage. Experimental results on a synthetic LS problem and on a real-world semantic segmentation task demonstrate that \algoname provides superior learning performance as compared to common scheduling strategies
by promoting a frugal use of computation and communication resources.

The present study focuses on the \acs{FL} algorithm, client mobility is assumed to be given and non-controllable. However, future smart and autonomous vehicles may be in the position of changing their planned route to favor ancillary tasks, such as the execution of an \acs{FL} algorithm or the transmission of data to roadside servers. Hence, we foresee scenarios where decision-making can be augmented via trajectory steering of (some of) the participating vehicles, to optimize even further the vehicle learning performance and their resource utilization.

    \def\section#1{\tmpsection#1}
	
    \arxiv{}{
	    \appendices
        %!TEX root = ../main.tex

\section{Analytical Derivation of Local Proxy~\eqref{eq:local-proxy}}\label{app:proxy}
We choose the proxy $\Theta^v_t\lr H_t^v\rr$ based on the convergence behavior of deterministic gradient descent when applied to least squares problems. For completeness, we now illustrate the explicit derivation of this proxy. Let $\loss(\param) = \sum_{i = 1}^S(\param[\top] x_i - y_i)^2+\lambda\norm{\param}^2$ be a regularized quadratic cost function with $\lambda >0$ (see \autoref{sec:exp-ls-dataset}). We denote by $g(\param) = \nabla\loss(\param)\in\mathbb{R}^n$ and $\Lambda = \nabla^2\loss(\param)\in\mathbb{R}^{n\times n}$ the gradient and the (constant) Hessian matrix of the cost function, respectively. Let $\param[*]$ be the unique minimizer of $\loss(\param)$. In least squares, the gradient can be written as $g(\param) = \Lambda(\param-\param[*])$~\cite{dal2022shed}. Recall that, for a constant step size $\alpha>0$, the gradient descent update is $\param[t+1] = \param[t] - \alpha g(\param[t])$ starting from the initial parameter $\param[0]$. Therefore, we can write
\begin{equation}
    \begin{aligned}
        g(\param[t+1]) &= \Lambda\lr\param[t+1]-\param[*]\rr = \Lambda\lr\param[t] - \param[*]- \alpha g\lr\param[t]\rr\rr 
        \\& = g\lr\param[t]\rr -\alpha \Lambda g\lr\param[t]\rr = \left(I - \alpha\Lambda\right)g\lr\param[t]\rr.
    \end{aligned}
\end{equation}
Hence, the rate of convergence of $g(\param[t])$ to zero is dictated by the largest eigenvalue (in magnitude) of $\left(I - \alpha\Lambda\right)$ that, denoting the eigenvalues of $\Lambda$ by $\lambda_1\geq \cdots \geq \lambda_n>0$, is minimized by choosing $\alpha = \frac{2}{\lambda_1 + \lambda_n}$ (see, \eg the proof in~\cite[Theorem~1]{dal2022shed}).
This yields a convergence factor of 
$\frac{\lambda_1 - \lambda_n}{\lambda_1+ \lambda_n}\leq 1-\frac{\lambda_n}{\lambda_1} = 1-\frac{1}{\kappa}$,
%with $q_t = 0$ and $\hat{H}_t = \rho_tI = I$) 
where $\kappa = \frac{\lambda_1}{\lambda_n}$ is  the condition number of the problem. Consequently, we get
\begin{equation}
    \begin{aligned}
        \left\|g\lr\param[t+1]\rr\right\| \leq \left(1-\frac{1}{\kappa}\right)^{t+1}\left\|g\lr\param[0]\rr\right\|,
    \end{aligned}
\end{equation}
from which we derive the proxy~\eqref{eq:local-proxy}.
        %!TEX root = ../main.tex

\section{Additional Experiments on Linear Regression}\label{app:lambda}

\review{Here, for completeness, we include some results illustrating the performance of VREM-FL under different values of the regularization parameter $\lambda$ in~\eqref{eq:FL-problem}. In particular, we show results for $\lambda = 10^{-3}, 10^{-4}, 10^{-5}$. The results consistently show the superior performance of VREM-FL under all the considered configurations of the regularization parameter.}
\begin{figure}
\centering \includegraphics[width=\columnwidth]{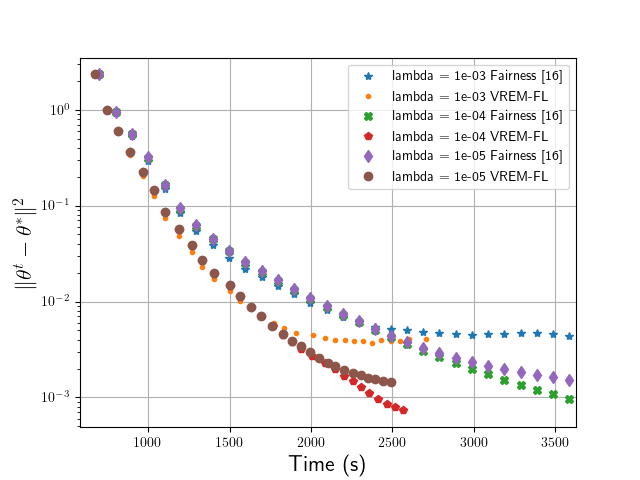}
\caption{\review{Comparison of VREM-FL against Fairness [16], for different values of the regularization parameter $\lambda$.}
}
\label{fig:lambda_1}
\end{figure}

\begin{figure}
\centering \includegraphics[width=\columnwidth]{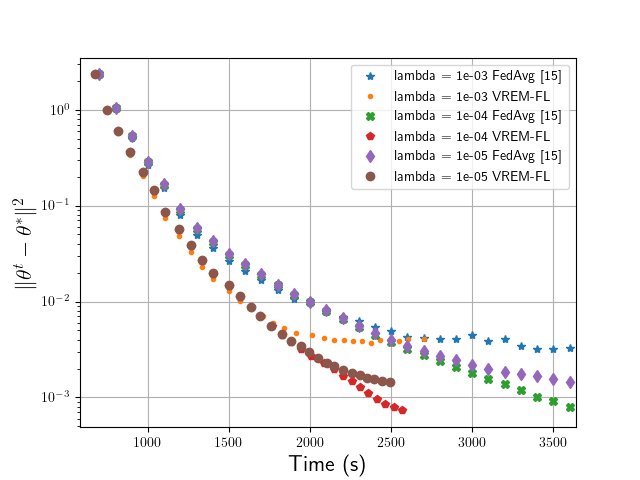}
\caption{\review{Comparison of VREM-FL against FedAvg [15], for different values of the regularization parameter $\lambda$.}
}
\label{fig:lambda_2}
\end{figure}

\begin{figure}
\centering \includegraphics[width=\columnwidth]{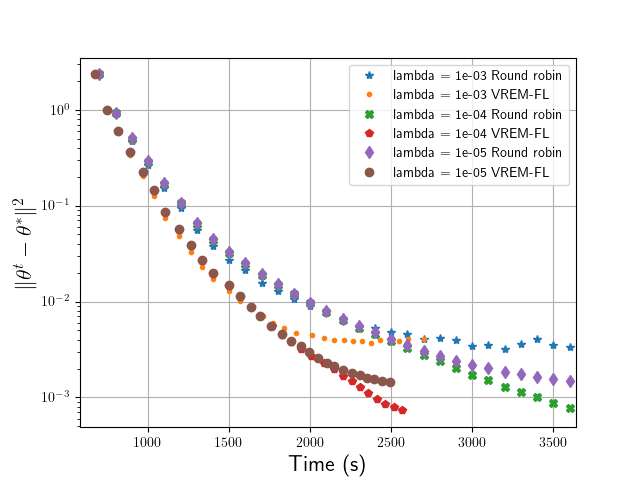}
\caption{\review{Comparison of VREM-FL against the round robin benchmark, for different values of the regularization parameter $\lambda$.}
}
\label{fig:lambda_3}
\end{figure}
    }
	
	\arxiv{
        \bibliographystyle{IEEEtran}
	    \bibliography{IEEEabrv,
            biblio
        }
    }{

	}
	
    \vskip -2\baselineskip plus -1fil
	%!TEX root = ../main.tex

\begin{IEEEbiography}[{\includegraphics[width=1in,height=1.25in,clip,keepaspectratio]{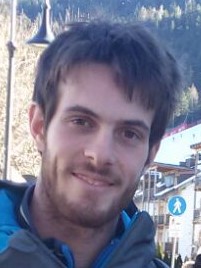}}]{Luca Ballotta}
	received the Master's degree in Automation Engineering and the Ph.D. degree in Information Engineering from the University of Padova, Italy, in 2019 and 2023, respectively.
He is currently a postdoctoral researcher at the Delft University of Technology, Delft Center for Systems and Control (DCSC).
He was Visiting Student at the Massachusetts Institute of Technology in 2020 and 2022.
He was awarded with the Young Author Prize at the 2020 IFAC World Congress and was finalist of the 2024 EECI PhD Award.
His research interests include control over networks under resource constraints, 
resilient consensus and distributed learning,
and safe control.

\end{IEEEbiography}
\vskip -2\baselineskip plus -1fil
\begin{IEEEbiography}[{\includegraphics[width=1in,height=1.25in,clip,keepaspectratio]{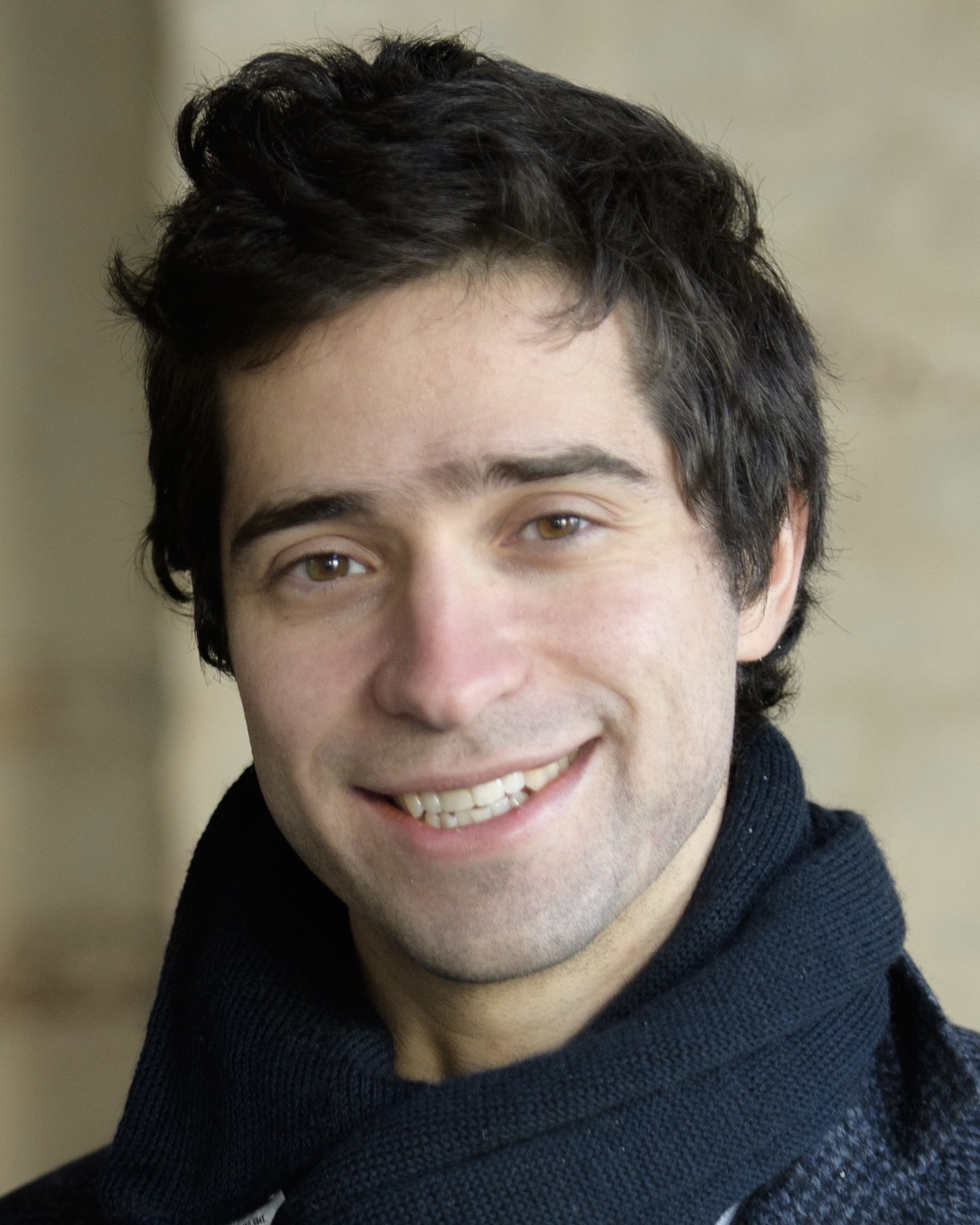}}]{Nicol\`o Dal Fabbro}
	obtained his Master's degree in Telecommunications Engineering and his 
PhD degree in Information Engineering from the University of Padova, Italy, in 
2020 and 
2023
, respectively
. 
He is currently a postdoctoral researcher at the University of Pennsylvania, USA, where he is with the department of Electrical and Systems Engineering. He is the recepient of the GTTI (italian telecommunications and information technology group) best 2024 PhD thesis award and of one best dataset award from the IEEE. His research interests are in federated learning, multi-agent reinforcement learning and wireless sensing.
\end{IEEEbiography}
\vskip -2\baselineskip plus -1fil
\begin{IEEEbiography}[{\includegraphics[width=1in,height=1.25in,clip,keepaspectratio]{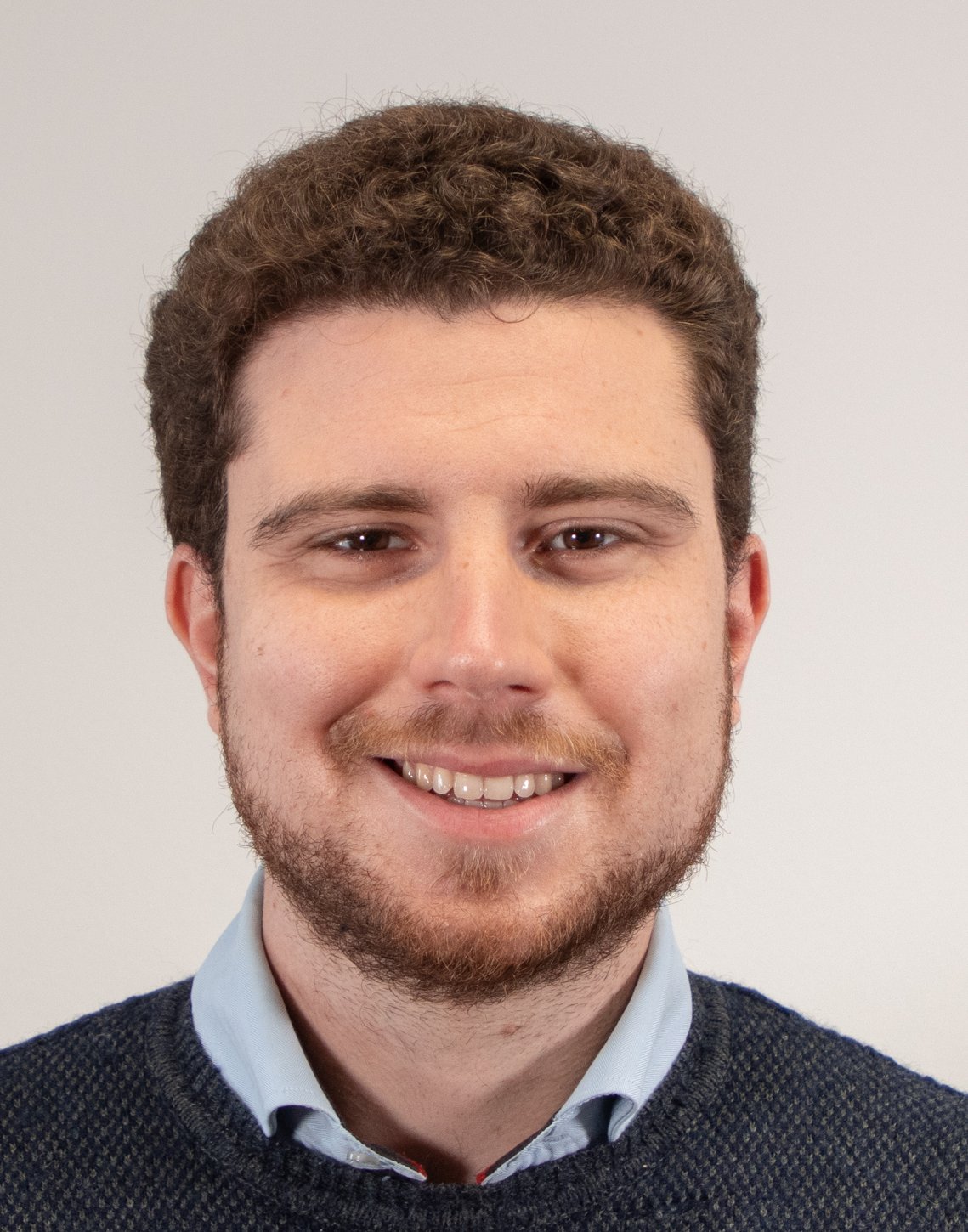}}]{Giovanni Perin}
	(Member, IEEE) received the M.Sc. degree (summa cum laude) in ICT for Internet and Multimedia and the Ph.D. degree in Information Engineering from the University of Padova, Italy, in 2019 and 2023, respectively, where he is currently a postdoc research fellow. In 2019, he was with the Deutsche Telekom Chair of Communication Networks, Technical University of Dresden, Germany, working on broadcast routing, while in 2022, he was a visiting scholar with the University of California at Irvine, USA, researching vehicular communications and edge computing. His research focuses on sustainable edge computing, distributed optimization and processing, and federated learning.
\end{IEEEbiography}
\vskip -2\baselineskip plus -1fil
\begin{IEEEbiography}[{\includegraphics[width=1in,height=1.25in,clip,keepaspectratio]{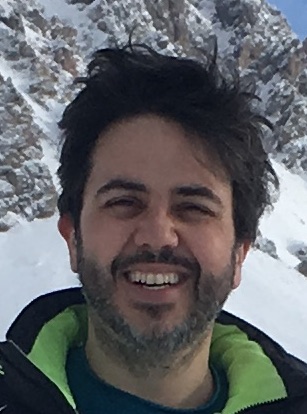}}]{Luca Schenato}
	(Fellow, IEEE) received the Dr. Eng. degree in electrical engineering from the University of Padova in 1999 and the Ph.D. degree in Electrical Engineering and Computer Sciences from the U.C. Berkeley, in 2003. He held a post-doctoral position in 2004 and a visiting professor position in 2013-2014 at U.C. Berkeley. Currently, he is a Full Professor at the Information Engineering Department at the University of Padova. His interests include networked control systems, multi-agent systems, wireless sensor networks, smart grids, and cooperative robotics. Luca Schenato has been awarded the 2004 Researchers Mobility Fellowship by the Italian Ministry of Education, University and Research (MIUR), the 2006 Eli Jury Award in U.C. Berkeley and the EUCA European Control Award in 2014, and IEEE Fellow in 2017. He served as Associate Editor for IEEE Trans. on Automatic Control from 2010 to 2014. He is Senior Editor for IEEE Trans. on Control of Network Systems and Associate Editor for Automatica.

\end{IEEEbiography}
\vskip -2\baselineskip plus -1fil
\begin{IEEEbiography}[{\includegraphics[width=1in,height=1.25in,clip,keepaspectratio]{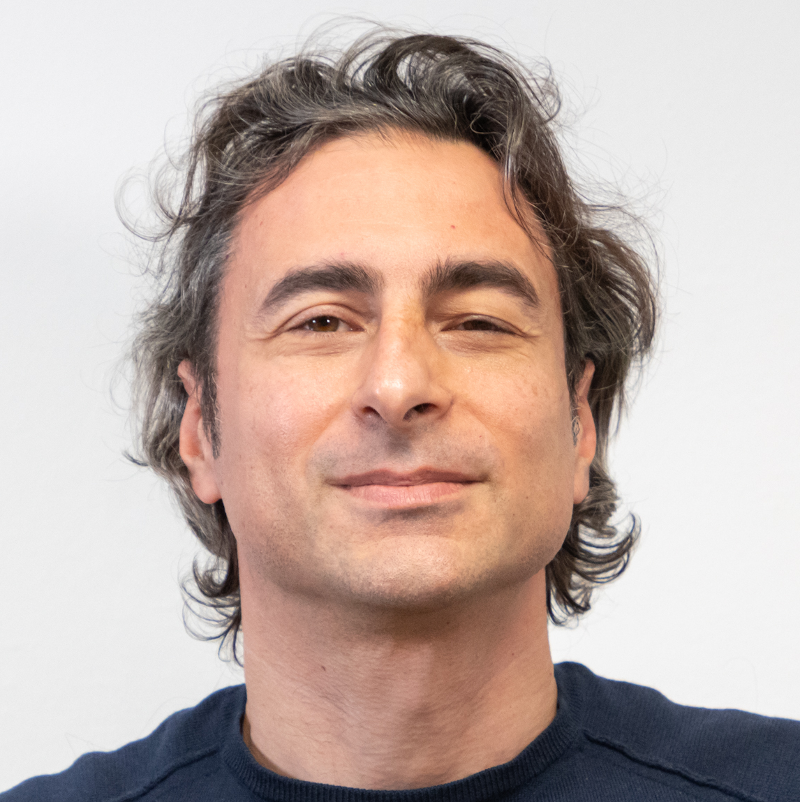}}]{Michele Rossi}
	(Senior Member, IEEE) is a full professor at the Department of Information Engineering (DEI), University of Padova, Italy. His research interests lie broadly in wireless sensing systems (joint communication and sensing for next generation wireless systems), green mobile networks (energy efficient operation and resource allocation), energy efficient machine and deep learning (lightweight architectures including spiking neural networks) for edge computing. Over the years, he has been involved in numerous European projects on wireless sensing and Internet of Things and has collaborated with major companies such as Ericsson, DOCOMO, Nokia, Samsung and INTEL. His research is currently supported by the European Commission through the projects GREENEDGE (GA no. 953775) on ``green edge computing for mobile networks'' (project coordinator) and ROBUST-6G on ``smart, automated and reliable security service platforms for 6G systems'' (GA no. 101139068). Prof. Rossi has been the recipient of six best paper awards and one best dataset award from the IEEE, and currently serves on the Editorial Boards of the IEEE Transactions on Mobile Computing.
\end{IEEEbiography}
\vskip -2\baselineskip plus -1fil
\begin{IEEEbiography}[{\includegraphics[width=1in,height=1.25in,clip,keepaspectratio]{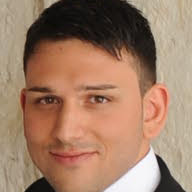}}]{Giuseppe Piro}
	(Member, IEEE) is a Full Professor in Telecommunications at the Polytechnic University of Bari (Italy) and co-responsible of the iTNT-NS laboratory developed in the context of the PNRR PE14 RESTART Partnership. His main research interests include mobile communication systems, integrated terrestrial and non-terrestrial networks, physical and network layer security, intent-based networking, Internet of Things, Software-Defined Networking, Information-Centric Networking, nano-scale communications, and network simulation tools. He is Principal Investigator for the Italian MIUR PRIN project INSPIRE and Local Investigator for the ISP5G+ project supported by the PNRR PE7 SERICS Partnership. He has been Local Investigator for the Italian MIUR PRIN project “Realtime Control of 5G Wireless Networks” and for two ESA projects SATIABLE and NB-IoT4Space. He has also been involved in EU H2020 projects (FANTASTIC-5G, BONVOYAGE, symbIoTe, GUARD) and Italian MIUR PON projects (Pico\&Pro, FURTHER, AGREED, RAFAEL). He founded 5G-air-simulator, LTE-Sim, and NANO-SIM open-source projects. He is regularly involved as a member of the TPC of many international conferences. He serves as Associate Editor for Internet Technology Letter (Wiley), Wireless Communications and Mobile Computing (Hindawi), and Sensors (MDPI).
\end{IEEEbiography}
	
\end{document}